\documentclass{emulateapj}
\usepackage{apjfonts}

\input psfig
\input colordvi

\newcommand{\etal}{et~al.\ }

\def\ts{\thinspace}
\def\gapprox{$_>\atop{^\sim}$} 
\def\lapprox{$_<\atop{^\sim}$}

\def\cl{\centerline}

\def\omit#1{}

\newdimen\sa  \def\sd{\sa=.1em  \ifmmode $\rlap{.}$''$\kern -\sa$
                                \else \rlap{.}$''$\kern -\sa\fi}

\newdimen\sb  \def\md{\sb=.04em \ifmmode $\rlap{.}$'$\kern -\sb$
                                \else \rlap{.}$'$\kern -\sb\fi}

\newdimen\sc  \def\degd{\sc=.04em \ifmmode $\rlap{.}$^\circ$\kern -\sc$
                                \else \rlap{.}$^\circ$\kern -\sc\fi}

\reversemarginpar

\begin{document}


\submitted{\vskip -15pt} 

\lefthead{The cD Halo of NGC 6166}

\righthead{Bender, Kormendy, Cornell, \& Fisher}

\centerline{\null}\vskip -140pt\centerline{\null}

\title{\kern -18pt The Cluster Velocity Dispersion of the Abell 2199 
                  \lowercase{c}D Halo of NGC 6166\altaffilmark{1}}

\author{
Ralf Bender\altaffilmark{2,3},
John Kormendy\altaffilmark{4,2,3},
Mark E. Cornell\altaffilmark{4,5}, and
David B.~Fisher\altaffilmark{4,6}
}

\altaffiltext{{\kern -5pt}1}{Based on observations obtained with the Hobby-Eberly Telescope, 
                 which is a joint project of the 
                 University of Texas at Austin, the Pennsylvania State University, Stanford University, 
                 Ludwig-Maximilians-Universit\"at M\"unchen, and Georg-August-Universit\"at G\"ottingen.
                 Submitted to ApJ.} 

\altaffiltext{2}{Max-Planck-Institut f\"ur Extraterrestrische Physik,
                 Giessenbachstrasse, D-85748 Garching-bei-M\"unchen, Germany; 
                 bender@mpe.mpg.de}

\altaffiltext{3}{Universit\"ats-Sternwarte, Scheinerstrasse 1,
                 M\"unchen D-81679, Germany}

\altaffiltext{4}{Department of Astronomy, University of Texas at Austin,
                 1 University Station C1400, Austin,
                 Texas 78712-0259; kormendy@astro.as.utexas.edu}

\altaffiltext{5}{Present{\ts}Address:{\ts}MIT{\ts}Lincoln{\ts}Laboratory,{\ts}ETS{\ts}Field{\ts}Site,{\ts}P.{\ts}O.{\ts}Box\ts1707,
                 Socorro, NM 87801; Mark.Cornell@ll.mit.edu}

\altaffiltext{6}{Present Address:{\ts}Centre for Astrophysics and Supercomputing, Swinburne University of 
                 Technology, Mail Stop H30, P.{\ts}O.{\ts}Box 218, Hawthorn, Victoria 3122, Australia; 
                 dfisher@astro.swin.edu.au}

\pretolerance=15000  \tolerance=15000

\begin{abstract} 

      Hobby-Eberly Telescope (HET) spectroscopy is used to measure the velocity dispersion profile of the
nearest prototypical cD galaxy, NGC 6166 in the cluster Abell 2199.  We also present composite surface 
photometry from many telescopes.  We confirm the defining feature of a cD galaxy; i.{\ts}e., a halo of 
stars that fills the cluster center and that is controlled dynamically by cluster gravity, not by the
central galaxy.  Our HET spectroscopy shows that the velocity dispersion of NGC 6166 rises from 
$\sigma \simeq 300$ km s$^{-1}$ in the inner $r \sim 10^{\prime\prime}$ to $\sigma = 865 \pm 58$ km s$^{-1}$ 
at $r$\ts$\sim$\ts100$^{\prime\prime}$ in the cD halo.  This extends published observations of an outward 
$\sigma$ increase and shows for the first time that $\sigma$ rises all the way to the cluster velocity 
dispersion of $819 \pm 32${\ts}km{\ts}s$^{-1}$.  We also observe that the main body of NGC 6166 moves at 
$+206 \pm 39$ km s$^{-1}$ with respect to the cluster mean velocity, whereas the velocity of the inner cD halo 
is  $\sim$\ts70 km s$^{-1}$ closer to the cluster velocity.  These results support our picture that cD halos 
consist of stars that were stripped from individual cluster galaxies by fast tidal encounters.  
\lineskip=-20pt \lineskiplimit=-20pt 

      However, our photometry does not confirm the widespread view that cD halos are identifiable~as~an~extra,
low-surface-brightness component that is photometrically distinct from the inner, steep-S\'ersic-function 
main body of an otherwise-normal giant elliptical galaxy.~Instead, all of the brightness profile of NGC\ts6166 
outside its core is described to $\pm$\ts0.037\ts$V${\ts}mag{\ts}arcsec$^{-2}$ by a single S\'ersic 
function~with~index $n \simeq 8.3$.  The cD halo is not recognizable from photometry alone.~This 
blurs the distinction between cluster-dominated cD halos and the similarly-large-S\'ersic-index halos 
of giant, core-boxy-nonrotating ellipticals.  These halos are believed to be accreted onto compact,
high-redshift progenitors (``red nuggets'') by large numbers of minor mergers.  They belong dynamically
 to their central galaxies.  Still, cDs and core-boxy-nonrotating Es may be more similar
than we think: Both may have outer halos made largely via minor mergers and the accumulation of tidal debris.  
\lineskip=-20pt \lineskiplimit=-20pt

      We construct a main-body\ts$+${\ts}cD-halo decomposition that fits both the brightness and dispersion profiles.  
To fit $\sigma(r)$, we need to force the component S\'ersic indices to be smaller than a minimum-$\chi^2$ 
photometric decomposition would suggest.  The main body has $M_V \simeq -22.8 \simeq$\ts30\ts\% of the total galaxy
light.  The cD halo has $M_V \simeq -23.7$, $\sim$\ts1/2 mag brighter than the 
brightest galaxy in the Virgo cluster.   A mass model based on published cluster dynamics and X-ray observations 
fits our observations if the tangential dispersion is larger than the radial dispersion at 
$r \simeq 20^{\prime\prime}$ to 60$^{\prime\prime}$.  The cD halo is as enhanced in $\alpha$ element abundances as 
the main body of NGC\ts6166.  Quenching of star formation in \lapprox1{\ts}Gyr suggests that the center of Abell 2199 
has been special for a long time during which dynamical evolution has liberated a large mass of
now-intracluster stars.

\end{abstract}

\section{Introduction}

\pretolerance=15000  \tolerance=15000

      Matthews, Morgan, \& Schmidt (1964) and Morgan \& Lesh (1965) introduced the 
cD class\footnote{The name ``cD'' has created some confusion.  It~has~been interpreted
                  to mean ``cluster dominant'' or ``central dominant'' or ``central diffuse''.  All are 
                  correct descriptions, but they are not the origin of the name.  Morgan (1958) introduced 
                  the ``D'' form classification for galaxies that are like ellipticals but with
                  distinct, outer halos with shallow brightness gradients.  The ``D'' class has 
                  not been as useful as Hubble classes (Hubble 1936; Sandage 1961), because it includes 
                  several different physical phenomena,
                  (a) S0 galaxies, in which the outer halo is the disk; 
                  (b) giant ellipticals with high S\'ersic (1968) indices $n \gg 4$, and 
                  (c) the subjects of this paper{\kern 1pt}:~giant ellipticals whose distinct outer halos  
                      consist of intracluster stars that have been stripped from cluster
                      galaxies.  Because this involves important physics, the name ``cD'' 
                      has survived even though the name ``D'' has not.  But ``c''~does~not~mean~``central'' 
                      or ``cluster''.  Rather, it is a historical anachronism that survives from stellar 
                      spectral classes that are no longer used.  Quoting Mathews \etal (1964): ``These very 
                      large D galaxies observed in clusters are given the prefix `c' in a manner similar to
                      the notation for supergiant stars in stellar spectroscopy.''}
of galaxies in the context of the optical identification of extragalactic radio sources.  Quoting the 
latter paper, ``Of the `strong' sources identified, approximately one-half are associated with galaxies 
having the following characteristics:
({\it a\/}) they are located in clusters, of which they are outstandingly the brightest and largest members;
({\it b\/}) they are centrally located in their clusters;
({\it c\/}) they are never highly flattened in shape; and
({\it d\/}) they~are of a characteristic appearance,
            having bright, \hbox{elliptical-like} [centers], surrounded by an extended amorphous envelope.
            These supergiant galaxies have been given the form-type class of cD in Morgan's [1958] 
            classification.''

\vskip -0.4pt

      This paper presents two new observational results:

\vskip -0.4pt

      1 -- Section 2 demonstrates that the velocity dispersion of the stars in the nearest, prototypical 
cD galaxy{\ts}--{\ts}NGC\ts6166 in the cluster Abell\ts2199{\ts}--{\ts}rises from values typical of giant
elliptical galaxies near the center to the cluster dispersion in the cD halo.~The halo also shifts
toward the velocity of the cluster, which is different from that of NGC\ts6166.  Thus the
halo shares the dynamics of individual galaxies~in~the~cluster.  We interpret this as evidence that the 
stars in the cD~halo~of NGC\ts6166 were stripped from the galaxies by fast collisions.

\vskip -0.4pt

      2 -- We measure the brightness profile of NGC\ts6166~to~make quantitative Morgan's point 
({\it d\kern 1.2pt}) that cDs consist of a central elliptical plus a distinct, shallow-brightness-gradient 
halo.  Photometry by Oemler (1976) suggested that NGC\ts6166 has such two-component structure.
Our ideas about cD halos are based in large part on this result.
However, we find that NGC 6166 is described by a single S\'ersic (1968) profile
at all radii outside the core.  The cluster-dominated halo that is obvious in the kinematics is
not obvious in the photometry.  We need to rethink our understanding of how we recognize cDs and of
whether cD galaxies are fundamentally different from other giant, core-boxy-nonrotating elliptical galaxies.  

\section{HET Spectroscopy:\\Velocity And Velocity Dispersion Profiles of NGC 6166}

\subsection{History and Motivation}

\pretolerance=15000  \tolerance=15000

\lineskip=-20pt \lineskiplimit=-20pt

      To distinguish between competing theories about the origin of cD galaxies (Section 8), a particularly 
powerful diagnostic is their internal kinematics.  Does the velocity dispersion $\sigma(r)$ profile increase 
to the cluster velocity dispersion as one looks farther out into the part of the halo that encompasses many 
non-central cluster members?  Is the systemic velocity of the halo similar to that of the central galaxy 
or is it similar to that of the cluster as a whole?  Are these velocities ever different?  This subject has
a long history, and partial answers to these questions have been known for several decades:

\vskip -15pt
\cl{\null}

\subsubsection{Systemic Velocities} 

Zabludoff, Huchra, \& Geller (1990) find that NGC\ts6166 has 
$(V_{\rm cD} - \bar V) = 378 \pm 99$ km s$^{-1}$ for galaxy and cluster velocities of $V_{\rm cD} = 9348 \pm 15$ km s$^{-1}$
and $\bar V = 8970 \pm 98$ km s$^{-1}$ (71 galaxies).  Zabludoff~et~al.~(1993) find that
$V_{\rm cD} = 9293 \pm 20$ km s$^{-1}$; $\bar V = 9063 \pm 104$ km s$^{-1}$; $(V_{\rm cD} - \bar V) = 230 \pm 106$ 
km s$^{-1}$ for 68 cluster galaxies.  Oegerle \& Hill (2001) get peculiar velocities of
258\ts$\pm$\ts69 to 346\ts$\pm$\ts73{\ts}km{\ts}s$^{-1}$, depending on how $\bar V$ is calculated
and on how far out in the cluster the ($\sim$\ts132) galaxies are counted.  The derived peculiar velocity 
gets smaller as more galaxies get averaged.  Among recent determinations,
Coziol \etal (2009) get $V_{\rm cD} = 9304${\ts}km{\ts}s$^{-1}$; 
$\bar V = 9143${\ts}km{\ts}s$^{-1}$; \hbox{$(V_{\rm cD} - \bar V)$} = 156{\ts}km{\ts}s$^{-1}$ for
471 cluster galaxies.  The most up-to-date study by Lauer et{\ts}al.\ts(2014) gets
$V_{\rm cD}$\ts=\ts9317\ts$\pm$\ts10{\ts}km{\ts}s$^{-1}$; 
$\bar V$\ts=\ts9088\ts$\pm$\ts38{\ts}km{\ts}s$^{-1}$; \hbox{$(V_{\rm cD} - \bar V)$} = 229 $\pm$ 39 
km s$^{-1}$ for 454 cluster galaxies.

      Many cDs are essentially at rest at their cluster centers (e.{\ts}g., 
Quintana \& Lawrie 1982;
Zabludov \etal 1990;
Oegerle \& Hill 2001).  Generally, cDs are more nearly at rest in their clusters
than are non-cD first-ranked galaxies (e.{\ts}g.,
Oegerle \& Hill 2001;
Coziol~et~al.~2009).
But a significant fraction move at several hundred km s$^{-1}$ with respect to their clusters, 
often in association with cluster substructure which suggests that a merger of two clusters is in progress (e.{\ts}g.,
Oegerle \& Hill 2001;
Pimbblet, Rosebloom, \& Doyle 2006; see also
Beers \& Geller 1983;
Zabludoff \etal 1990, 1993).
Proof of concept is provided by the Coma cluster.  It is in the process of a cluster merger 
(White, Briel, \& Henry~1993;
Briel \etal 2001;
Neumann \etal 2001,\ts2003;
Gerhard \etal 2007;
Andrade-Santos \etal 2013;
Simionescu \etal 2013).
The NGC\ts4839 group is falling into the main Coma cluster, which itself has
two central galaxies, NGC\ts4874 and NGC\ts4889, with different velocities (by about 680 km s$^{-1}$)
and their own X-ray halos.  NGC\ts4889 has a velocity of $\sim +430${\ts}km{\ts}s$^{-1}$
with respect to the Coma cluster.  Only NGC\ts4874 is within \lapprox \ts250 km s$^{-1}$ of the
cluster velocity.  NGC\ts4874 and NGC\ts4889 are weak cDs, and NGC 4839 also shows signs of cD structure.  

      NGC\ts6166's velocity with respect to Abell\ts2199 is typical.
{\it The diagnostic question is:~Does the halo of 
NGC\ts6166 have the same systemic velocity as its central galaxy~or~as~its cluster?  We find 
that the cD halo shows velocities between that of the galaxy and that of the cluster.  
The observation that NGC 6166 is not centered in 
velocity in its cD halo is evidence that that halo does not belong dynamically to the galaxy.}

\vskip -15pt
\cl{\null}

\subsubsection{Velocity Dispersion Profiles} 

In a paper that fundamentally shaped our concept of cD galaxies,
Dressler (1979) pushed measurements of velocity dispersions to then-unprecedented~low~surface~brightnesses 
and showed that $\sigma(r)$ for IC\ts1101,~the~brightest galaxy in Abell\ts2029, rises with increasing 
radius~$r$ from $\sim$\ts375{\ts}km{\ts}s$^{-1}$ at the center to \gapprox\ts500{\ts}km{\ts}s$^{-1}$ at 
$r$\ts$\simeq$\ts71{\ts}kpc.~(The distance has been converted to the WMAP~5-year~cosmology distance
scale, Komatsu{\ts}et{\ts}al.{\ts}2009; NED.)~Thus the dispersion rises toward
but does not reach the cluster $\sigma$ of $1160$ km s$^{-1}$ (Coziol et{\ts}al.\ts2009) or $1222 \pm 75$ km s$^{-1}$
(Lauer{\ts}et{\ts}al.\ts2014).
Dressler interpreted this in the context of suggestions 
(Gallagher \& Ostriker 1972;
White 1976;
Ostriker \& Tremaine 1975;
Richstone 1976; 
Merritt 1983;
Richstone \& Malumuth 1983) 
that cD halos consist of accumulated debris of stars stripped from cluster members by tidal encounters~and
by dynamical friction against the growing halo.  Thus~a~cD consists of ``a luminous but normal elliptical 
galaxy sitting in a sea of material stripped from cluster galaxies'' (Richstone 1976; Dressler 1979).  
Dressler concludes: ``The results of this study confirm an [outward] {\it increase in 
velocity dispersion, which is a necessary (but not sufficient) condition in the proof of the stripped 
debris hypothesis}''.  Sembach \& Tonry (1996) and Fisher \etal (1995) confirm these results.

      Among the nearest galaxies, M{\ts}87 is marginally a cD in the sense of having extra light at large 
radii with respect~to an $n \simeq 9^{+2}_{-1}$ S\'ersic fit (Figure\ts50 in Kormendy~et~al.~2009;
hereafter KFCB).  This is a normal S\'ersic index for a core-boxy-nonrotating elliptical, but
the amount of extra light is small, and in fact, an $n = 11.8^{+1.8}_{-1.2}$ S\'ersic function fits the
whole profile outside the core.  This S\'ersic index is outside the range normally observed for 
core-boxy-nonrotating Es.  Nevertheless, a cD halo cannot securely be identified as an outer 
component that is photometrically distinct from the main body of the galaxy.  At the same time, it is 
clear that the Virgo cluster does contain intracluster stars, from broad-band surface photometry 
(Mihos 2011; 
Mihos \etal 2005, 2009), 
from spectroscopy of individual stars (Williams \etal 2007b), and
from the detection of intracluster globular clusters (Williams \etal 2007a) and 
planetary nebulae 
Arnaboldi \etal 1996, 2002, 2004; 
Castro-Rodrigu\'ez \etal 2009; see
Arnaboldi \& Gerhard 2010 and 
Arnaboldi 2011 for reviews).
The intracluster light is irregular in its spatial distribution and defined largely
by (tidal?)~streams.  It is reasonable to conclude that the intracluster light is in early stages of
formation.   Nevertheless, it pervades the cluster and must feel the cluster gravitational 
potential.  And the outer halo of M{\ts}87 merges seamlessly with this intracluster light (Mihos papers).  
Do we observe that the velocity dispersion of stars in M{\ts}87 increases toward the 
cluster dispersion?  

      The answer -- tentatively -- is yes.  The integrated light shows an outward
drop in $\sigma$ from $\sim 360$ km s$^{-1}$ in the central few arcsec to $\sim 300$ km s$^{-1}$ at 
$20^{\prime\prime}$ \lapprox \ts$r$ \lapprox $100^{\prime\prime}$ and then an outward rise to $\sim 340$ 
km s$^{-1}$ at $r \sim 250^{\prime\prime}$ (Murphy \etal 2011, 2014).  This is subtle~and not easily
interpreted. But the upward trend in $\sigma$ continues~in the globular cluster population,
which reaches $\sigma$\ts$\simeq$\ts400--470{\ts}km{\ts}s$^{-1}$ by $r \sim 30^{\prime\prime}$ (Wu \& Tremaine
2006; see C\^ot\'e \etal 2001 for earlier~results).  Planetary nebula data in Doherty \etal (2009) reveal
both M{\ts}87 halo and intracluster stars, but the data are too sparse to determine a $\sigma(r)$ profile.  
Also, though they do not overlap greatly in radial leverage, stellar dynamical models and mass profile 
measurements from the X-ray gas give essentially consistent results (e.{\ts}g., 
Gebhardt \& Thomas 2009;
Churazov \etal 2008).
Thus, M{\ts}87 is the nearest galaxy where various test particles have been used to probe 
the dynamics of a marginal cD from its center out to radii where the cluster dominates.  The problem
is that the test particles are heterogeneous enough and the statistics for point particles are poor
enough so that we cannot securely see the transition from the galaxy's main body to any halo that is
controlled by cluster gravity.  Nevertheless, as a proof of concept,~M{\ts}87 is important.  And it
provides a hint that proves to be prescient: The dispersion profile starts to rise at $r \sim
100^{\prime\prime} \sim 8$ kpc, well interior to the radii where any plausible argument identifies
the beginning of a cD halo based on photometry alone.

     Ooutward $\sigma$ rises in cD or cD-like galaxies are reported~by 
\hbox{Carter{\ts}et{\ts}al.\ts(1981,\ts1985)~and~by~Ventimiglia{\ts}et{\ts}al.\ts(2010).~Still}, 
the only prototypical cD in which the 
velocity dispersion of the stellar halo is robustly seen to rise toward larger radii by several authors 
is NGC{\ts}6166 in the cluster Abell{\ts}2199. From a central velocity dispersion of 
$\sigma \sim$\ts300{\ts}km{\ts}s$^{-1}$, the dispersion first drops outward and then rises to 
$\sigma \sim$\ts400{\ts}km{\ts}s$^{-1}$ (Carter, Bridges, \& Hau 1999) 
at about 30$^{\prime\prime}$ and $\sigma\sim$\ts600\ts$\pm$\ts100{\ts}km{\ts}s$^{-1}$ (Kelson \etal 2002) at 
$\sim$\ts50$^{\prime\prime}$--\ts60$^{\prime\prime}$.  {\it No velocity dispersion measurments of any cD
galaxy reach large enough radii to show that $\sigma$ increases all the way up to the cluster dispersion.}

      {\it The first purpose of this paper is to push the measurements of NGC 6166 far enough out in radius
to see whether or not $\sigma(r)$ reaches the cluster dispersion.}

\vskip 6.18truein

\includegraphics{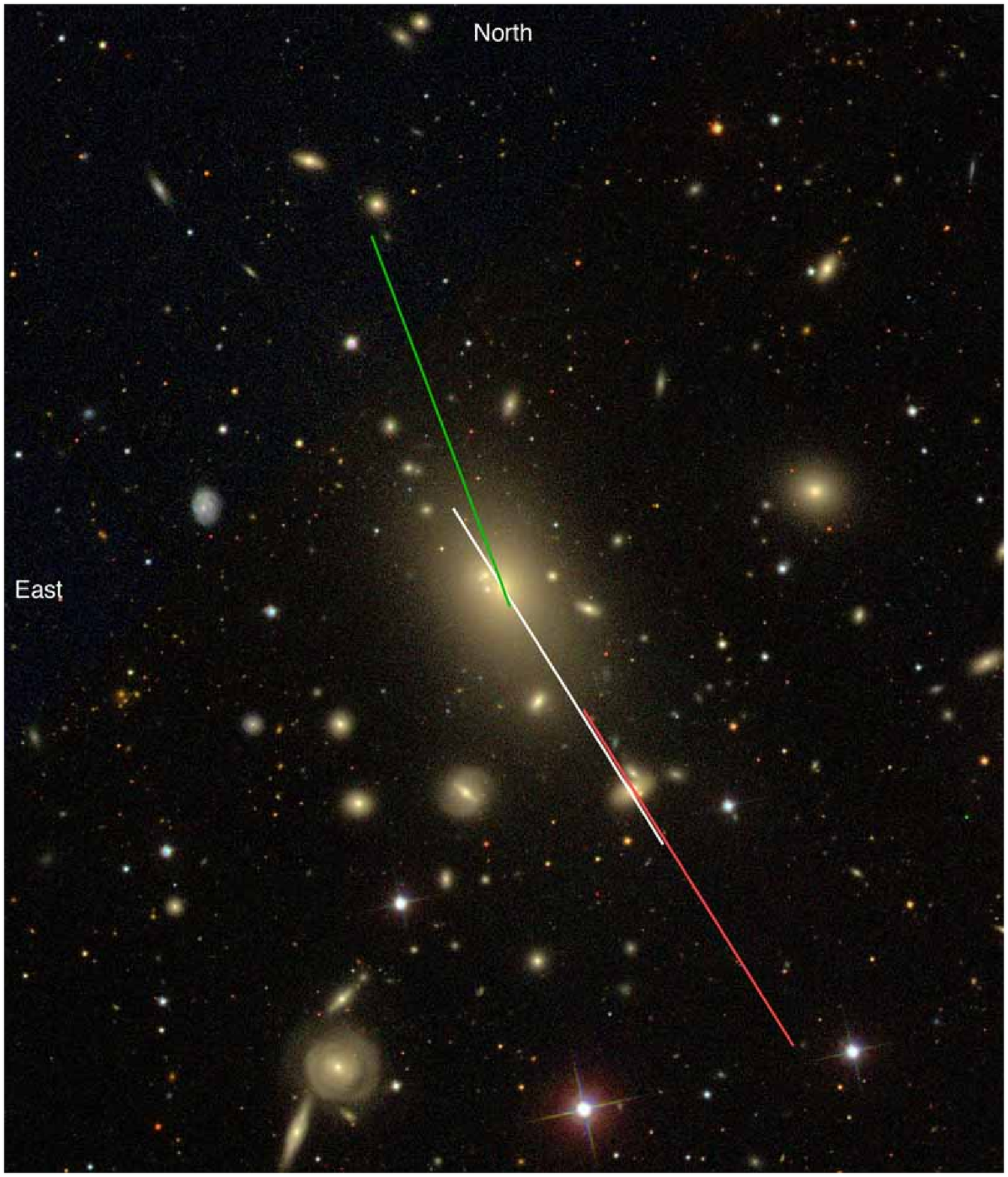}

\includegraphics{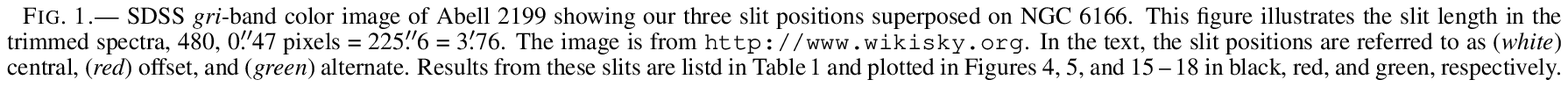}

\eject

\subsection{HET Spectroscopy}

      We obtained spectra at three slit positions (Figure\ts1)~along and near the major axis of NGC\ts6166
with the 9.2{\ts}m Hobby-Eberly Telescope (HET) and Low Resolution Spectrograph 
(LRS:~Hill~et~al.~1998).~~The~slit~width~was~1\sd5,~the~reciprocal
dispersion~was~116~km~s$^{-1}$~pixel$^{-1}$,~and~the~resolution expressed as a velocity 
dispersion was $\sigma_{\rm instr} = 125$ km s$^{-1}$.
The slit positions had exposure times of 8\ts$\times$\ts900~s
(``center'', with NGC\ts6166 centered well inside the slit), 4\ts$\times$\ts900 s (``offset'' position 
along the major axis, centered on the bright, elongated galaxy NGC\ts6166A visible in Figure\ts1), and
6\ts$\times$\ts900{\ts}s~$+$\ts1\ts$\times$\ts800{\ts}s~(``alternate''~position~offset~by~11\degd5
from the major axis but on the other side of the center, positioned to miss star and galaxy images).  
All individual exposures were taken on different nights.  The standard star spectrum used is a 
combination of HD74377 and HR2600.  Our experience is that this combination fits old elliptical-galaxy 
stellar populations well and give kinematic results that are relatively free from template bias.
In any case, the kinematics were measured with Bender's (1990) Fourier correlation quotient method, 
which is designed to eliminate template bias.
\pretolerance=15000  \tolerance=15000 

\vskip 5.42truein

\clearpage
\clearpage

\cl{\null}

\vskip 4.4truein

\includegraphics{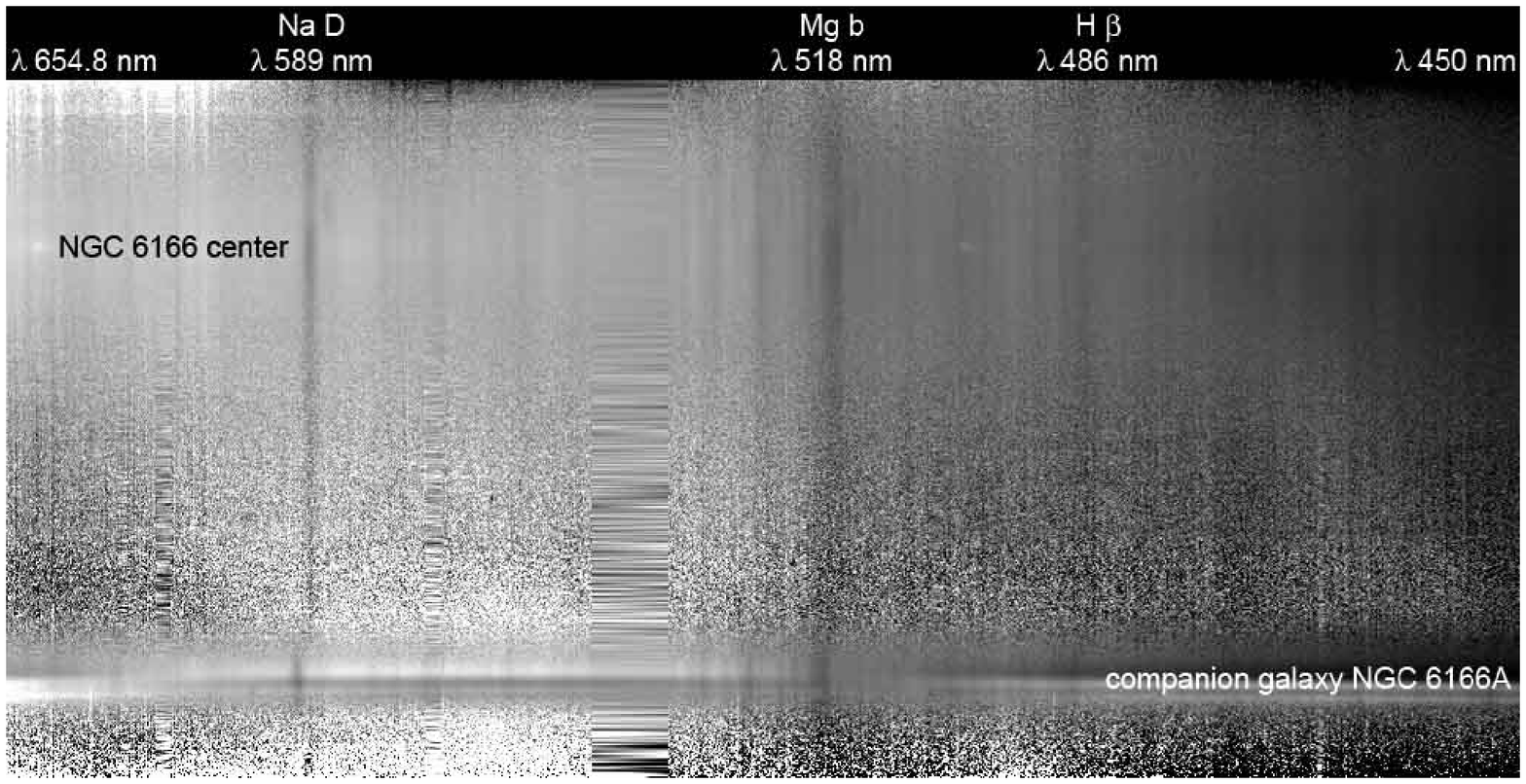}

\includegraphics{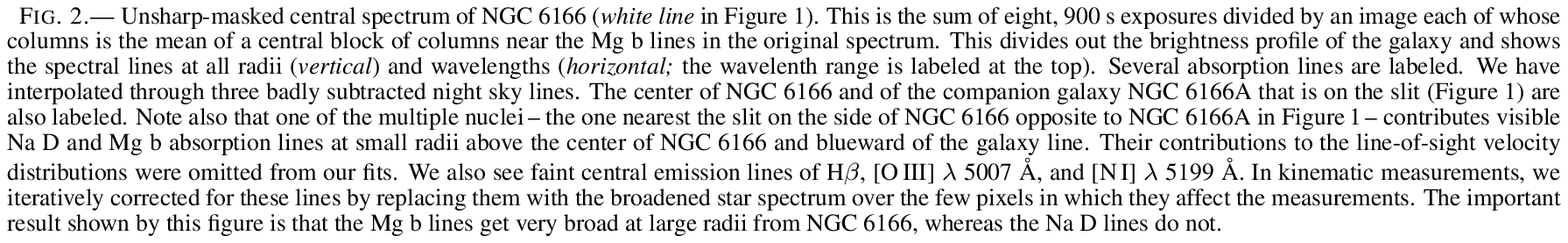}

\vskip 30pt

      Figure 2 shows an unsharp-masked version of the sum of the best spectra along
the central slit position ({\it white line\/} in Figure 1).  By dividing out the brightness
profile of the galaxy, we can see absorption lines and qualitatively judge $S/N$ from the center
out to the largest radii.  The strongest lines in NGC 6166, Mg b, Na D, and H$\beta$, are visible
all the way to the companion galaxy on the slit.  Even Fe $\lambda$5270\ts\AA~and 
5335\ts\AA~are visible quite far out (see also Figure 3).  They are used in Section 6 to measure 
[$\alpha$/Fe] overabundance out into the part of the halo where the velocity dispersion is large. 
Most important, Figure 2 already shows that all lines except Na D get very wide in the cD halo
of NGC 6166.  

      The Na D line is narrow at all radii and shows little gradient in velocity.~It
gives a dispersion of $\sigma_{\rm Na{\ts}D} \simeq 300$ km s$^{-1}$ at all radii.
We assume that the line is produced by interstellar gas and do not include it in the
wavelength region from redward of the iron lines to blueward of H$\beta$ that we use for $V$ and $\sigma$
measurements.  Dust is seen near the center in Figure 8.  There may be a more
smoothly distributed ISM at larger radii, as suggested also by the fact that H$\beta$ 
absorption in our spectra is significantly weaker than even a very old stellar
population would show.  However, it is not obvious that its kinematics should be a simple as we
measure with the Na D line.  Interpretation of this line in the context of the X-ray gas halo of 
the galaxy is beyond the scope of this paper.

      The{\ts}offset{\ts}and{\ts}alternate{\ts}slit{\ts}positions{\ts}yielded{\ts}poorer{\ts}spectra.
We discard one spectrum taken with too much moonlight, so the alternate slit position has only 6 good spectra.
Of these, one is fainter than normal by $\sim$\ts14\ts\% and two more are fainter by $\sim$\ts23\ts\%,
presumably due to clouds.  (The observations are \phantom{000000000000} 

\cl{\null} \vskip 4.66truein

\noindent queue-scheduled, so we cannot personally monitor the observing 
conditions.  However, we checked that the galaxy was centered on the slit.  Seeing is relatively unimportant.)  

      Offset{\ts}sky{\ts}spectra{\ts}were{\ts}taken{\ts}after{\ts}all{\ts}NGC\ts6166{\ts}exposures.  
For the center slit position, these were cleaned of bad pixels and averaged to give high $S/N$ and then
used for all sky subtractions.  Each spectrum was individually sky-subtracted before the spectra were added.
The sky subtraction of the central slit spectra is good (Figure 2).  However,
for the other two slit positions, most sky spectra could not be used for sky subtraction because too many
night sky emission lines changed in strength in the short time between exposures.  For the offset sky
positions, the sky was measured as far from the galaxy as possible; since even the NGC 6166 end of the
slit is far from the galaxy (Figure~1), these sky spectra should be essentially free of galaxy light.
However, for the alternate slit position, sky spectra taken from the galaxy images do subtract a little
halo light.  For this reason{\ts}--{\ts}as well as problems with moonlight~and~with~clouds{\ts}-- the alternate
slit position does not reach as far out as the primary slit position illustrated in Figure 2.  In addition, we
found that we got the best results to the largest radii in the alternate slit position by using only the
four best spectra.

      Figure 3 shows sample spectra for five radial bins in NGC 6166 and for the optimized template star.  
This binning is used in Section\ts6 to measure line strengths for the Mg b and Fe lines.  Reliable line
strength measurements are possible out to the bin at $r = 59^{\prime\prime}$.  Velocity dispersions are easier --
they are measureable for the $r = 87^{\prime\prime}$ bin and for several others at large radii in the center,
alternate, and offset slit positions.

\vfill \eject

\cl{\null}\vfill

\includegraphics{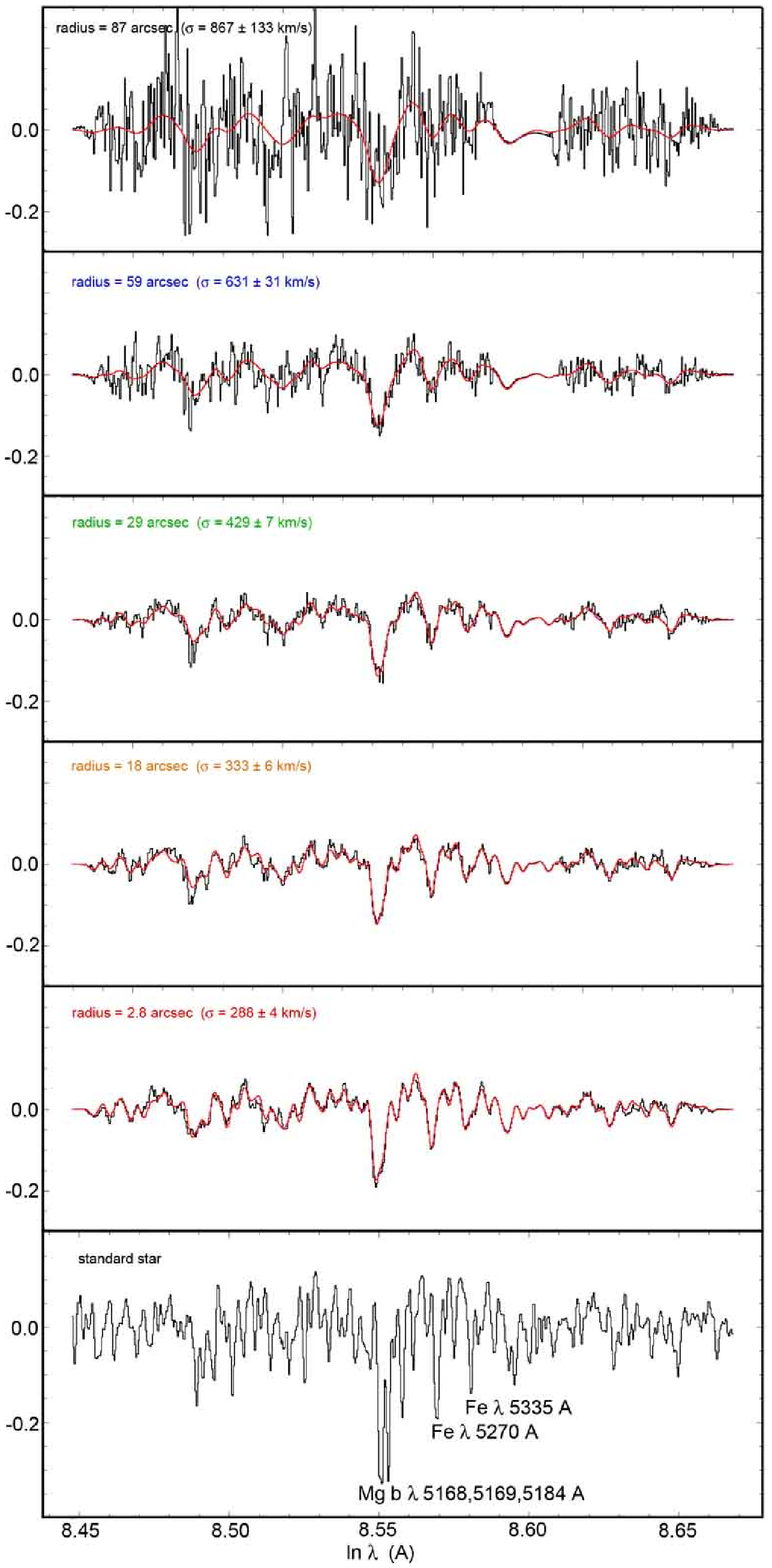}

\includegraphics{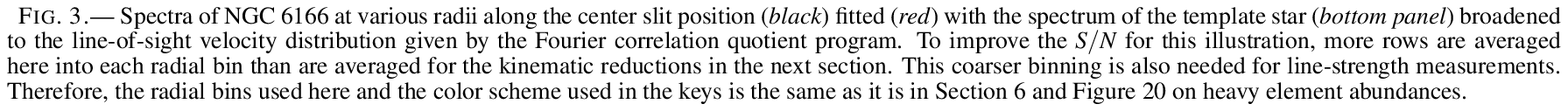}

\clearpage

\cl{\null}

\vfill

\includegraphics{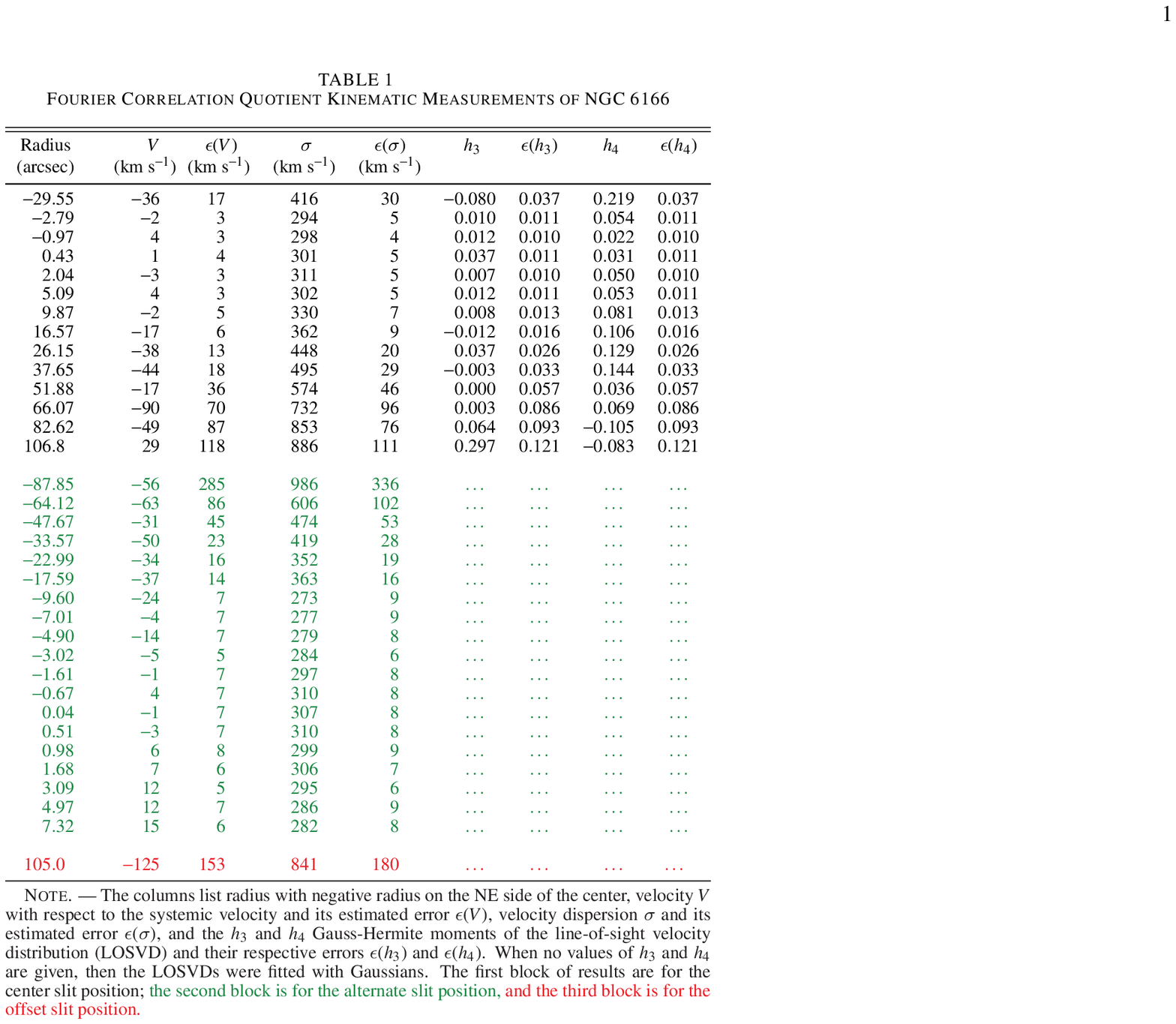}

\vfill

\cl{\null}

\clearpage

\subsection{Kinematic Results}

      The summed center, alternate, and offset spectra were reduced with the Fourier correlation quotient
program of Bender (1990).  This gives velocity~$V$, velocity dispersion $\sigma$, the higher-order Gauss-Hermite 
coefficients $h_3$ and $h_4$, and nonparametric line-of-sight velocity distributions (LOSVDs).  At some radii 
near $r \sim -12^{\prime\prime}$ (see Figure 2), the LOSVDs show a main peak at the systemic
velocity of NGC\ts6166 and smaller peak in its wings associated with another of the multiple nuclei.  We omitted
the corresponding velocity bins from the LOSVD fit.  Since neither the center nor the radii where $\sigma$
starts to climb are affected, this cleaning does not affect our conclusions.  However, many published $V$
and $\sigma$ measurements show contamination from the multiple nuclei.

      The instrumental velocity dispersion was measured in our reduced spectra to be 
$\sigma_{\rm instr} = 125$ km s$^{-1}$, easily
adequate for the galaxy dispersions $\sigma$ \gapprox \ts300 km s$^{-1}$ studied in this paper.

      The kinematics are listed in Table 1 and shown in Figure 4.

\vfill

\includegraphics{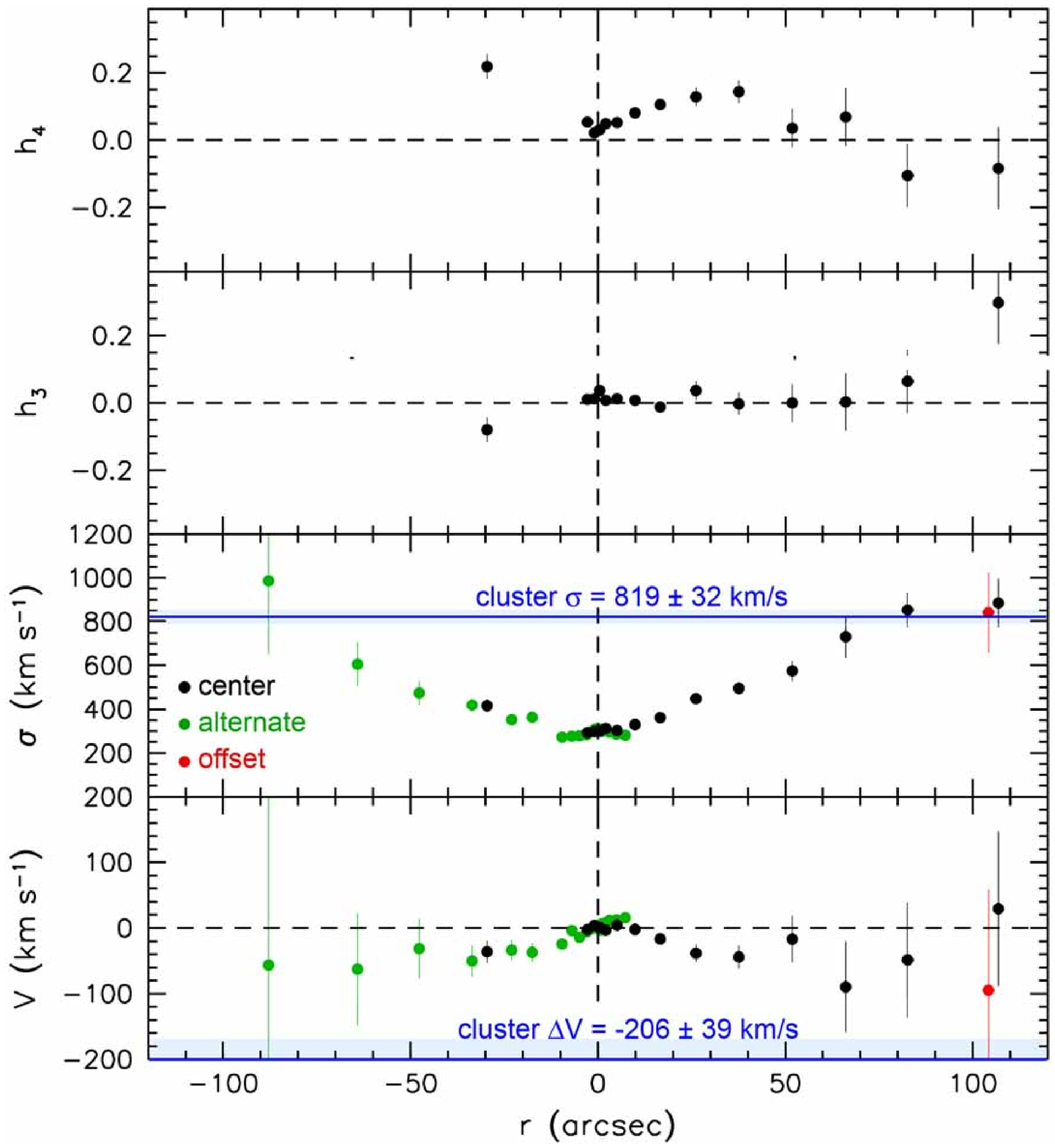}

\includegraphics{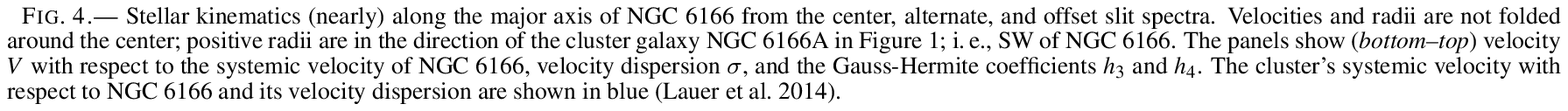}

\eject

\subsection{The Velocity Profile of NGC 6166}

      The systemic velocity of NGC 6166 is $206 \pm 39$ km s$^{-1}$ higher than the velocity
$9088 \pm 38$ km s$^{-1}$ of 494 cluster galaxies (Lauer \etal 2014).~Here we use our measure~of~the 
systemic velocity of NGC\ts6166, $V_{\rm cD} = 9294 \pm 10$ km s$^{-1}$.  It is consistent within
errors with values in Zabludoff \etal (1993) and in Coziol \etal (2009).  Other, inconsistent 
published measurements may be affected by contamination by the multiple nuclei.  
Using our $V_{\rm cD}$, NGC 6166 moves at $(0.25 \pm 0.05)$\ts$\sigma$, typical of the values found by 
Lauer~et~al.~(2014).
\lineskip=-20pt \lineskiplimit=-20pt

      If the cD halo consists of tidal debris, then we expect that its systemic velocity should
shift toward that of the cluster at the radii where $\sigma$ rises toward the cluster value.  Figure 4 shows
that the velocity at large radii on both sides does decrease from $V_{\rm cD}$ toward the cluster velocity.  
The average of the large-radius points is only $\sim$\ts$-70$ km s$^{-1}$.  Still, the inner part of the
cD halo of NGC 6166 is -- as far as we can measure it -- more nearly at rest within the cluster
than is the central galaxy.

\vfill\eject

\subsection{The Velocity Dispersion Profile of NGC 6166}

      Figure 5 compares our kinematic results on NGC 6166 with published dispersion profiles.
Carter \etal (1999) and Kelson \etal (2002) observed much of the rise in $\sigma$ to the cluster value.  
However, our observations are the first to reach deep enough to see $\sigma$ for the intergrated starlight in
a cD halo rise all the way to the cluster dispersion in any galaxy cluster.

      The Carter \etal (1999) data are not shown in Figure 5, because they did not publish a table of
their results.  Their outermost measurements at radii of 30$^{\prime\prime}$ to 36$^{\prime\prime}$ are 
$\sigma \simeq 390$, 361, and 438 km s$^{-1}$.  These are consistent with our results and with Kelson's.
(However, Carter \etal 1999 derive velocities that increase as $r$ increases; they interpret
this as ``modest major-axis rotation''.  Kelson \etal 2002 also see ``systematic rotation 
[$V/\sigma \approx 0.3$] in the intracluster stars beyond 20 kpc''.  We do not see rotation; 
rather, the halo velocity decreases toward the cluster velocity on both sides of the center.)

      Tonry (1984, 1985) measured the multiple nuclei of NGC\ts6166 but did not reach far enough out to see an 
outward increase in $\sigma$.  Similarly, Fisher, Illingworth, \& Franx (1995) and Loubser \etal (2008) measured 
only a slight outward drop in $\sigma$ in the main body of the galaxy.

       Figure 5 illustrates the most important result in this paper: {\it The velocity dispersion
in NGC\ts6166 increases outward to a weighted mean of $\sigma = 865 \pm 58$ km s$^{-1}$ for the four data
points at $r = 83^{\prime\prime}$ to 107$^{\prime\prime}$.  This equals the
velocity dispersion $\sigma = 819 \pm 32$ km s$^{-1}$ for 454 galaxies in Abell 2199 (Lauer \etal 2014).
The rise in $\sigma$ to the cluster velocity dispersion is seen in all three of our slit positions.
This result is the strongest evidence that the cD halo of NGC 6166 is made of stars that have been 
accreted in minor mergers or stripped from cluster galaxies by dynamical harassment.}

\figurenum{5}

\vskip 3.8truein

\begin{figure}[hb!]

\includegraphics{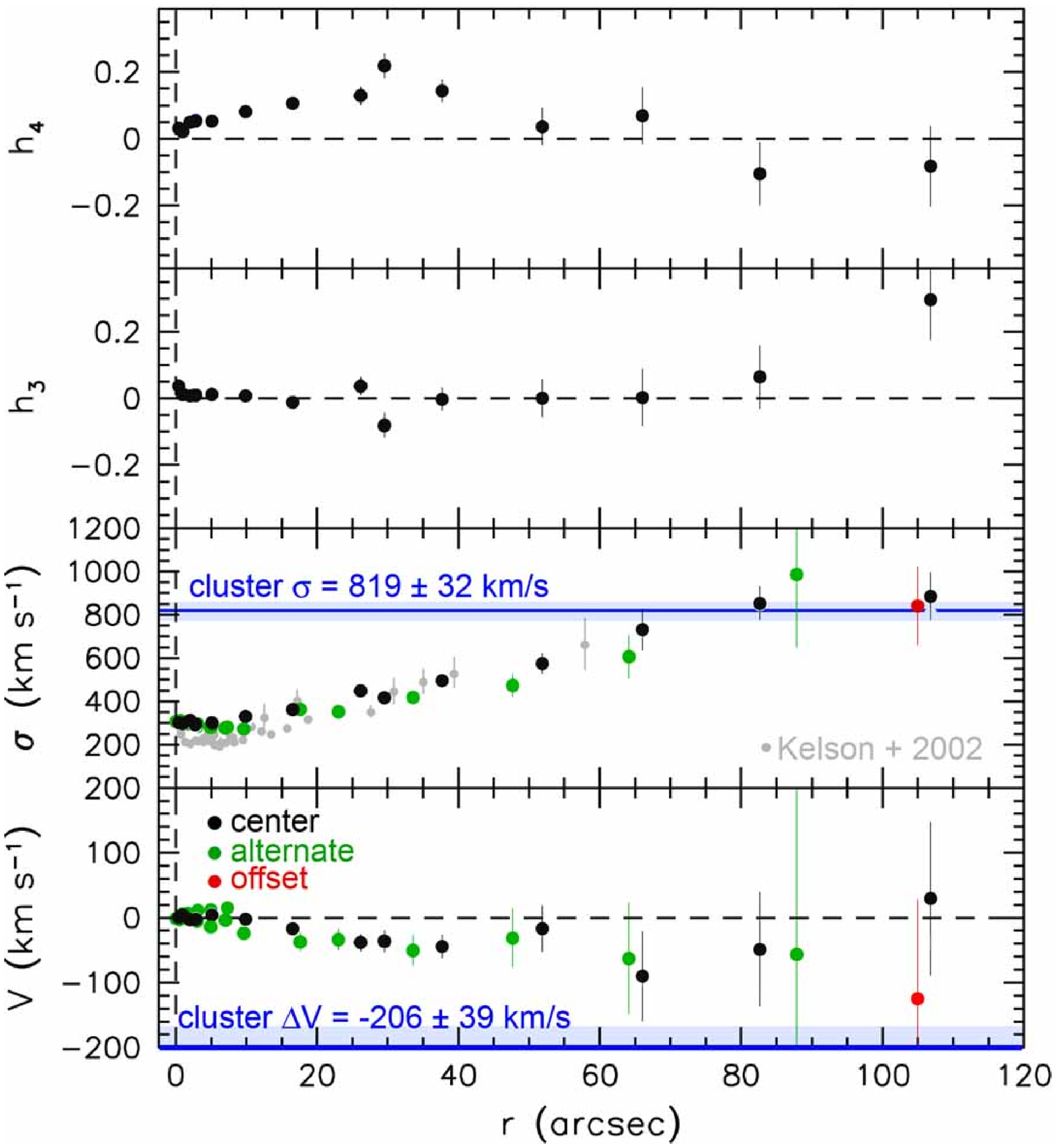}

\figcaption[]
{NGC 6166 kinematic measurements from Figure 4 folded in radius around the center and compared with the
 velocity dispersion profile obtained by Kelson \etal (2002).  Note that, whereas $r$ has been replaced 
 with $|r|$, the sign of $V$ has not been changed.  Again, the cluster systemic velocity with respect to
 NGC 6166 and its velocity dispersion are shown in blue.}
\end{figure}

\vskip 10pt

\vfill\eject

\section{Surface Photometry:\\Does NGC 6166 Have a Photometrically Distinct Halo?}

\pretolerance=15000  \tolerance=15000

\subsection{The Standard Picture of cD Halos}


      Our standard picture of the nature of cD halos and the way in which we identify cD galaxies are 
based in large part on photometry of NGC 6166 and other cD galaxies
by Oemler (1976).  Oemler's procedures and conclusions were later made quantitative 
by Schombert, as discussed~below.  But the iconic, two-component structure suggested by
Oemler's photometry of cDs{\ts}--{\ts}particularly NGC\ts6166{\ts}--{\ts}firmly cemented in our minds
the notion that cDs consist of an elliptical-galaxy-like central body plus a {\it photometrically
distinct\/}, shallower-surface-brightness halo that is not present in normal giant ellipticals.
Oemler's profile of NGC 6166 -- augmented by {\it Hubble Space Telescope\/} (HST) photometry 
to improve the central spatial resolution -- is shown in Figure 6. 

      The clearly two-humped profile in Figure 6 decisively quantifies Morgan's description of his
visual impression of two-component structure.  Other cDs in Oemler (1976), in Schombert (1986, 1987, 1988),
and in other~papers~from the same era behave similarly.  The picture of cD halos that has been in our
minds ever since is made still more concrete using modern profile analysis machinery by decomposing
the profile into two S\'ersic (1968) functions.  Several recent papers have done this and suggested
that the inner components are normal ellipticals whereas the cD halos have exponential profiles 
(Seigar, Graham, \& Jerjen 2007; 
Donzelli, Muriel, \& Madrid 2011).
In fact, the S\'ersic-S\'ersic decomposition in Figure 6 requires that the cD halo 
have $n \simeq 0.77$, between an exponential ($n = 1$) and a Gaussian ($n = 0.5$) in its outer cutoff. 
A worrying hint is that the inner profile has $n = 1.62$, smaller than we have found for any other 
elliptical (KFCB).  Note that, in making this fit, we have been very conservative about excluding the 
inner, shallow-power-law core (see
Kormendy \etal 1994;
Lauer \etal 1995, and
KFCB 
for the definition of cores and 
Gebhardt \etal 1996;
Kormendy 1999, and
Lauer \etal 1995
for a demonstration that they are features of the unprojected and not just the projected profiles).
We also omit the central AGN from the fit.  About 2/3 of the light of the profile in Figure 6
is in the cD halo.

\figurenum{6}

\begin{figure}[hb!]

\vskip 2.65truein

\includegraphics{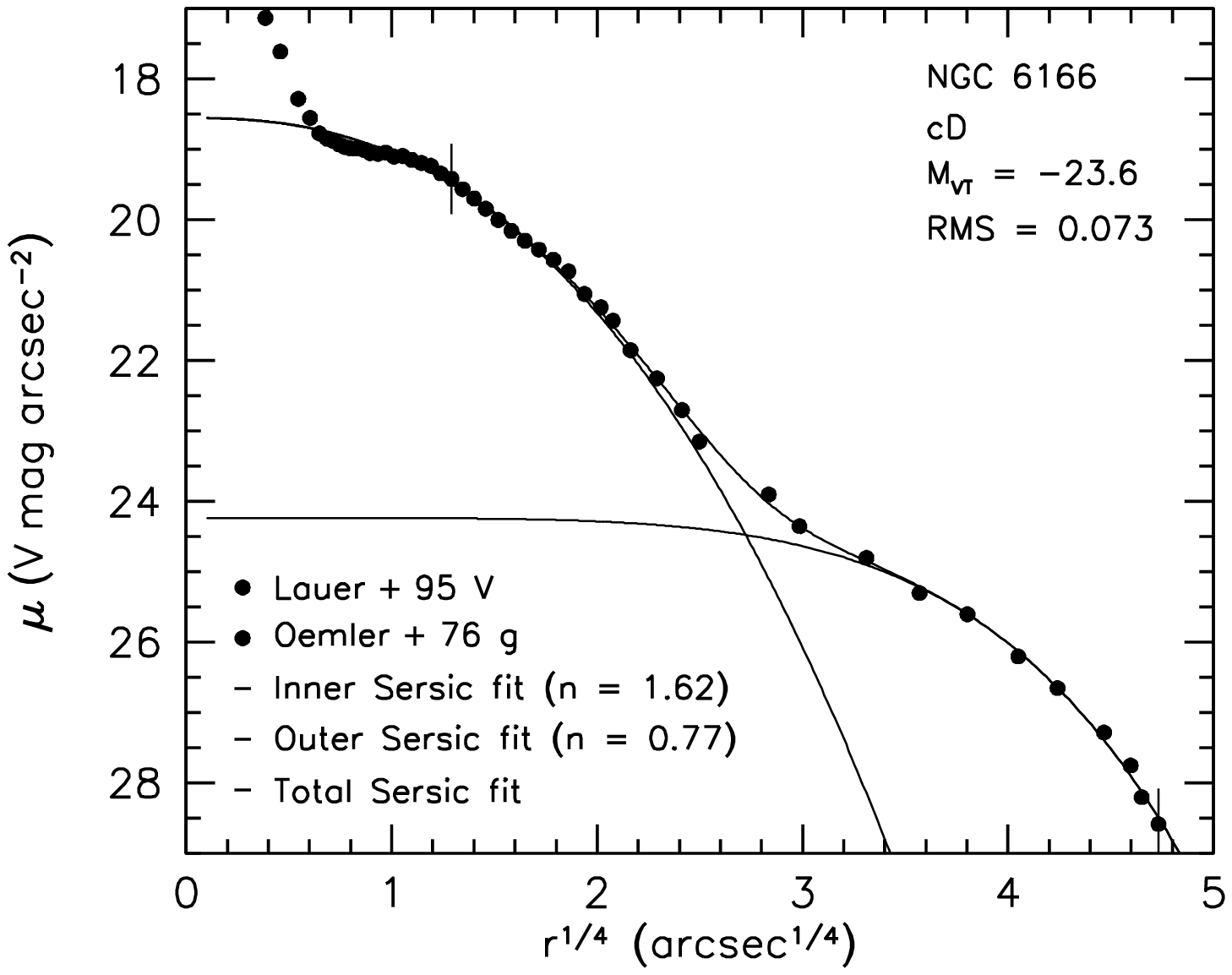}

\figcaption[]
{Circles show an average of the major-axis profile of NGC 6166 measured with HST by Lauer \etal
(1995) and the outer profile measured by Oemler (1976).  The lines show a photometric decomposition 
 into two S\'ersic functions in the radius range shown by the vertical dashes across the profile.
 The S\'ersic indices and fit RMS in mag arcsec$^{-2}$ are given in the key.}
\end{figure}

\vskip 12pt

      The ideas summarized above were made more quantitative by Schombert (1988).
Schombert (1986, 1987) measured average surface brightness profiles of non-first-ranked ellipticals 
as functions of galaxy absolute magnitude $M_V$ in seven $M_V$ bins from $-17$ to $-22.5$ 
($H_0 = 50$ km s$^{-1}$ Mpc$^{-1}$). Schombert (1988) then used these template profiles to define 
cD galaxies.  First, the template profile is found that best matches the inner profile of the 
candidate galaxy over the largest possible radius range.  If this profile fits all of the candidate galaxy
to within the scatter seen among the individual profiles that were used to make the template, then 
this galaxy is an ordinary elliptical.  In contrast (Figure 1 in Schombert 1988; cf.~Figure 6 here), 
if the galaxy in question has a giant outer halo above the template profile fitted to the inner parts, 
then the galaxy is a cD and the integrated difference between its observed profile and the best-fitting 
template is the cD halo.  This definition is similar in spirit to one used by Oemler (1976) but has the 
advantage of allowing the profiles of ellipticals to depend on luminosity.  And it has the virtue of being 
nonparametric -- it does not depend on describing the inner profile with an analytic fitting function.

      The profile decomposition shown in Figure 6 is nothing more nor less than Schombert's procedure
in parametric~form, using S\'ersic functions for the inner and outer components.  Much experience
in recent years has shown that S\'ersic functions are excellent fits to elliptical-galaxy profiles (see
KFCB for data and review) and hence also to Schombert's template profiles.  However:

      {\it We find a problem with our canonical picture of cD halos (\S\ts3.2). The photometry shown
in Figure 6 is in error.  Our composite profile measurements of NGC 6166 are very well fitted by a 
single S\'ersic function at all radii outside the central core.  In contrast to our kinematic results, 
there is no photometric hint of two-component structure.}

\cl{\null} \vskip -0.99truein 
\cl{\null} \vskip  0.8truein

\subsection{Composite $V$-Band Brightness Profile of NGC 6166}

      We have measured the $V$- and $I$-band surface brightness profiles of NGC 6166 using CCD images from
four ground-based telescopes and four cameras (WFPC1 PC, WFPC2 WF, ACS, and NICMOS2) on HST.  Parameters
of the images are listed in Table 2.  This section discusses the $V$-band profile.

\vskip 3.4truein

\includegraphics{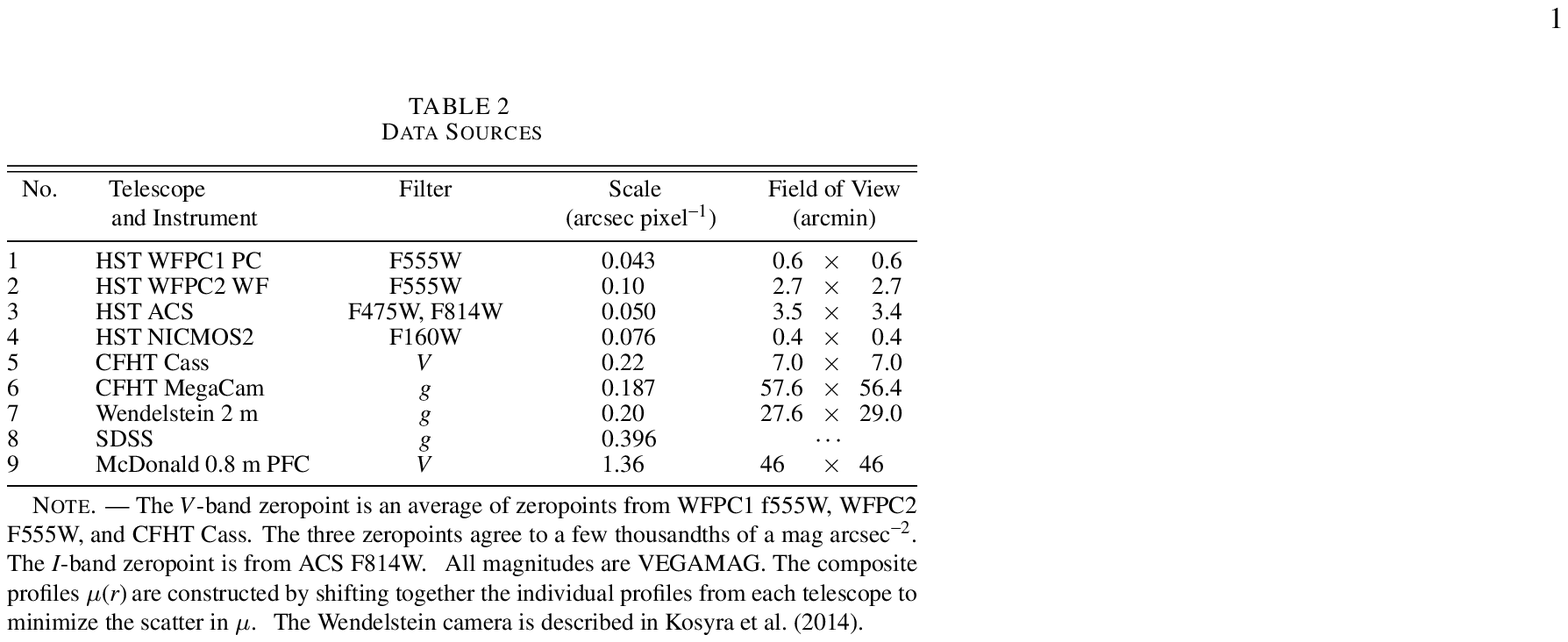}

      The central profile is from an HST WFPC1 measurement by Lauer \etal (1995), from our measurement
of an HST WFPC2 F555W image (GO program 7265; D.{\ts}Geisler,{\ts}P.{\ts}I.), and from our high-resolution (Gaussian 
dispersion radius $\sigma_* = 0\sd32$) $V$-band image from the Canada-France-Hawaii~Telescope Cassegrain camera.  
The CFHT observing run is discussed in KFCB.  The three images give independent $V$-band zeropoints 
that agree (fortuitously) to much better than $\pm 0.01$ mag arcsec$^{-2}$.  The three zeropoints have been averaged.

      Similar in resolution to the CFHT Cassegrain image is a $g$ image from the CFHT Megacam.  We 
also include photometry of an $r$ image from SDSS; it is used over a larger radius range to derive 
the $I$-band profile in the next subsection, but it is used here to help to tie together small and large radii, 
and it helps to measure the ellipticity and PA profile.  The outer profile is obtained using a $g$-band image
from the Wendelstein Observatory's new 2 m Fraunhofer Telescope (FTW) and a $V$-band image from the McDonald 
Observatory 0.8 m telescope.  The latter profile reaches $r = 416^{\prime\prime}$, where 
$\mu = 27.28$ $V$ mag arcsec$^{-2}$.  The $V$-band profile of NGC 6166 is similar in accuracy and limiting 
surface brightness to the data in KFCB.

      Figure 7 shows the raw profiles.  Three kinds of profiles are shown.  Most are based on isophote fits as in
Bender (1987), 
Bender \& M\"ollenhoff (1987), and 
Bender, D\"obereiner, \& M\"ollenhoff  (1987, 1988). 
The algorithm fits ellipses to the galaxy isophotes; it calculates the ellipse
parameters surface brightness, isophote center coordinates $X_{\rm cen}$ and $Y_{\rm cen}$, 
major and minor axis radii, ellipticity $\epsilon$, and position angle PA of the major axis. 
Radial deviations of the isophotes from the ellipses are expanded in a Fourier series
in the eccentric anomaly $\theta_i$, 
\vskip -10pt
$$ 
\Delta r_i = \sum_{k=3}^{N} 
             \left [ a_k \cos (k \theta_i) + b_k \sin (k \theta_i ) 
             \right ].\eqno{ (1) } 
$$
\vskip 5pt
\noindent The most important parameter is $a_4$, expressed in the
figures as a percent of the major-axis radius $a$.  If $a_4 > 0$,
the isophotes are disky-distorted; large $a_4$ at intermediate~radii would
indicate an S0 disk.  If $a_4 < 0$, the isophotes are boxy.
The importance of these distortions is discussed in Bender (1987, 1988);
Bender \etal (1987, 1988, 1989, 1994); Kormendy \& Djorgovski (1989); Kormendy \& 
Bender (1996), KFCB, Kormendy (2009), and below.

\vfill\clearpage

      Some profiles were measured using Lauer's (1985) program 
{\tt profile\/} in the image processing system {\tt VISTA} (Stover 1988).  The
interpolation algorithm in {\tt profile\/} is optimized for high
spatial resolution, so it is best suited to our high-$S/N$ images of the core of NGC 6166.
The isophote calculation is Fourier-based, so it not well suited to measuring the 
outer parts of NGC 6166, where masking of other galaxies in the cluster results in very
incomplete isophotes.

      Finally, as discussed further below, we use a major-axis, (0\farcs2-) two-pixel-wide cut profile
to verify that the ellipse fitting was not adversely affected by the companion galaxies. 

      Seriously discrepant data in the profiles at small radii (usually because of inadequate spatial
resolution) and at large radii (usually because of spatial variations in sky brightness) were pruned 
out before final averaging.  Two additional complications require discussion:

\vskip 6.9truein

\includegraphics{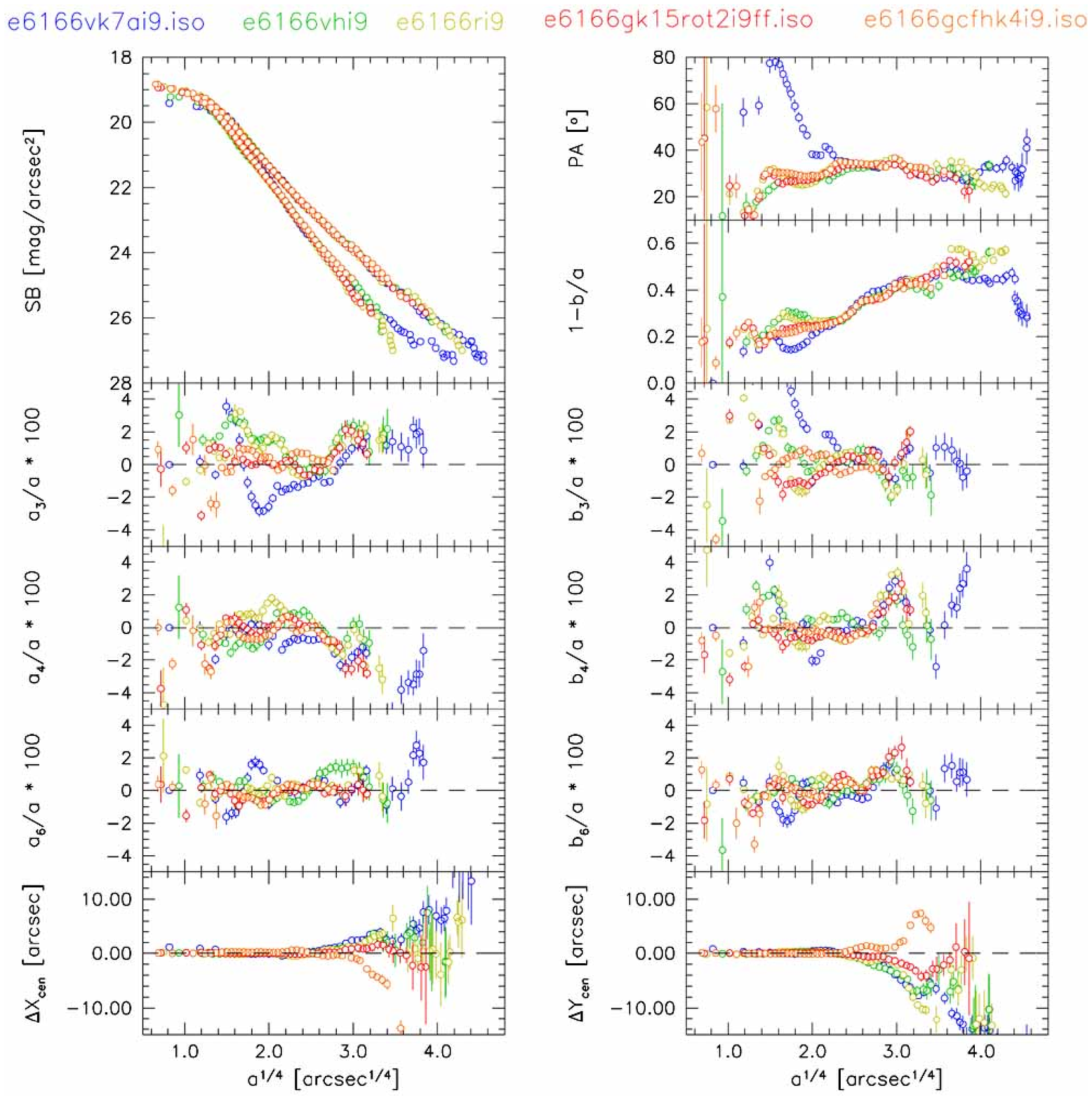}

\includegraphics{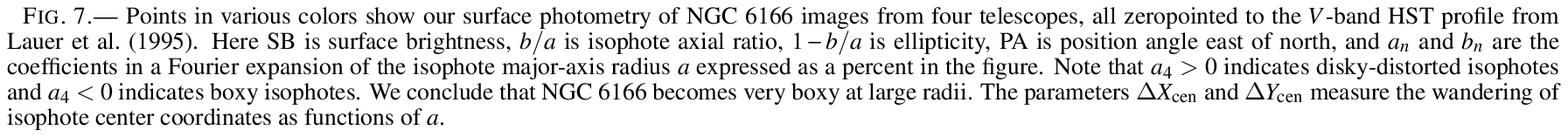}

\eject

      (1)~Three additional cluster galaxies lie in projection close to the center of NGC 6166 (e.{\ts}g.,
Minkowski 1961;
Burbidge 1962;
Tonry 1984, 1985).~Profile calculations need to correct for the light of these galaxies.
Lauer (1986; see also Lachi\`eze-Rey \etal 1985) decomposed the four galaxies using ground-based images and
concluded that the two large companions are relatively undistorted, consistent with the hypothesis
that they are not strongly interacting with NGC 6166.  It was already known
that the brighter two companions differ in velocity from NGC 6166 by $-1520$ and $+570$ km s$^{-1}$
(e.{\ts}g., Minkowski 1961); these velocity differences are consistent with true separations that are
similar to the projected ones, but they do not clearly establish a close physical relationship.  We follow 
Lauer and assume that NGC\ts6166 itself is not affected by the companions.  We therefore calculate its
profile by masking out the companions.

(2) There is patchy dust absorption near the galaxy center.  We take this into account next.

\vfill

\clearpage

      Figure 8 illustrates both problems.  The~top~image~shows isophotes at average major-axis radii of
7\farcs9, 11\farcs7, 18\farcs6, and 24\farcs9.   Above the center, all contours except the one at 24\farcs9 
are substantially affected by the closest companion.~Various strageties were used to correct for the 
companions.~For some profiles, the companions were masked; for others, contaminated pixels were replaced 
by pixels from the opposite side of the galaxy center.  The same strategy was used on the dust contamination; 
the most reliable results were obtained by interpolating through the dust in the right-hand quadrants and 
then replacing the most strongly affected pixels

\figurenum{8}

\begin{figure}[hb!]

\vskip 439pt

\includegraphics{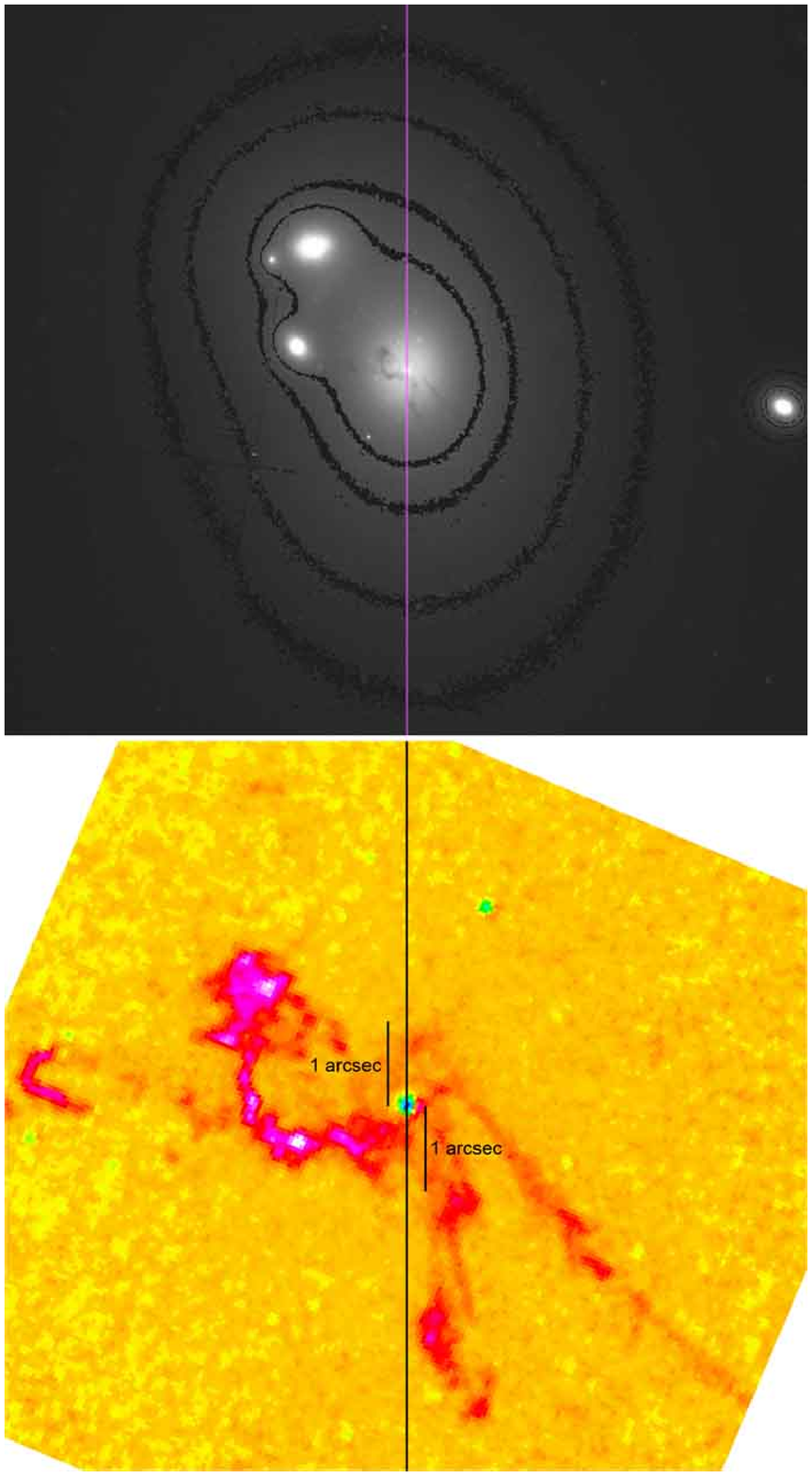}

\figcaption[]
{Central dust distribution in NGC 6166 shown ({\it top\/}) in the HST WFPC2 F555W $V$-band image and ({\it bottom\/}) 
by the ratio of an HST ACS F814W $I$-band image to an ACS F475W $g$-band image.  The scale of the top image is 
0\farcs1 pixel$^{-1}$'; the scale of the bottom image is 0\farcs05 pixel$^{-1}$.  Both images are rotated by
$\Delta PA = -23^\circ$ so that the inner major axis is vertical (see Figure 1).  In the bottom image, $g$- to
$I$-band flux ratios $f_g/f_I$ are color coded as follows:
yellow corresponds to $f_g/f_I = 1.0$;
red    corresponds to $f_g/f_I = 0.8$, and
white  corresponds to $f_g/f_I = 0.6$
The central blue pixel has $f_g/f_I \simeq 2.0$:~the central source has at least some contribution from an AGN.
Long vertical lines show the position of the 2-pixel-wide, $V$-band cut profile measured in the top image.  
This image also includes four contour levels to show how companion galaxies affect the isophotes.  We use only
these parts of the cut profile that are as unaffected by companions as possible. 
\vskip -10pt 
}
\end{figure}

\vskip 20pt

\noindent in the left quadrants by pixels from the opposite side of the center.  All these procedures are 
somewhat vulnerable, because isophote fitting requires many pixels that need correction.~So, as a check 
on the isophote fitting, we derive a major-axis cut profile along the vertical line in Figure\ts8.  The cut 
is 2 pixels = 0\farcs2 wide in the F555W WF image.  The lower part of Figure\ts8 shows that the cut is minimally 
affected by dust (a few pixels were corrected).  More importantly, we 
used pixels only from the bottom half of the image at radii where
the top half is affected by the companions shown and only from the top half of the image at much larger radii
where a companion not illustrated in the figure begins to be important.  

\figurenum{9}

\begin{figure}[hb!]

\vfill

\includegraphics{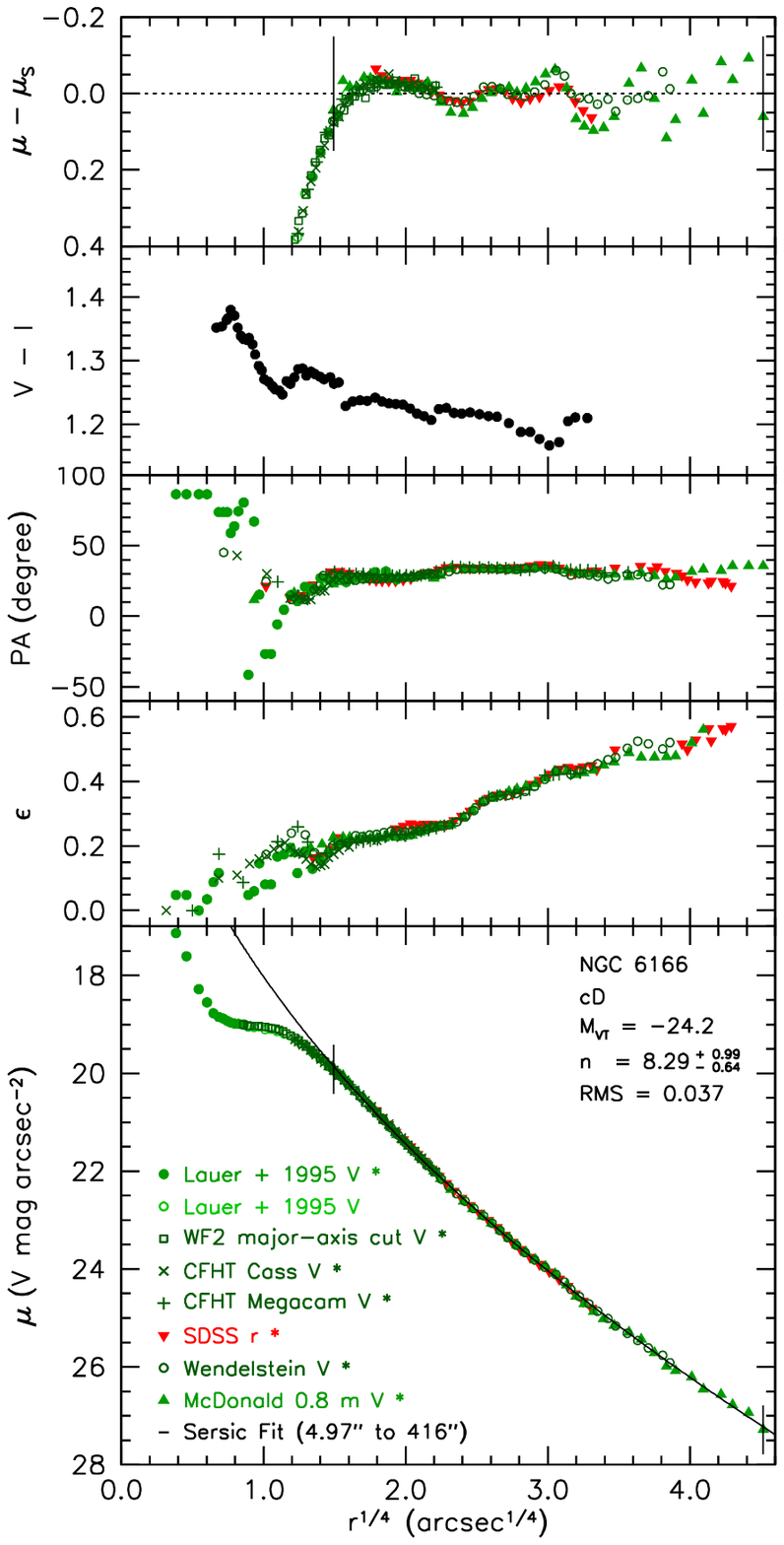}

\figcaption[]
{Major-axis profile measurements of NGC 6166:~those labeled~* in the key are used to calculate the average 
profile used in the analysis.  The curve is a S\'ersic fit in the radius range shown by the vertical dashes.
The fit RMS = 0.037 mag arcsec$^{-2}$; the residuals are shown in the top panel.  The next panel downward
shows the $V - I$ color profile via the $I$-band profile from the next section.  The brightness
profile shows no sign of two-component structure; i.{\ts}e., the cD halo is not distinguishable using 
photometry alone.
}
\end{figure}

\clearpage

      Figure 9 shows that the average $V$-band composite profile is robustly determined.
We have enough different data~sets with different problems (e.{\ts}g., non-flatness of the sky
brightness) so that agreement among data sets reliably identifies problem points.~They are pruned.~Near 
the center, the profiles that are corrected with Lucy (1974) - Richardson (1972) deconvolution --
i.{\ts}e., the ones from Lauer \etal (1995) and from the CFHT Cassegrain camera -- agree with the much
higher-resolution WF profile.  In fact, since the $V$-band cut profile is most free of dust effects, it
is used at~radii~near~1$^{\prime\prime}$ in preference to the Lauer \etal (1995) data.  (The difference
is only a few hundredths of a mag arcsec$^{-2}$ -- see Figure\ts11.)  Most~important:~{\it The major-axis 
cut profile agrees with the isophote fit profiles to \lapprox \ts0.02 $V$ mag arcsec$^{-2}$.  
The success of this check is important to our confidence in the final profile. } \vskip 0.5pt

      The average $V$-band photometry is tabulated in Table 3. 

\subsection{Composite $I$-Band Brightness Profile of NGC 6166}

      An $I$-band composite profile is derived in Figure\ts10, albeit from few sources.  We need it
primarily as another check of the $V$-band profile, including the ellipticity and position angle.
The central profile and VEGAMAG zeropoint are from an HST ACS F814W image (GO program 9293; 
H.{\ts}Ford,{\ts}P.{\ts}I.).  It helps that dust is less important at $I$ band.  However, we
can go further: the availability of an ACS F475W image (GO program 12238, W.~Harris, P.~I.)~allows 
us to make a dust-corrected image, as follows.

\figurenum{10}

\begin{figure}[hb!]

\vskip 4.5truein

\vfill

\includegraphics{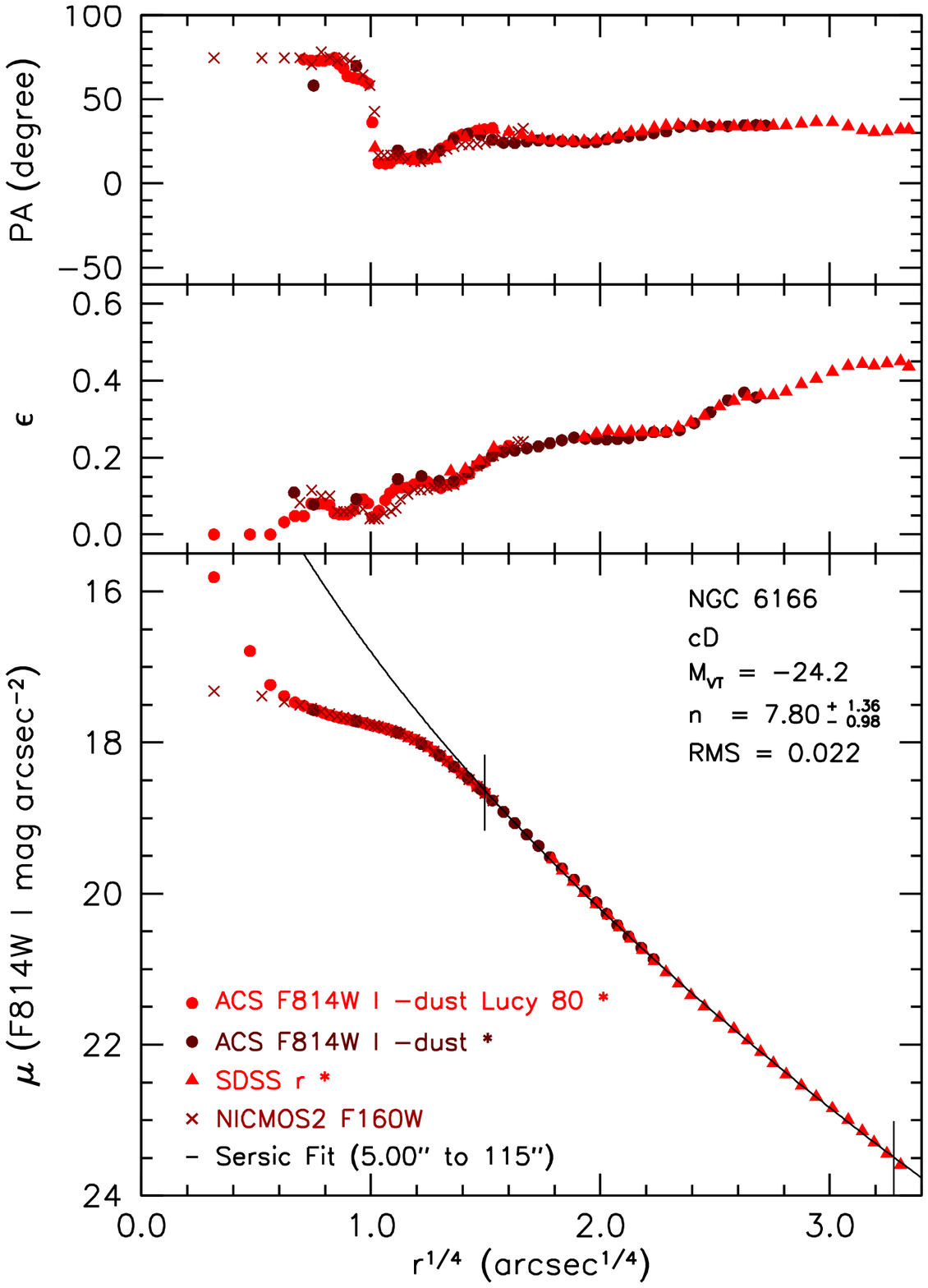}

\figcaption[]
{Major-axis $I$-band profile measurements of NGC 6166.  Profiles labeled * are averaged to make the
mean profile used in the analysis.  The curve is a S\'ersic fit in the 
radius range shown by the vertical dashes; the fit RMS = 0.022 mag arcsec$^{-2}$.  Again, there is 
no sign of two-component structure:~the cD halo is not distinguishable via photometry alone.
}
\end{figure}

      First, the F475W $g$-band image was rotated and registered to $\sim 0.2$-pixel accuracy with the
F814W $I$-band image.  Then a dust-corrected $I$-band image was derived using the procedure described in 
Nowak \etal (2008, Appendix A) and summarized here.  In the following, $f_g$ and $f_I$
are the F475W and F814W surface fluxes per square arcsecond; no subscript indicates magnitudes
or fluxes as observed, a subscript `0' refers to an extinction-corrected quantity. 
From the relation,

$$ A_I \equiv I-I_0 = \alpha E(g-I)~, \eqno{(2)} $$

\noindent where $A_I$ is the $I$ absorption and $E(g - I) \equiv (g - I) - (g - I)_0$ is
the reddening in the color $(g - I)$, it follows that:

$$ f_{I,{\kern 1pt}0} 
= \frac{f_I^{\alpha+1}}{f_g^{\alpha}} \frac{f_{g,{\kern 1pt}0}^{\alpha}}{f_{I,{\kern 1pt}0}^{\alpha}}~~. \eqno{(3)}$$

\vskip 7pt
 
\noindent If the stellar population gradient in the inner regions of NGC\ts6166 is negligible,
then $f_{g,{\kern 1pt}0}/f_{I,{\kern 1pt}0}$\ts$\approx${\ts}constant and thus:

\vskip 3pt

$$ f_{I,{\kern 1pt}0} \propto \frac{f_I^{\alpha+1}}{f_g^{\alpha}}~~.  \eqno{(4)} $$ 

\vskip 6pt

\noindent The parameter $\alpha$ is determined by

\vskip -3pt

$$ \alpha = (A_g/A_I -1)^{-1} \approx 1.0\ts,  \eqno{(5)}$$ 

\noindent where we have assumed a standard extinction curve to obtain the numerical
value for the filters considered here (e.{\ts}g. Savage and Mathis 1979).

      The correction is not perfect, because it is based on the assumption that all of the dust is 
in a screen in front of the image.  In NGC 6166, most of the dust is near the middle of the galaxy, 
in front of only about half of the stars.  Then Equation (5) overcorrects for the dust.  Better results 
are obtained if we adopt a smaller value for $\alpha$ (a value of $0$ would imply no correction).
After some experimenting,~we~adopt $\alpha=0.6$, which yields the smoothest appearance of the isophotes.  
Explicitly,

$$
I_{\rm dust-corrected} = I_{\rm observed}^{1.6} / g_{\rm observed}^{0.6} .  \eqno{ (6) }
$$

\noindent The residual dust contamination is small.  

      Then the brown circles in Figure 10 are derived from the dust-corrected image using Bender's isophote 
fitting program.  The red points are derived using {\tt VISTA} {\tt profile} on the dust-corrected image after
80 iterations of Lucy-Richardson deconvolution and after further cleaning of dust as discussed in \S\ts3.2.  
These profiles agree essentially perfectly.

      A final check is possible using an HST NICMOS2 F160W image (GO program 7453, J.~Tonry, P.~I.).
There is no star in the field of view, so we do not attempt PSF deconvolution.  But dust is essentially
unimportant.  The core profile calculated from this image also agrees very well with the $I$-band
results, when PSF blurring is taken into account.  In particular, the F160W profile confirms that
the core profile is cuspier at red wavelengths than it is in $V$ band.

\vfill\eject

\cl{\null}

\vfill

\includegraphics{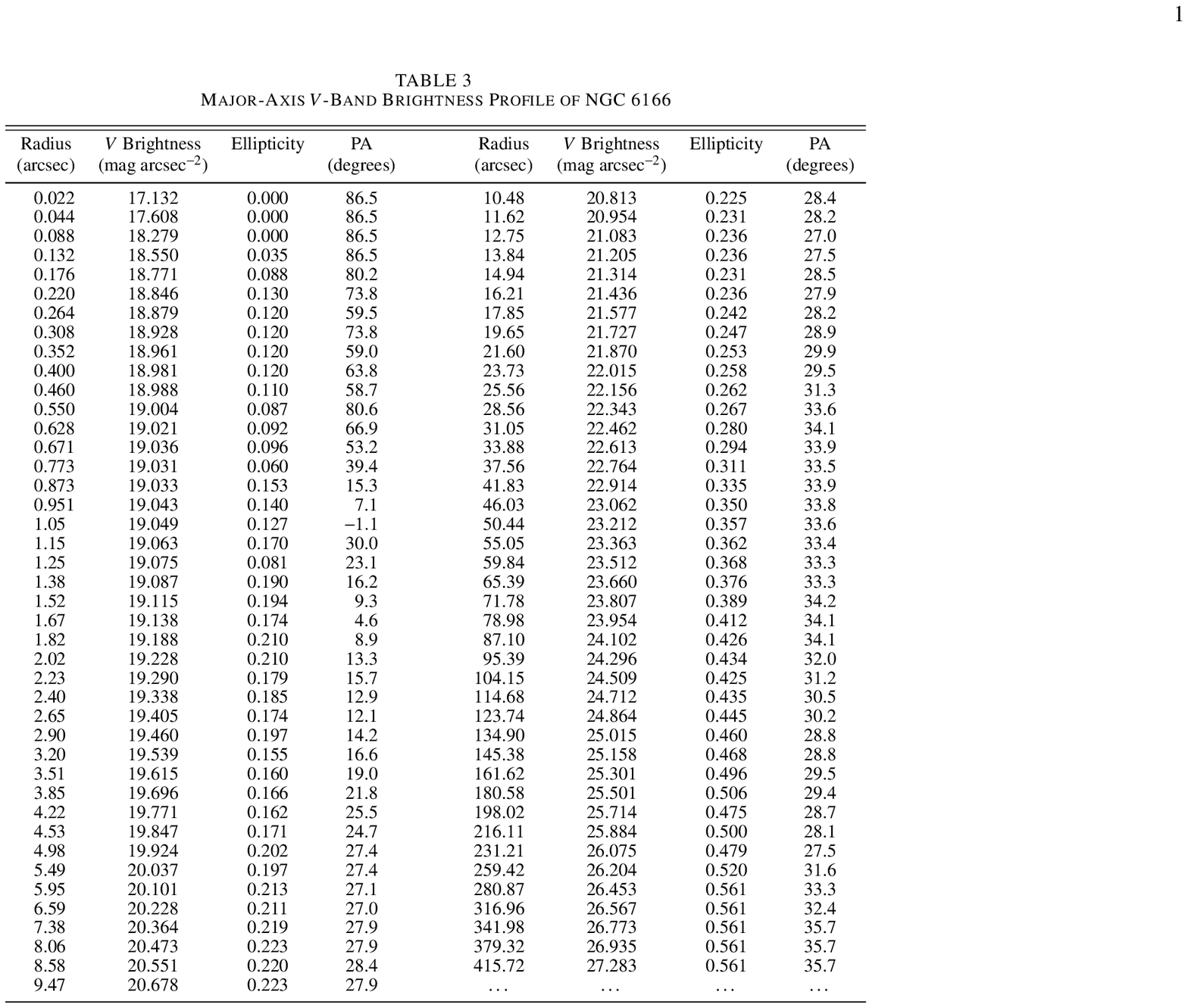}

\vfill

\cl{\null}

\clearpage

\subsection{Photometry Results. I. The Profile in the Core}

      Figure\ts11{\ts}illustrates{\ts}our{\ts}\S\ts3.3{\ts}conclusion:~{\it The{\ts}core{\ts}profile{\ts}of 
NGC\ts6166 is cuspier at red wavelengths than~it~is~in~$V$~band.  We suggest that the difference is caused by 
$V$-band absorption over the entire central arcsec of the galaxy.}  Clear hints of widespread, low-level 
absorption are visible in Figure 8.

      It is difficult to measure the power-law cusp slope~far~inside the profile break
radius $r_b = 2\farcs41$ (Lauer \etal 2007).  The reason is that the nuclear source is spatially
resolved and has an unknown profile.  Whether it consists of stars or an AGN or some combination,
we cannot subtract it robustly.  However, the shallowest $I$-band slope at $r \sim 0\farcs5$ to 1\farcs0  
corresponds to a Nuker function (Lauer et~al.~1995) $\gamma \simeq 0.13$.  This agrees with $\gamma = 0.12$
obtained in Lauer \etal (2007; any correction for the nuclear source is not discussed).~~Previous~estimates,
$\gamma = 0.08$ (Lauer et~al.~1995) and $\gamma = 0.081$ (Byun et~al.~1996), were determined from the Lauer
\etal (1995) $V$-band PC1 profile shown below.~Our $V$-band cut profile is even flatter than Lauer's 
profile{\ts}--{\ts}it is less affected by {\it patchy\/} dust{\ts}--{\ts}so our composite $V$-band profile is
even less cuspy than $\gamma \simeq 0.08$.

      The cuspiness of the central profile affects no conclusions of this paper.  But it will
be important to use the appropriate, dust-free profile if in future we obtain stellar kinematic data that
allow a dynamical search for a supermassive black hole.


\vfill

\figurenum{11}

\begin{figure}[hb!]

\vskip 4.2truein

\includegraphics{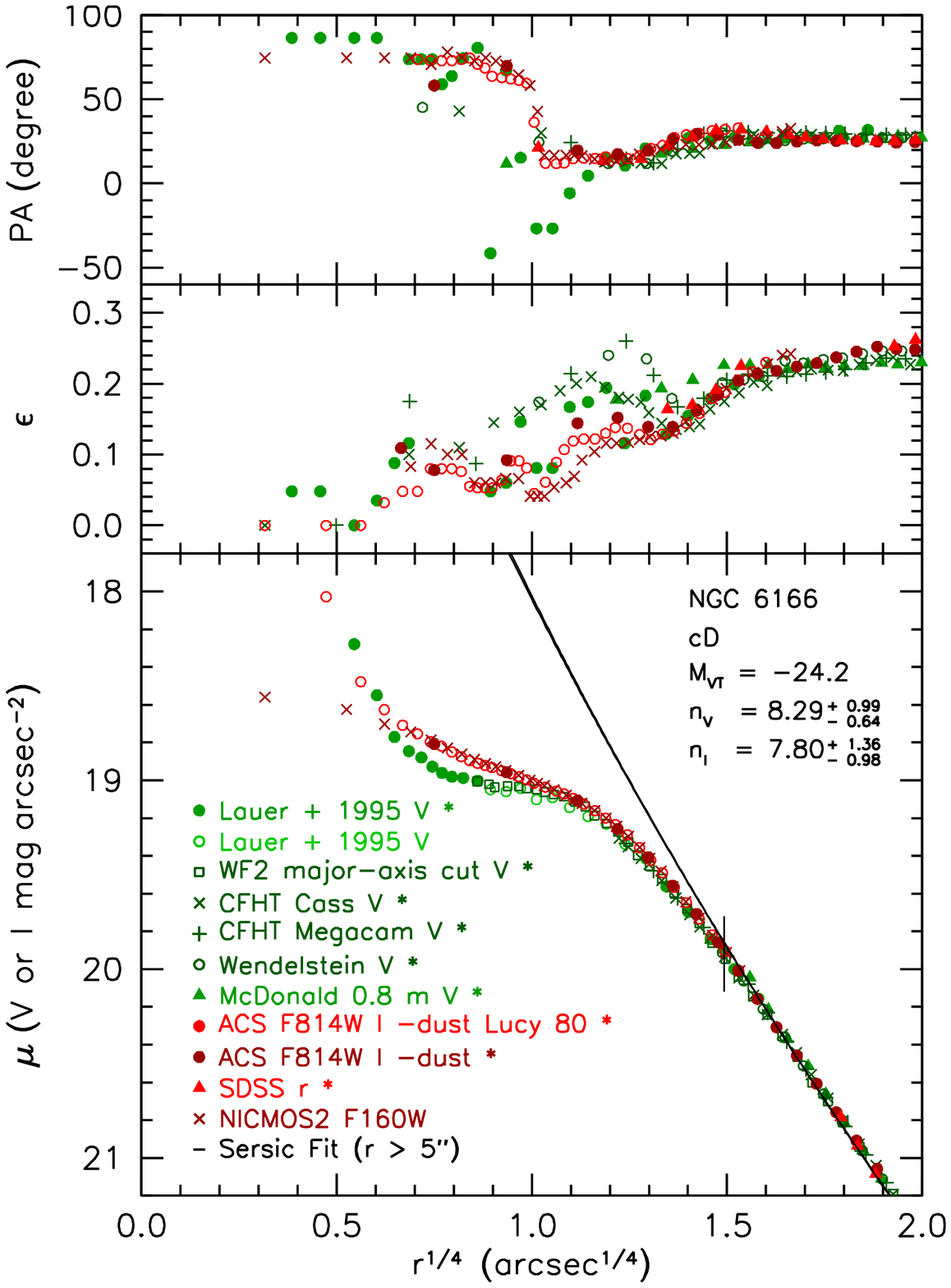}

\figcaption[]
{Major-axis $V$- and $I$-band profiles of NGC 6166 fitted together outside the core ($V - I = 1.24$).  
Both S\'ersic fits are also shown.  The purpose of this figure is to show that the core profile is
robustly cuspier in $I$ band than in $V$ band.  This is probably due to dust absorption in $V$ band in the
central $r \sim 1^{\prime\prime}$, as suggested also by Figure 8.  It is not due to PSF smearing; 
even 80 iterations of Lucy-Richardson deconvolution have essentially no effect on the shallow core profile.
Also supporting our interpretation is the observation that an
F160W HST NICMOS2 profile agrees with the $I$-band data as well as can be expected, given PSF blurring.
The difference between the $V$ and $I$ core profiles affects no conclusions of this paper,
but it should be kept in mind in making dynamical models to look for any central supermassive black hole.
\vskip -23pt
}
\end{figure}

\eject

\subsection{Photometry Results. II. The cD Structure of NGC\ts6166 is \hbox{Not Recognizable from the Shape of the
            Brightness Profile}}

      Our profile measurements in Figures\ts9 and 10 do not show the two-component structure that is 
so obvious in Figure 6.  We believe that Oemler (1976) profile is in error; the most likely reason is the
difficulty of correcting for the many cluster galaxies that overlap the cD halo.  Modern 
ellipse-fit software copes more robustly with incomplete isophotes.

      A single S\'ersic (1968) function fits the complete profile of NGC 6166 outside the cuspy core. 
Both this result and the S\'ersic~index, $n$\ts$=$\ts$8.3^{+1.0}_{-0.6}$ in $V$ band or $7.8^{+1.4}_{-1.0}$ 
in $I$ band, are completely normal for core-boxy-nonrotating ellipticals.  Figure 12 compares NGC\ts6166's 
profile shape with the sample of elliptical galaxies studied by KFCB.  They found that $n$ ranges from 
$5.4 \pm 0.3$ to 9\ts$\pm$\ts1~for~their core ellipticals (red profiles in Figure 12).~NGC\ts6166 is virtually
indistinguishable from these galaxies; indeed, many core ellipticals have shallower outer profiles
$\log{I(r/r_b)}$ than does NGC\ts6166.~It is especially interesting to contrast NGC\ts6166 with M{\ts}87.
M{\ts}87 is by all arguments a more marginal cD than NGC 6166.  But a 
S\'ersic fit to its overall profile gives $n = 12^{+2}_{-1}$, larger than $n \simeq 8$ in NGC\ts6166.  
Plausible allowance for a cD halo in M{\ts}87\ts--{\ts}i.{\ts}e., exclusion of the outermost profile points -- 
gave a marginally better fit with $n = 9^{+2}_{-1}$, consistent with our fit to NGC 6166 but with only a
little extra light in the cD halo of M{\ts}87.  Such a halo is 
less{\ts}--{\ts}not more{\ts}--{\ts}obvious in NGC\ts6166.

\vskip 3.00truein

\figurenum{12}

\begin{figure}[hb!]

\vfill

\includegraphics{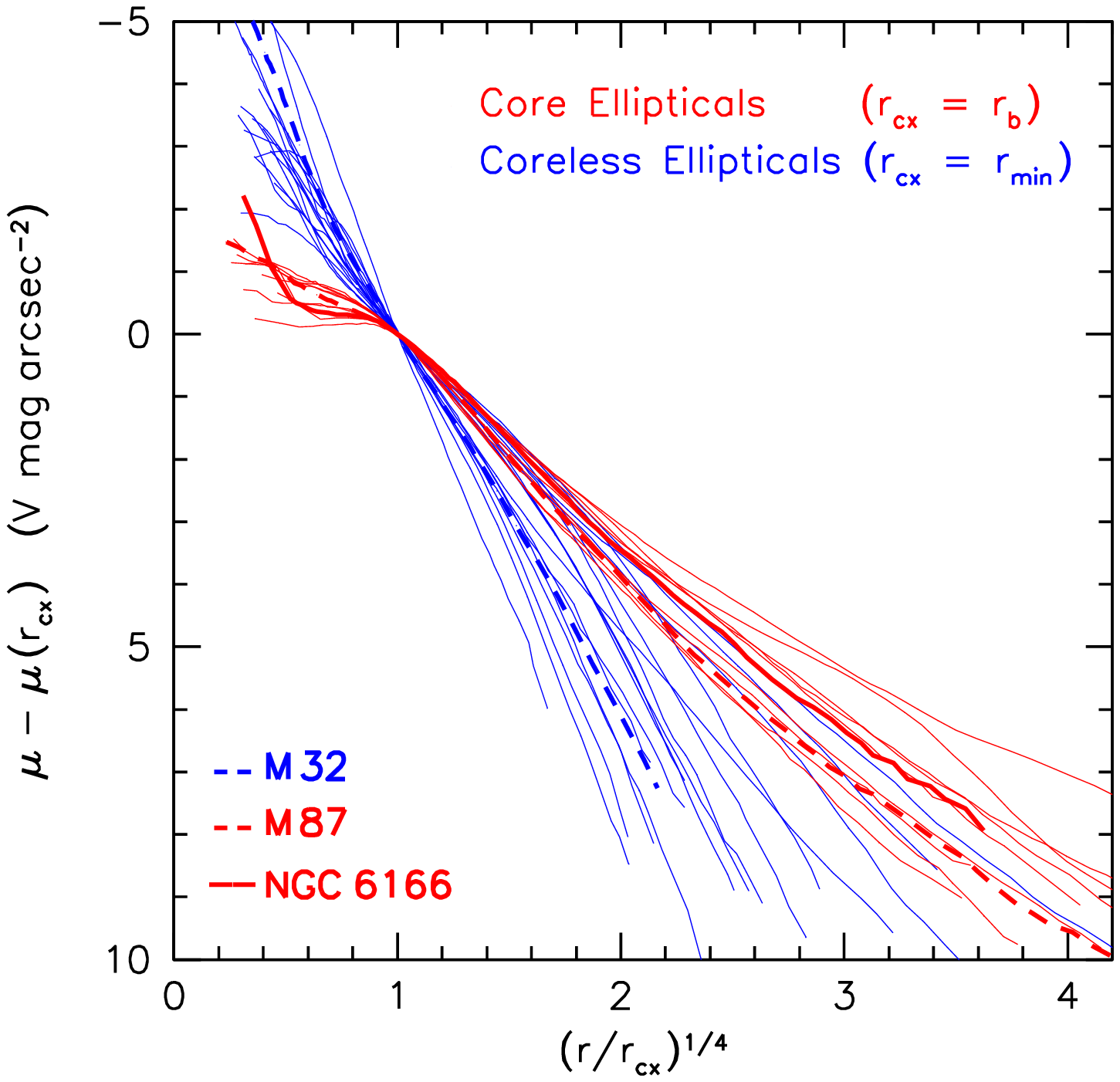}

\figcaption[]
{Major-axis profiles of all KFCB elliptical galaxies scaled together in radius and surface brightess.  
Core ellipticals are scaled at $r_{cx} = r_b$,
the break radius given by the Nuker function fit in Lauer \etal (2007).  Coreless ellipticals
are scaled at the minimum radius $r_{\rm min}$ that was used in the KFCB S\'ersic fits; inside
this radius, the profile is dominated by extra light above the inward extrapolation of the 
outer S\'ersic fit.
NGC 6166 and the fiducial galaxies M{\ts}87 and M{\ts}32 are plotted with thick lines.
There is no sign of two-component structure in NGC 6166; its profile resembles those
of other core galaxies.  I.{\ts}e., the cD halo is not distinguishable using photometry~alone.
}
\end{figure}

\phantom{00}

\vskip -25pt

\vskip 5pt

      A two-component, S\'ersic-S\'ersic decomposition is allowed by our data (\S\ts4), but
the fit is not significantly better than the one-component decomposition.  There is no reason to 
believe that we detect two components from photometry alone.

      This is a surprising result.  We plan but have not yet carried out similar photometry of
other cD galaxies.  We therefore do not know that the present results on NGC 6166 apply
more generally to all cD galaxies.   Nevertheless: 

      We arrive at an ironic situation: {\it The spectroscopy results resoundingly
confirm our standard picture that the cD galaxy NGC\ts6166 in Abell\ts2199 has an outer halo that
consists of debris from member galaxies.  The halo stars are dynamically controlled
by the cluster, not the central galaxy, and they have the kinematics (i.{\ts}e., more nearly the
systemic velocity and the velocity dispersion) of the other galaxies in the cluster,
even when the cD is dynamically colder and in motion with respect to the sea of 
background stars.  But the supposedly much easier task of recognizing the presence of
a cD halo from two-component structure in the surface brightness distribution turns
out to fail dramatically in the nearest, most prototypical cD galaxy, NGC 6166.}

\subsection{Photometry Results. III. Recognizing NGC 6166 as a cD\\Galaxy via Quantitative 
            Differences in Structural Parameters}
      
     Is it possible to recognize cD galaxies by photometry~alone?  A photometric technique is 
desirable, because spectroscopy to look for an outward rise in velocity dispersion is expensive.
Our results suggest a partial answer: The cD nature of NGC\ts6166 can be recognized via quantitative 
differences in structural parameters and parameter correlations.  This helps but is not entirely satisfactory.  
Parameter distributions 
for cD galaxies and non-cD ellipticals overlap.  There may be physics in this.  The physical differences 
between cDs and core-boxy-nonrotating ellipticals may be smaller than we have thought.  Figures 13 and 14
illustrate these points.

      Figure 13 compares the brightness profile of NGC 6166 to the Virgo cluster elliptical galaxies. Radii are
plotted in kpc.  NGC\ts6166 has a larger and fainter core than any elliptical in Virgo, including
M{\ts}87.~~~And its outer profile is shallower and it reaches larger radii than that of any elliptical in Virgo, 
including M87.  Quantitatively, the 
extreme cD NGC 6166 is distinguishable from normal core ellipticals.  However, the marginal cD M{\ts}87 
(see KFCB) overlaps with other core ellipticals in its profile properties.

\begin{figure}[hb!]

\figurenum{13}

\vskip 3.2truein

 \includegraphics{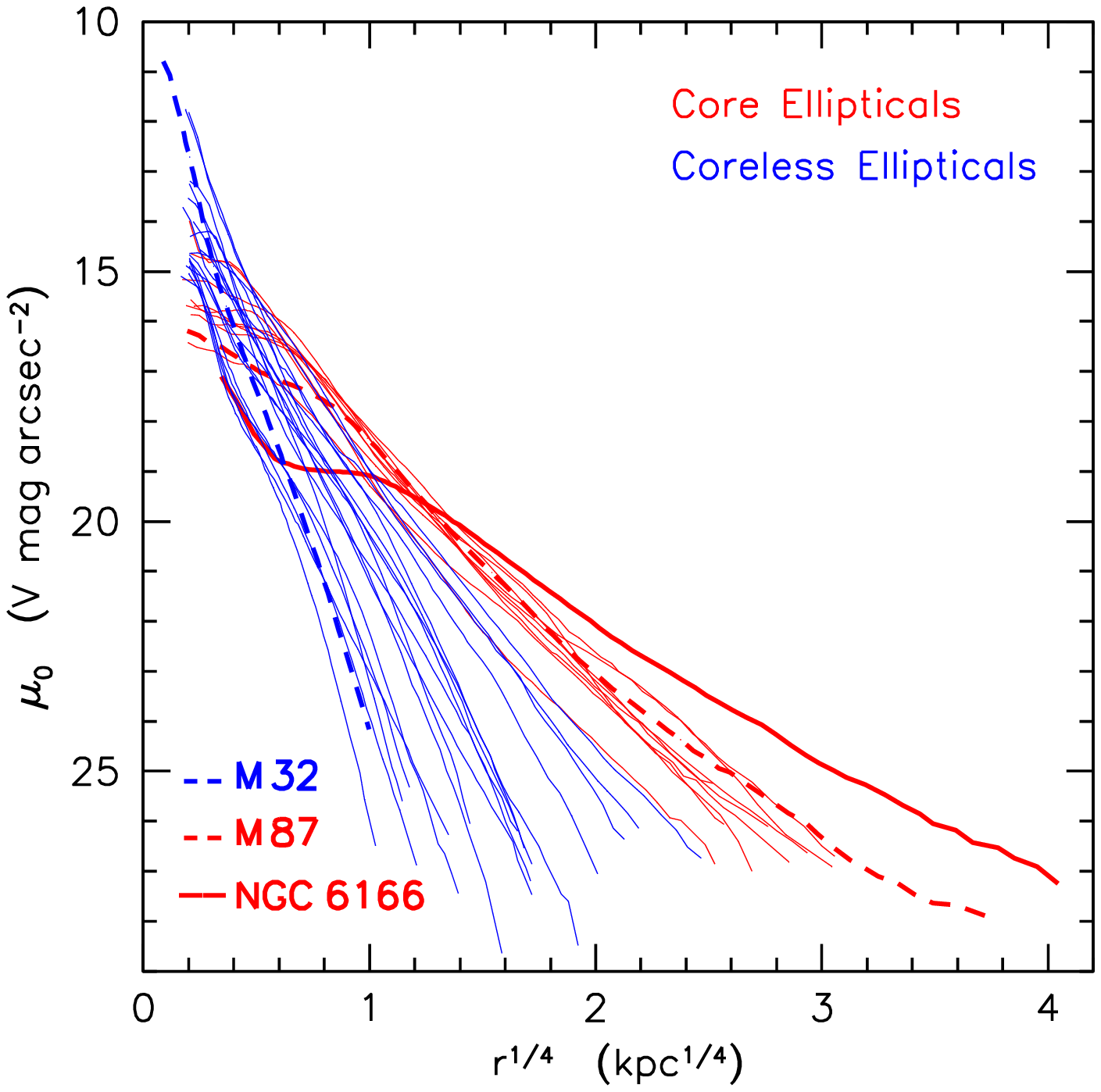}

\figcaption[]
{Major-axis profiles of all KFCB elliptical galaxies scaled so that radius is in kpc.  
The brightness profiles are corrected for Galactic absorption as in Schlegel \etal (1998).  
NGC 6166 is added; it and the fiducial galaxies M{\ts}87 and M{\ts}32 are plotted with thick lines.
NGC 6166 is not distinguished from the other galaxies by profile shape, but its parameters
are extreme.  That is, the cD halo is distinguishable quantitatively via the shallow outer 
profile and the consequently large effective radius.
}
\end{figure}

      Figure 14 compares the structural parameters of NGC\ts6166 with parameter correlations from KFCB 
and from Kormendy \& Bender (2012).  These are projections of the ``fundamental plane'' correlations 
(Djorgovski \& Davis 1987; 
Faber \etal 1987;
Dressler \etal 1987; 
Djorgovski \etal 1988; 
Djorgovski 1992; 
Bender \etal 1992, 1993),
between the effective radius $r_e$ that encloses half of the light of the galaxy, the effective brightness
$\mu_e$ at $r_e$, and (in this case) total absolute magnitude.

      NGC\ts6166 parameters are based on an assumed distance of $D = 130.8$ Mpc (NASA/IPAC Extragalactic Database ``NED''
$D$(Local Group) for cluster Abell\ts2199 and the WMAP~\hbox{5-year} cosmology parameters, Komatsu \etal 2009).  
NGC\ts6166 is plotted twice in Figure\ts14:

      To get the less extreme point, we integrate the brightness and ellipticity profiles (that is, the 
two-dimensional isophotes) to the outermost data point in Figure\ts9, i.{\ts}e., $r = 416^{\prime\prime}$ where
$\mu_V = 27.28$ $V$ mag arcsec$^{-2}$.  This gives $V = 11.75$, $M_V = -23.86$, $r_e = 71\farcs2 = 45.2$ kpc, and
$\mu_e = 23.76$ $V$ mag arcsec$^{-2}$.

\centerline{\null}

\figurenum{14}

\vskip 0.2truein

\vskip 4.95truein

\includegraphics{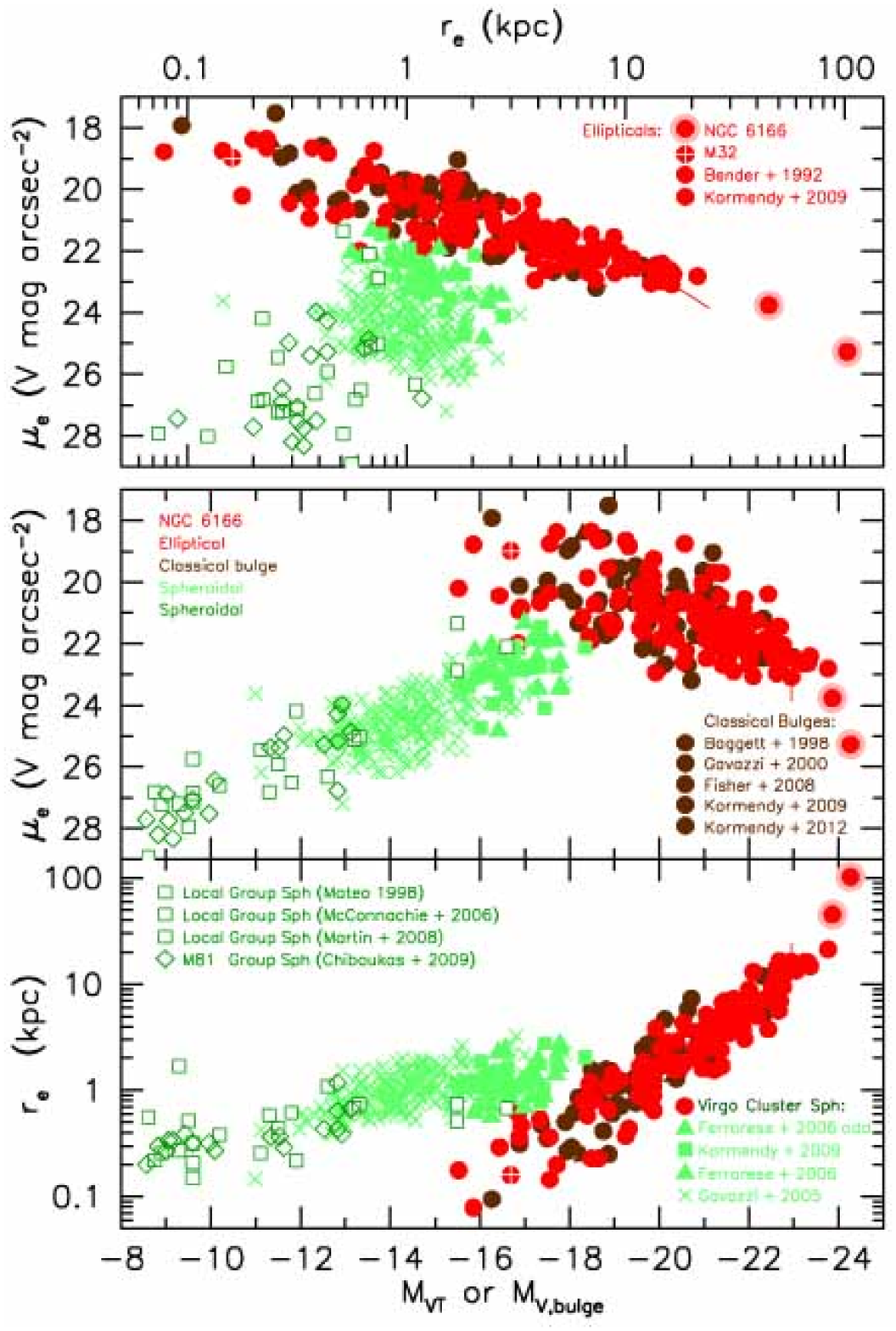}

\figcaption[]
{Structural parameter correlations for elliptical and spheroidal galaxies.  Major-axis effective
radii $r_e$ and effective surface brightnesses \hbox{$\mu_e \equiv \mu(r_e)$} are calculated by integrating 
isophotes with the observed brightness and ellipticity profiles out to half of the total luminosity.  
S- and S0-galaxy bulge parameters are from S\'ersic-S\'ersic or (when appropriate) S\'ersic-exponential
photometric decompositions into bulge and disk components.  The galaxy sample is from KFCB and from 
Kormendy \& Bender (2009).   
When necessary, mean-axis parameters are corrected to the major axis.  NGC 6166 is plotted twice; the
smaller-$r_e$ point is for the integral of the surface brightness distribution out to the last data point
in Table 3.  To derive the outer point, the observed profile is extended to 30.9 $V$ mag arcsec$^{-2}$ 
using the S\'ersic fit and keeping the ellipticity fixed at the value at the largest radii observed. 
Again, the cD nature of NGC 6166 together with its cluster-sized halo is evident quantitatively from 
the structural parameters.
\vskip -10pt
}

\centerline{\null} \vskip -22pt

\noindent Galactic{\ts}absorption{\ts}corrections{\ts}are{\ts}from{\ts}Schlegel{\ts}et{\ts}al.\ts(1998).  
This point in Fig.\ts14 is consistent with a slight extrapolation to higher luminosity of the 
correlations for other ellipticals.  

      The more extreme point is derived by extending the profile to $r \simeq 2000^{\prime\prime}
\sim 1.3$ Mpc using the overall S\'ersic fit and keeping the outer ellipticity constant at the last observed value.
The limiting surface brightness is 30.9 $V$ mag arcsec$^{-2}$; this is an ``integration to infinity''
similar to those discussed in KFCB.  Then $V_T = 11.35$, $M_{VT} = -24.27$, $r_e = 162^{\prime\prime} =
103$ kpc, and $\mu_e = 25.27$ $V$ mag arcsec$^{-2}$.  Within the scatter, this point is consistent with a larger
extrapolation of the correlations for normal ellipticals.  It deviates slightly from linear correlations in
having larger $r_e$ and fainter $\mu_e$, but slightly curved fits to normal ellipticals would not show NGC 6166 as deviant.

      We conclude that NGC\ts6166 is more extreme than the ellipticals in the combined sample in Figure 14
in the sense expected for a cD: It has larger effective radius and fainter effective brightness.  In this
sense, the cD structure is recognizable quantitatively in the parameter correlations.  

      cD and non-cD galaxies overlap in parameter distributions (Schombert 1986, 1987).
And yet, the cD NGC 6166 is qualitatively different~from~\hbox{non-cD} ellipticals, 
even brightest cluster galaxies.  This important, because cD and brightest cluster galaxies are
often considered to be equivalent.  But NGC 6166 is surrounded by an immense halo
of stars that are controlled dynamically by the cluster potential, not by the central galaxy.
Isolated ellipticals cannot have such halos, and observations of velocity dispersion profiles
in non-cD core ellipticals show no rise in $\sigma$ at large radii (e.~g.,\
Kronawitter \etal 2000;
Proctor \etal 2009;
Weijmans \etal 2009;
Foster \etal 2011;
Raskutti, Greene, \& Murphy 2014).

\vskip 4.76truein

\includegraphics{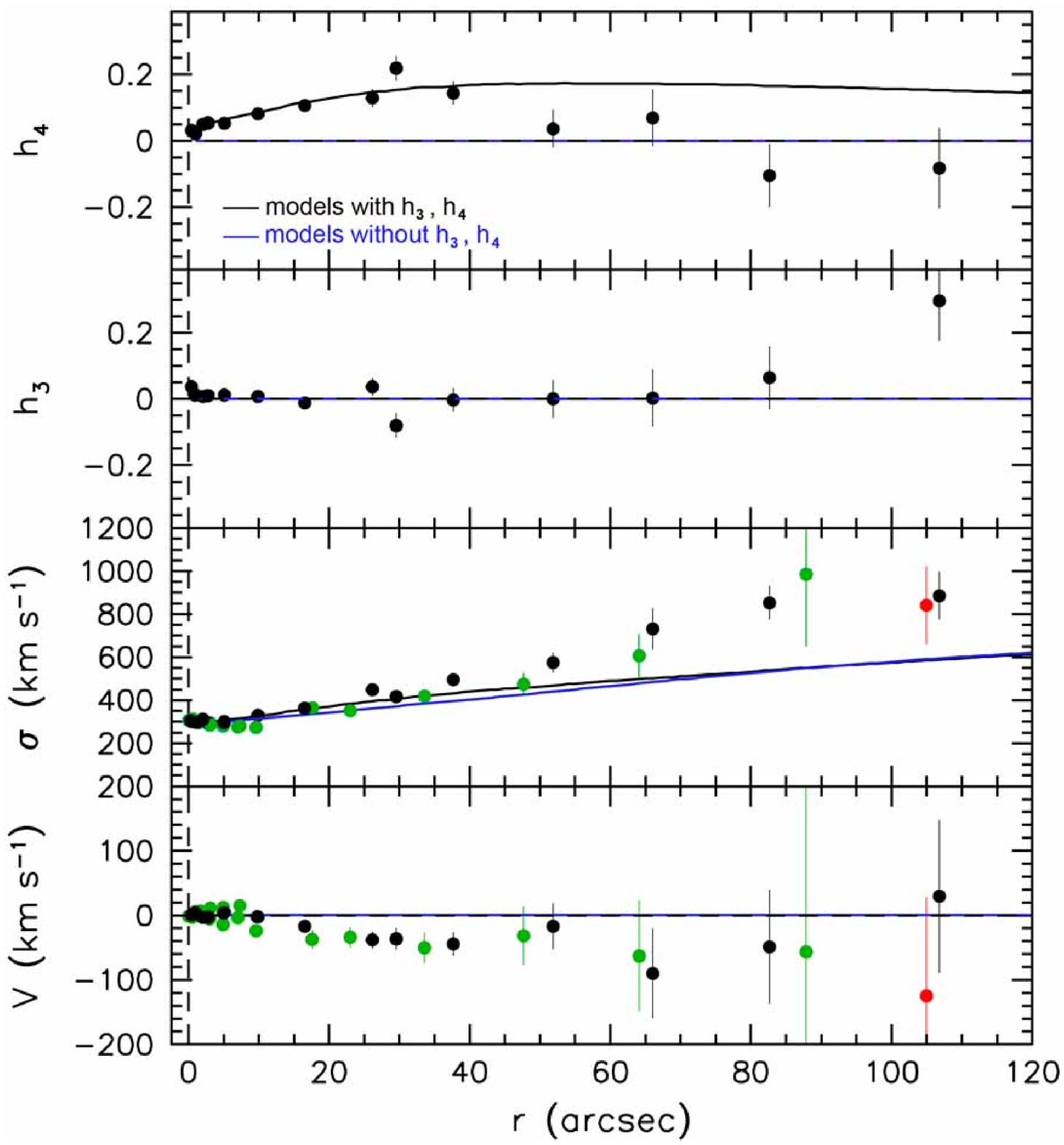}

\includegraphics{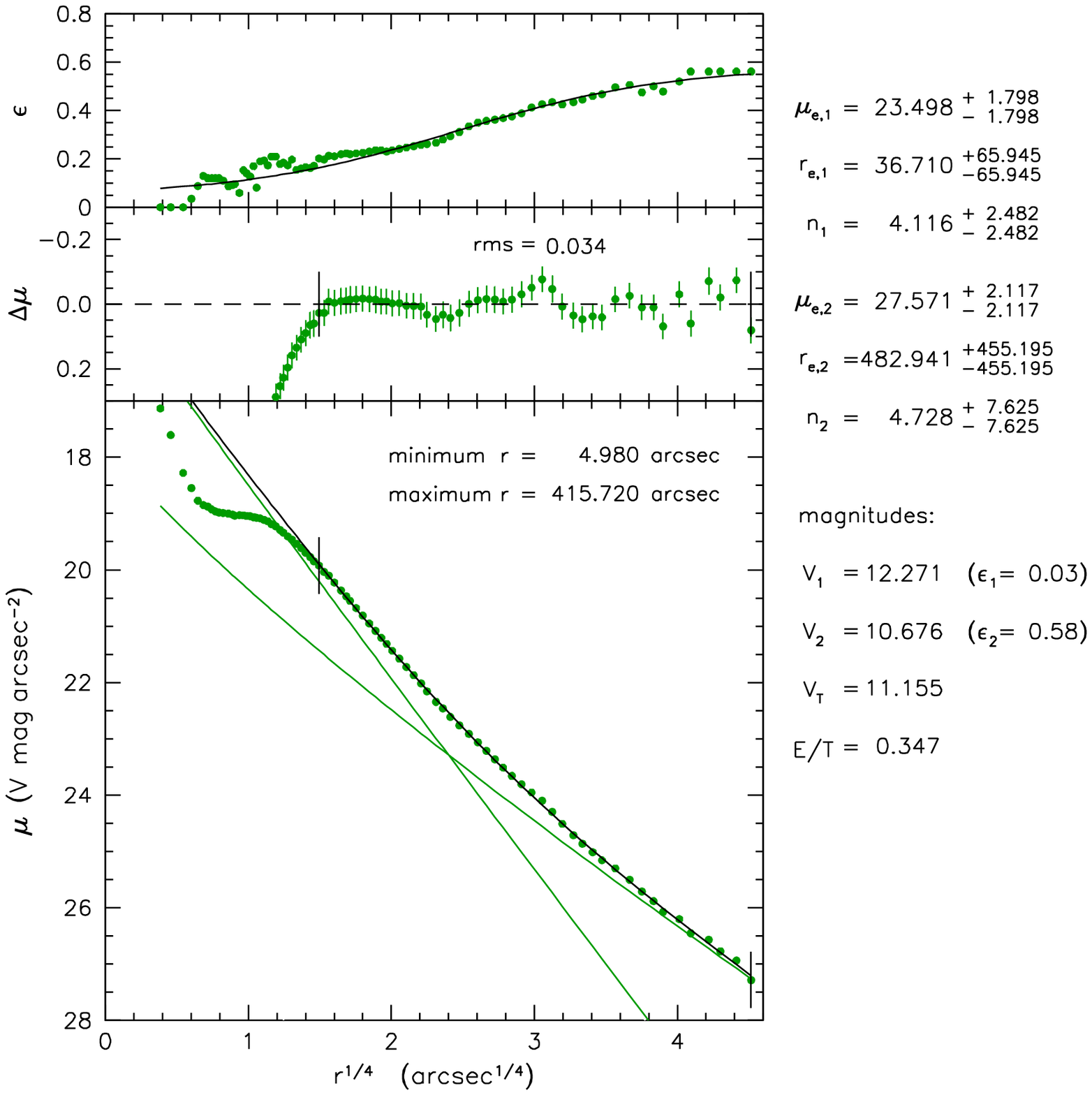}

\includegraphics{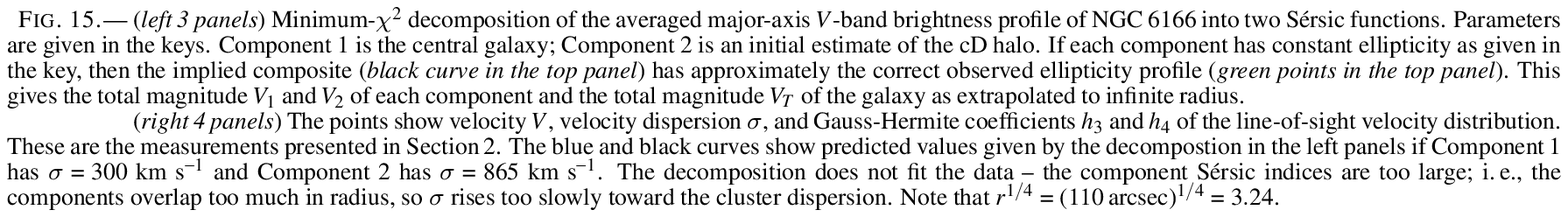}

\eject
      We conclude (1) that cD structure is real and distinct from non-cD ellipticals but 
(2) that it is difficult to recognize the difference photometrically.  Extreme structural parameters help (Figure 14).  
But in less extreme cases -- and, to be certain, even in NGC 6166 -- velocity dispersion data are required 
to identify cluster halos reliably.  The fact that cD classification is difficult is our problem, not the galaxy's.  


\section{A Photometric and Kinematic Decomposition of\\NGC 6166 Into an Elliptical Galaxy 
         Plus a \lowercase{c}D Halo}

      \hbox{~~~This section presents a decomposition of the inner,~\hbox{E-galaxy}} \noindent part of 
NGC\ts6166 and its cD halo that accounts for both the photometry and the velocity dispersion profile
of the galaxy.

      The best-fit two-component S\'ersic-S\'ersic decomposition is illustrated in the left part of Figure 15.
We emphasize: the RMS deviations 0.034 $V$ mag arcsec$^{-2}$ of the profile from the fit within the fit
range ({\it vertical dashes across the $\mu$ and $\Delta\mu$ profiles\/}) are not significantly better
than the deviations (Figure\ts9 RMS = 0.037 $V$ mag arcsec$^{-2}$) of a single-S\'ersic~fit.  

      The decomposition in Figure 15 is similar~to~those~in Huang \etal (2013a, b)\ts--{\ts}it minimizes
$\chi^2$~for~two~S\'ersic~components.  Huang and collaborators interpret such decompositions as supporting
a \hbox{two-phase} scenario of elliptical~galaxy~formation
(Oser \etal 2010; 
Johansson \etal 2012) 
in which wet mergers rapidly~\hbox{build~high-$z$,} compact ``red nuggets''
(Buitrago \etal 2008;
van Dokkum \etal 2010; 
Papovich \etal 2012; 
Szomoru \etal 2012)
that later grow high-S\'ersic-index halos via minor mergers.  The inner component(s) in the decomposition
are interpreted as descendent(s) of the red nuggets, and the outer component is interpreted as a later-accreted
debris halo.  Such a picture may be correct.  But (1) it is not compellingly supported by the
conclusion that two components fit the data better than one, and more importantly, (2) NGC 6166, with its
cD halo, is a clearcut example of essentially the above processes, and in it, a two-component 
decomposition made by minimizing $\chi^2$ fails to explain the kinematics.  As follows:

\vskip 4truein

\eject

      The observed dispersion profile implies that the central galaxy contributes most of the light along
the line of sight out to $r \simeq 50^{\prime\prime} \sim 32$ kpc ($ = 130.8$ Mpc).  The brightness profile 
extends out to $r \simeq 416^{\prime\prime} \simeq 260$ kpc~in~the~cD~halo.  In the transition region, 
we look through a short line of sight through the galaxy and a long line of sight~through~the~halo.  This suggests 
a simple procedure to capture the essence of the $\sigma(r)$ profle.  We assume that the components have
independent Gaussian LOSVDs.~To~keep~things~simple, we assume that the galaxy has the brightness profile
of Component\ts1 in Figure 15 and that it has $\sigma \sim 300$ km s$^{-1}$ at all radii.  We assume that
the cD halo has the brightness profile of Component 2 and $\sigma \sim 865$~km s$^{-1}$ at
all radii.  This is an oversimplification.  But if the decomposition in Figure 15 is
approximately correct, then it should approximately~fit~the~dispersion~profile.~~It~fails.  The components
overlap too much in radius; i.{\ts}e., the inner component contributes too much light at large radii for the
dispersion profile to increase outward as quickly as we observe toward $\sigma \sim 850$ km s$^{-1}$.  
Modifying the assumed inner and outer dispersions does not help.

      {\it So a two-S\'ersic-component photometric decomposition that minimizes $\chi^2$ fails to explain
the velocity dispersion profile of NGC\ts6166.~This argues for caution in the increasingly popular practice 
of making minimum-$\chi^2$, \hbox{S\'ersic-S\'ersic} decompositions of elliptical galaxies based on photometry 
alone.~It does not work in NGC\ts6166, where the~$\sigma(r)$~profile provides physically motivated guidance in
how to interpret the results.~This does not argue for confidence in decompositions of giant-boxy-coreless
ellipticals that are well fit by single S\'ersic functions and in which monotonically decreasing $\sigma(r)$ 
profiles provide no guidance about which decompositions measure something that is physically meaningful.}

\centerline{\null}

\vskip 4.47truein

 \includegraphics{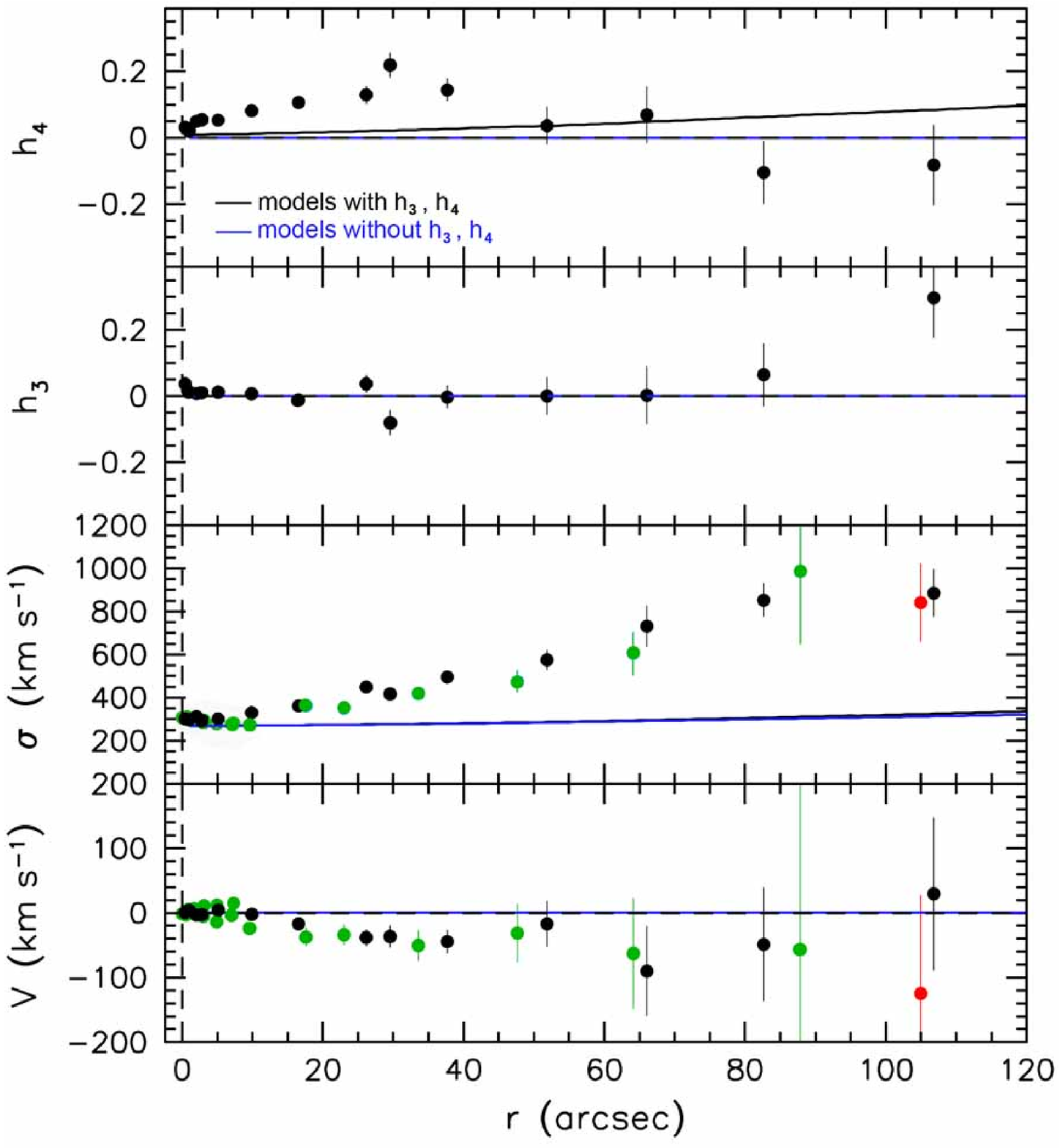}

 \includegraphics{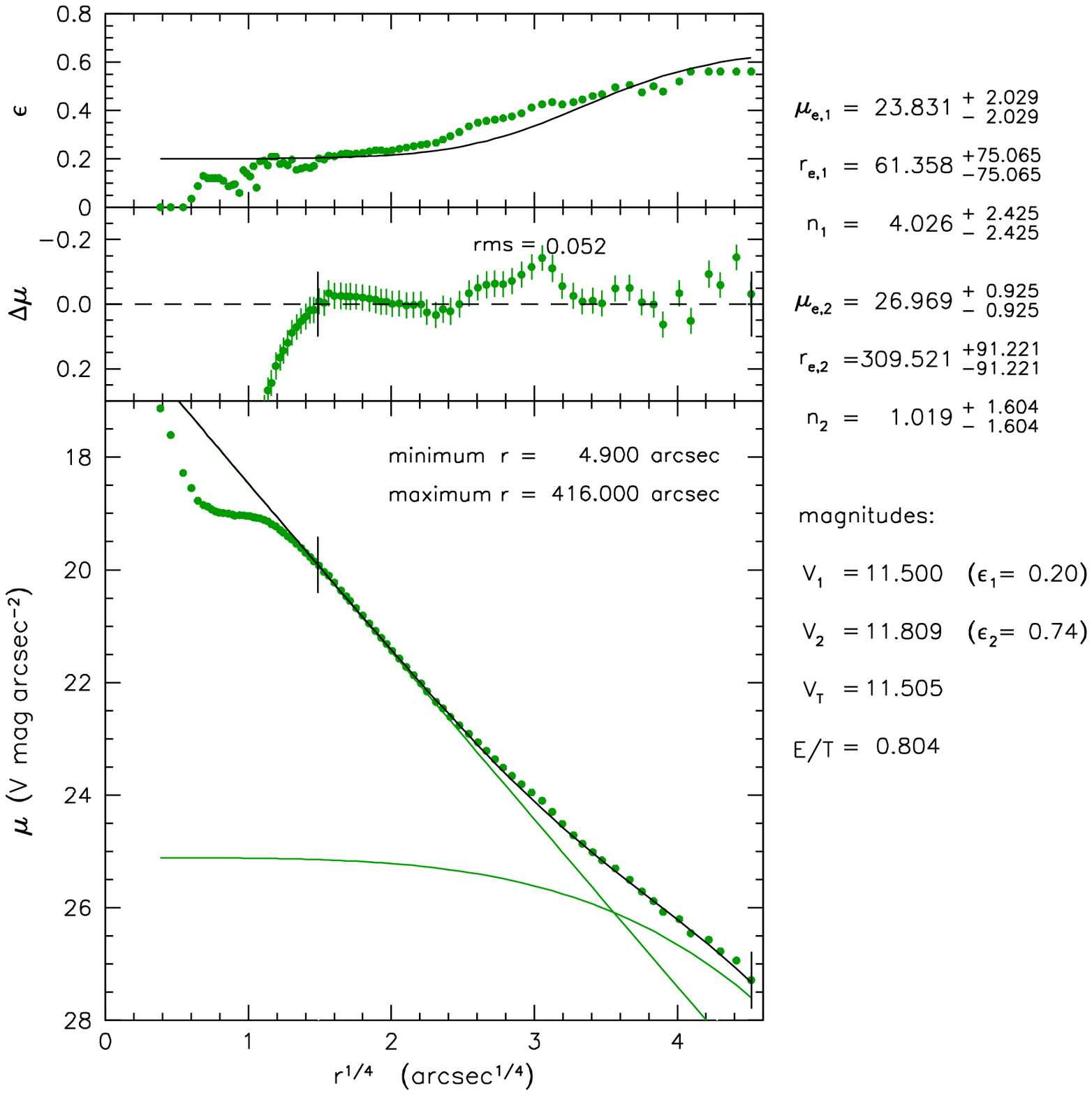}

 \includegraphics{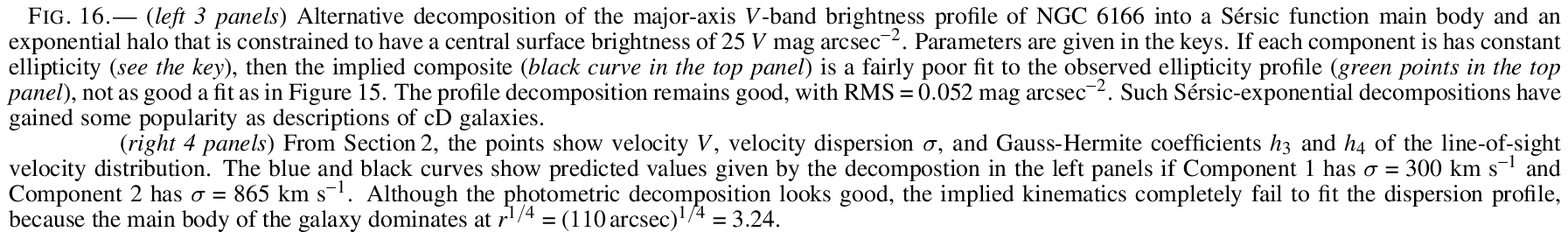}

\vskip 0pt

     \hbox{~~~~~~Figure\ts16 tries a different kind of photometric decomposition} \noindent that has been used to estimate 
the properties of cD halos.  E.{\ts}g., Seigar \etal (2007) and Donzelli \etal (2011) fit cD halos with exponential profiles.
Since NGC 6166 is well fitted by a single S\'ersic function, a S\'ersic-exponential decomposition has a larger $\chi^2$ 
with respect to the photometric observations.  It is therefore necessary to apply some additional constraint to force 
the program to find an exponential halo.~We~tried various decompositions in which the central surface brightness 
was constrained.  All such decompositions behave similarly if we require that the RMS of the fit be consistent with 
measurement errors.  Figure 16 shows an example in which the exponential is forced to have a central surface brightness 
of 25 $V$ mag arcsec$^{-2}$.  The fit RMS = 0.052 $V$ mag arcsec$^{-2}$ is worse than RMS = 0.037 $V$ mag arcsec$^{-2}$
in Figure\ts9 but is not excluded by the data.   However, this halo is much too faint.  The main galaxy contributes
essentially all the light at radii where we have kinematic data, so the dispersion profile fails to rise significantly
toward the outer observed value.

      {\it Again, we conclude that S\'ersic-exponential decompositions of cD galaxies -- at least in the case of NGC\ts6166 --
are not well constrained physically using photometry alone.}

      The ``cure'' is to make the two components be as separate as possible by decreasing both S\'ersic 
indices.  The resulting best fit gets worse  -- gets, in fact, increasingly {\it inconsistent\/} with the
photometric measurement errors -- but the fit to the dispersion profile gets better.  Figure 17 
shows the decompositions (two of many that we tried) that best~fit~$\sigma(r)$.  Given the crude assumptions, 
it makes no sense to look for further improvement; the way to get a better fit is to make
a full~Schwarzschild~(1979,~1982) model of the photometry and the kinematics.  We save this exercise for a 
future paper.  Here, we conclude that NGC\ts6166 and its cD halo are more distinct than a
minimum-$\chi^2$ photometric decomposition suggests. 

\vfill\eject\clearpage

      Figure 17 shows that, to fit the $\sigma(r)$ profile of NGC\ts6166, we need to make a 
photometric decomposition that does not minimize $\chi^2$.  This is no disaster: We chose S\'ersic 
functions for each component, and our experience that they fit non-cD ellipticals well (KFCB) may not be relevant here.  

\vskip 4.0truein

 \includegraphics{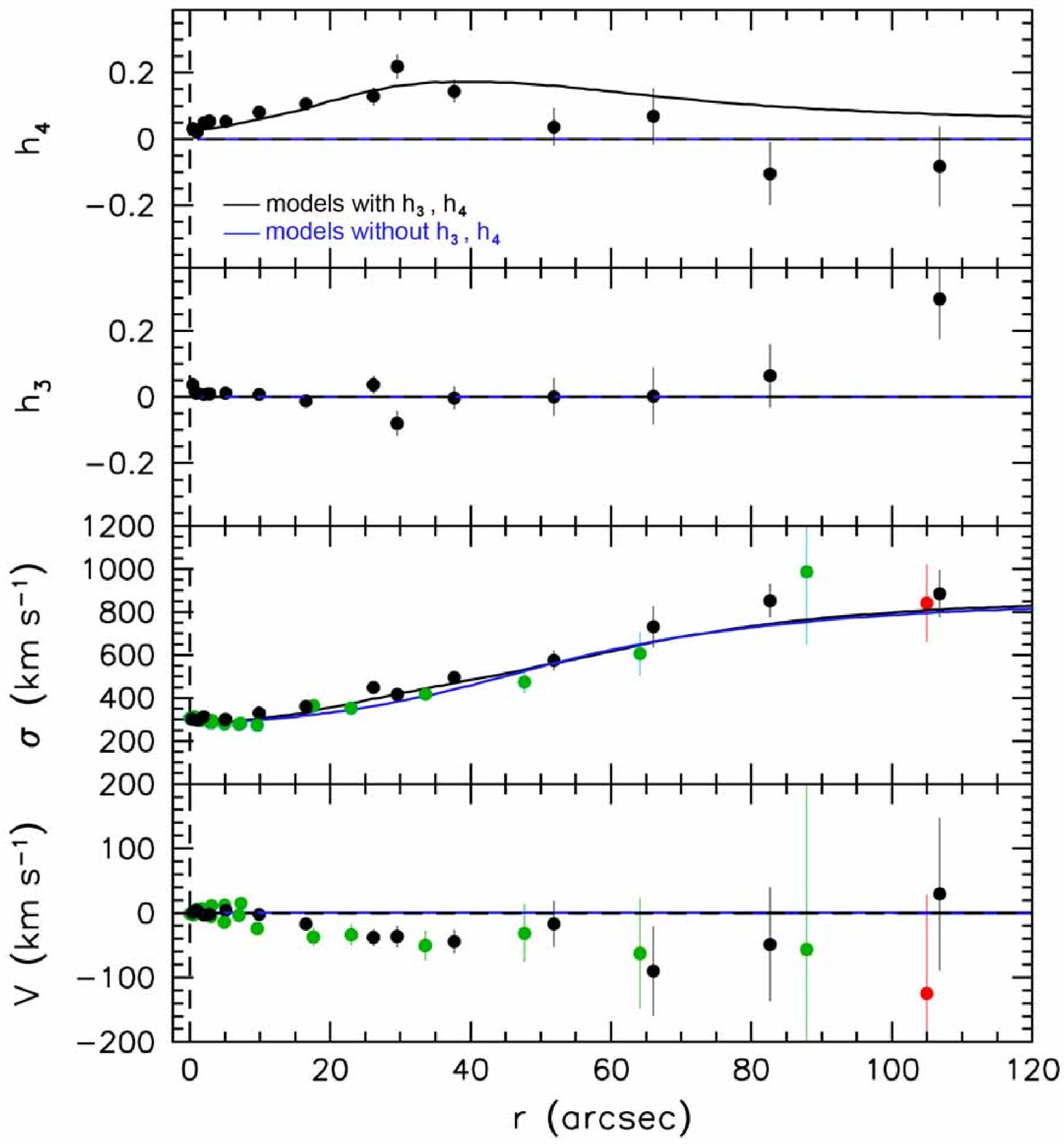}

 \includegraphics{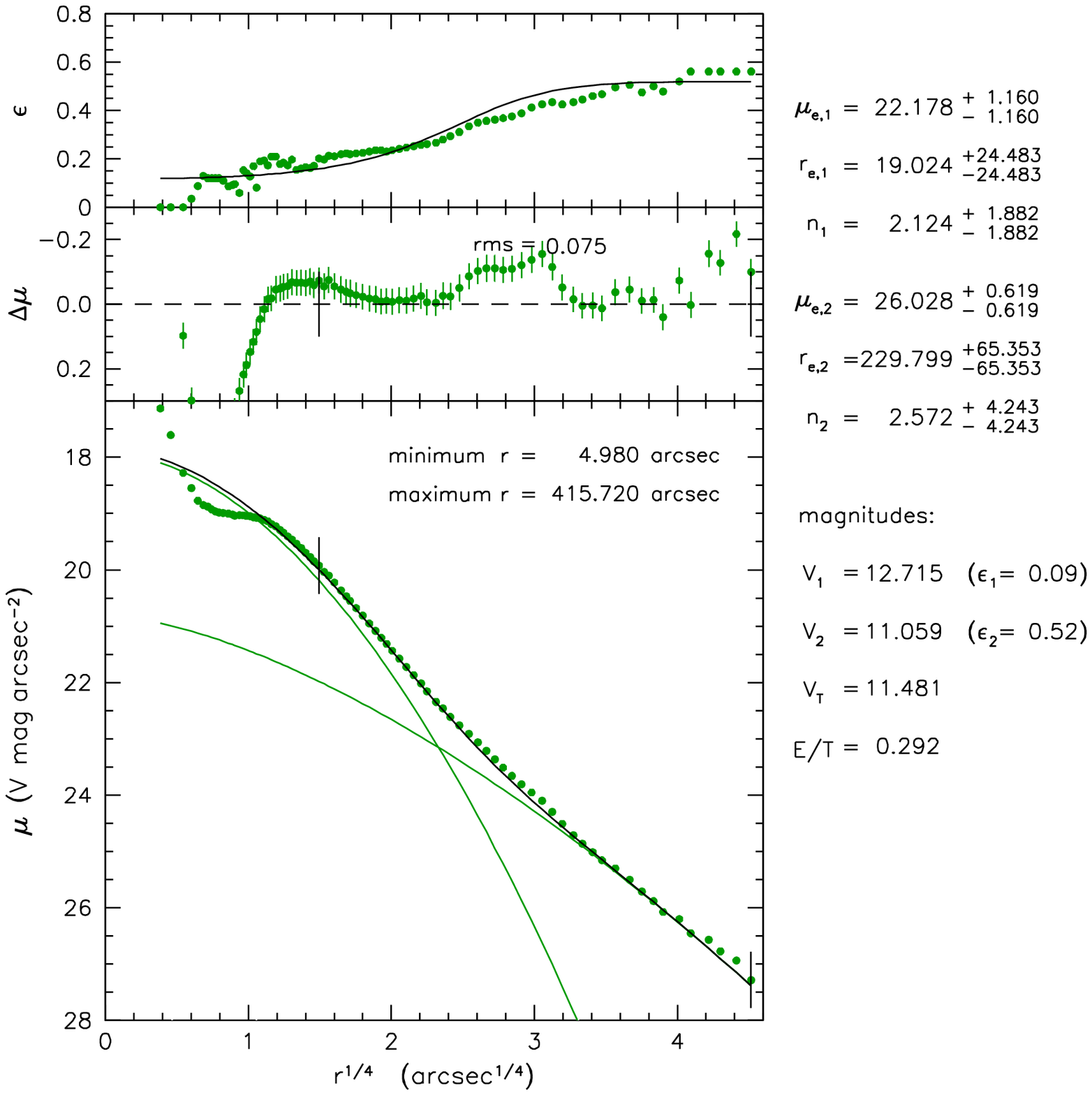}

\vskip 4.5truein

 \includegraphics{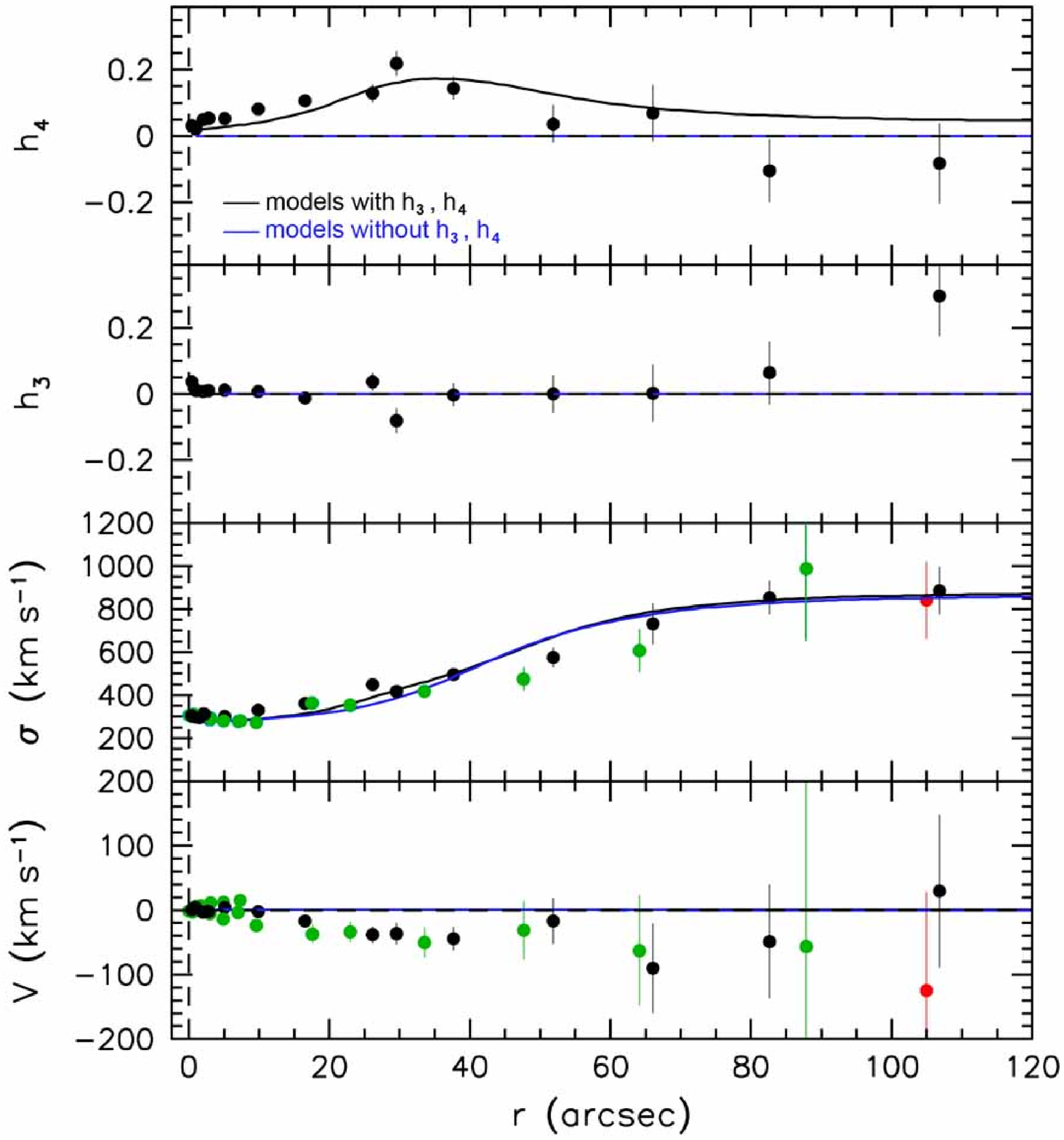}

 \includegraphics{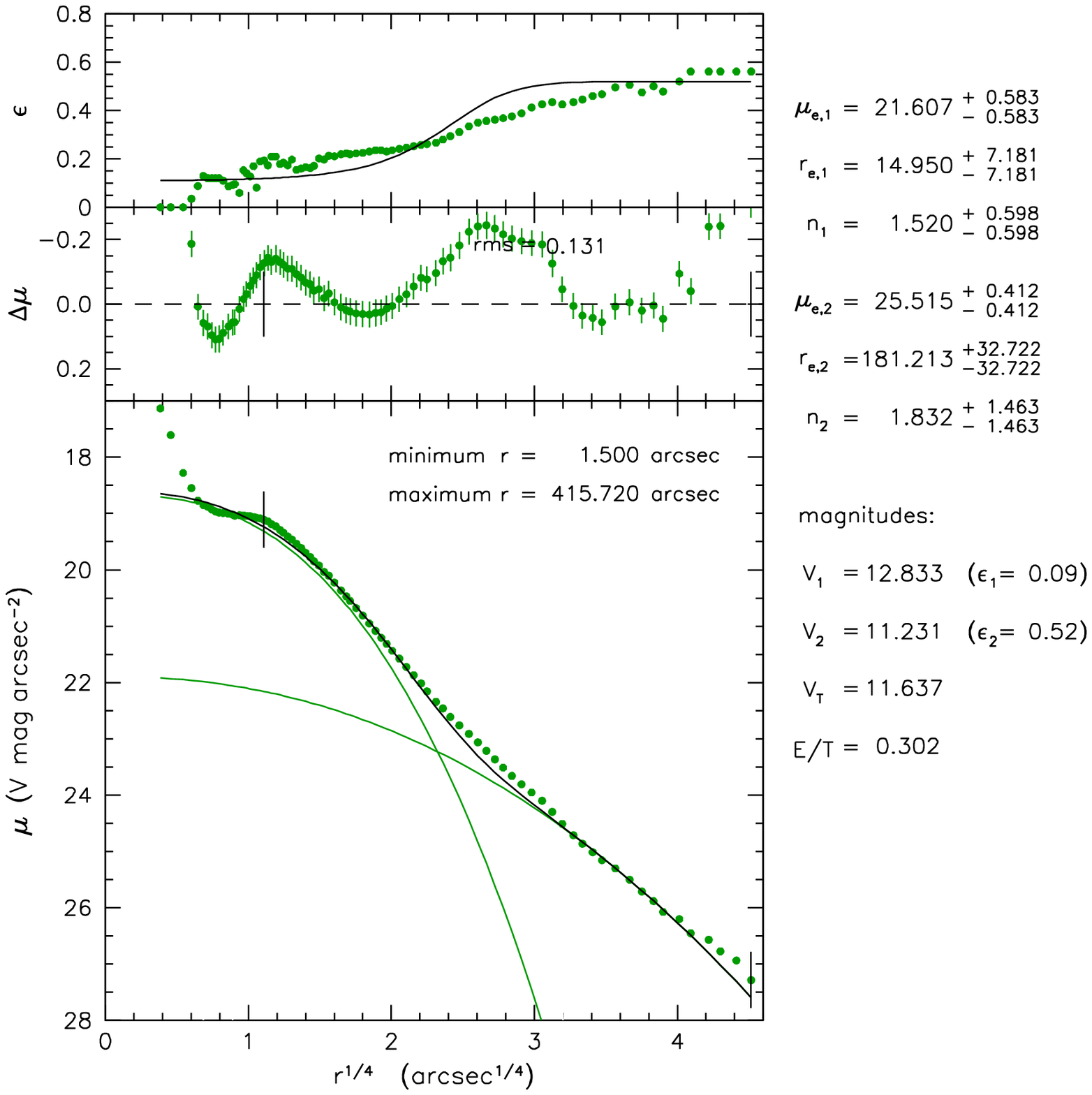}

 \includegraphics{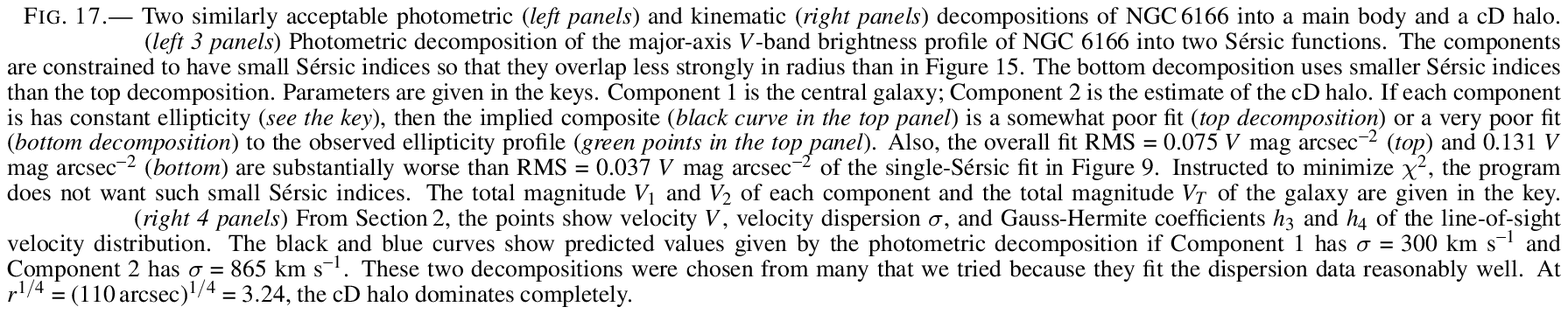}

\vskip 0pt

      Support for our photometric\ts$+${\ts}kinematic decomposition~is provided by the result in
Figures\ts15 and 17 that the predicted $h_4 > 0$ agrees with the observations at radii $r \lesssim 40^{\prime\prime}$.
At larger radii, the predicted $h_4$ remains positive but trends toward zero.  The spectra there are too noisy 
to provide reliable constraints.

      \hbox{~~~Fig.\ts15--17 suggest that the main body of NGC\ts6166 contains} \noindent $\sim$\ts30\ts$\pm$2\ts\% 
and the cD halo contains $\sim$\ts70\ts$\pm$\ts2\ts\% of the total luminosity.  The formal error is probably 
an underestimate.

\vfill\eject\clearpage

      For the assumptions made in \S\ts3.6 to get total absolute magnitudes of $M_V,{\rm tot} = -23.86$ out to 
the last photometric data point or $-24.27$ extrapolated to infinity, the main body of NGC 6166 has $M_V \simeq -22.6$ 
or $-23.0$.
These are essentially identical to the absolute magnitudes of M{\ts}87 and NGC\ts4472 in the Virgo cluster (KFCB).  
The cD halo of NGC\ts6166 has $M_V \simeq -23.5$ or $-23.9$, i.{\ts}e., 0.3 -- 0.6 mag brighter than the brightest 
galaxy in the Virgo cluster.

\omit{(Parenthetical note: If{\ts}--{\ts}as in Fig.\ts15 -- 17{\ts}-- we interpret the total profile as
the sum of several components, then the S\'ersic indices of the individual components are smaller
than $n \simeq 8.3$ for the single-component fit in Figure 9.  In fact, the inner component shown in Figure 15 has
$n = 4.1$ and the outer component has $n = 4.7$.  Whenever one increases the number of S\'ersic
components, their indices necessarily get smaller.  The extreme version of this is multi-Gaussian
decomposition\ts-- when the number of components is made very large, then $n$ is forced to be very
small.  A Gaussian $n = 0.5$ is convenient for numerical reasons but is not unique.  Such
decompositions can be useful, precise representations of the photometry, as long as no attempt
is made to attach physical significance to individual components.)
}

\section{SPHERICAL JEANS MODELS}

      Our kinematic measurements allow a detailed study of the velocity distribution of the 
galaxy plus stellar halo and of the total mass distribution  including X-ray gas and dark matter.  
Orbit-superposition models (Schwarzschild 1979, 1982) are postponed to a future paper. 
Here, we explore the stellar velocity anisotropy using spherical Jeans models.

\vskip 0pt

      Figures 18 and 19 show results for Jeans-model fits to our photometry and $\sigma$ data.~We assume that 
dark matter (``DM'', including X-ray gas) is distributed as a 
non-singular pseudo-isothermal $\rho \propto (1+(r/r_c)^2)^{-1}$ (Kormendy\ts\&{\ts}Freeman\ts2015) or as an
NFW density profile (Navarro, Frenk, \& White 1996, 1997).  We choose the 
outer, circular-orbit rotation velocity $V_{\rm circ} = 1160$ km s$^{-1}$ of massless test particles in the halo 
to be consistent with the cluster dispersion of $819 \pm 32$ km s$^{-1}$.  Next, we assume that the stars have a 
Kroupa (2001) initial mass function with mass-to-light $M/L_V = 4$, based on the metallicity and age estimated 
in the next section and on
stellar population models of Maraston \etal (2003).  Then the only free parameter left is the scale length $r_s$
of the NFW profile or the core radius $r_c$ of the isothermal.  We vary this scale length until the mass density 
profile matches the one derived from the X-ray gas by Markevitch \etal (1999).  In this way, we derive a density 
profile over the full radius range (Figure 19) without yet using our kinematic data on NGC 6166.  Finally,
we vary the velocity anisotropy as a function of radius ({\it middle panel\/} of Figure 18) until we reproduce the 
observed velocity dispersion profile ({\it bottom panel\/} of Figure 18).  Although the isothermal sphere and the NFW 
DM profiles are quite different, especially at $r \leq 16$ kpc, the anisotropy profiles are qualitatively
similar.  That is, the total density profile and the dispersion profile together determine the anisotropy profile.

     The important result is observed at radii $r \sim 20^{\prime\prime}$ to
$70^{\prime\prime}$, where $\sigma$ rises from the galaxy value of 300 km s$^{-1}$ to the cluster
value of $>$ 800 km s$^{-1}$.  In this radius range, the tangential velocity dispersion is larger
than the radial~one, $\sigma_t > \sigma_r$. We were unable to change this result by varying the DM profile.  
The observed dispersion rises so rapidly that it is necessary to ``boost'' the line-of-sight component by
increasing $\sigma_t$. Our conclusion that $\sigma_t > \sigma_r$ in the inner part of the cD halo of
NGC 6166 is consistent with the suggestion that cD halo stars are the debris torn off of individual cluster
galaxies by fast collisions (see, e.{\ts}g., Puchwein \etal 2010). 

      In recent years, the growth of S\'ersic $n > 4$ halos of giant-core-boxy-nonrotating 
elliptical galaxies (Kormendy et al.~2009) has also been attributed to accumulated debris 
from minor mergers (e.{\ts}g.,
Naab \etal 2009;
Hopkins \etal 2010;
Oser \etal 2010, 2012;
Hilz \etal 2012;
Hilz, Naab, \& Ostriker 2013).
The relationship between these $n > 4$ halos\ts-- which manifestly belong to the galaxy -- and the
$n \simeq 8$ halo of NGC 6166\ts--{\ts}which manifestly belongs to the cluster\ts-- is a
puzzle addressed in the following sections.

      At large $r$, the data hint that $\sigma_r > \sigma_t$.  This as a preliminary result.
If it is correct, it could be a sign that even at $\sim$\ts100 kpc, we reach radii where 
infall from the filaments of the cosmic web affect the velocity distribution (cf.~Biviano \etal 2013
and Wu \etal 2014).  

      The isothermal halo parameters derived here, $r_c = 20${\ts}kpc and $\rho_0 = 6.2 \times 10^{-2}$
$M_\odot$ pc$^{-3}$ for $M_B \simeq -23$, deviate~from~the DM parameter correlations found by Kormendy \&
Freeman (2015).  The DM halo of NGC\ts6166 is more compact (e.{\ts}g., higher in projected surface density)
than expected from halos of late-type galaxies.  However, it is consistent with scaling relations for
cluster halos (Chan 2014), and its parameters agree with those derived for Abell 2199 by Chen \etal (2007).

\vfill

\begin{figure}[ht!]

\figurenum{18}

\vskip 5.02truein

\includegraphics{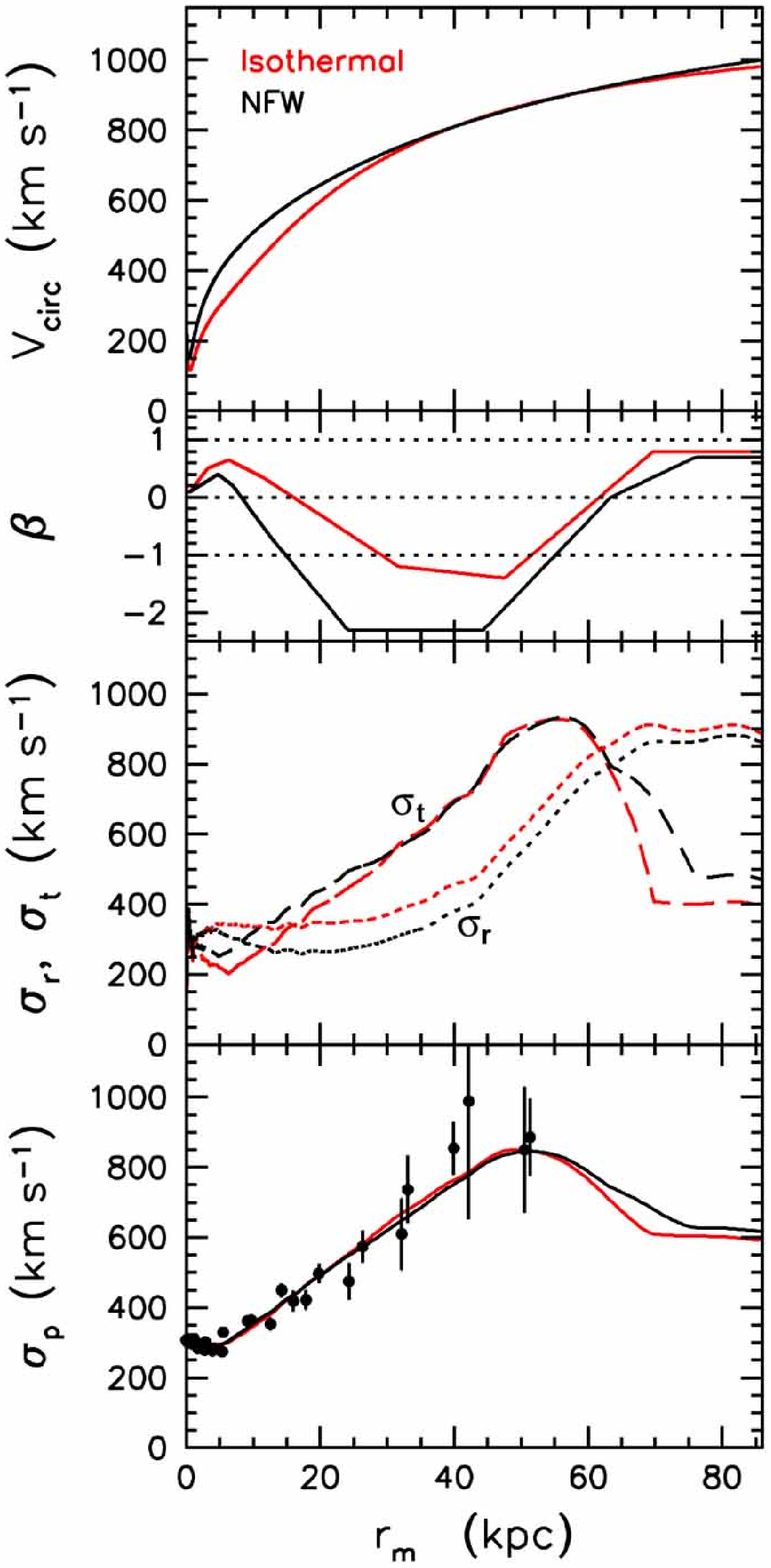}

\figcaption[]
{Kinematics of our best-fitting spherical Jeans model of the mass distribution.
The bottom panel compares to our data the model projected line-of-sight velocity dispersion 
of stars as a function of radius.  The next panel upward
shows the radial and tangential components $\sigma_r$~and~$\sigma_t$, respectively,
of the unprojected velocity dispersion.  For readers who prefer to express the velocity anisotropy as 
$\beta \equiv 1 - \sigma_t^2/\sigma_r^2$, $\beta(r)$ is shown in the
third panel.  The top panel shows the circular-orbit rotation velocity of 
massless test particles embedded in the mass distribution.  Results are shown for
two dark matter (DM) halo models, the nonsingular isothermal and the Navarro, 
Frenk, \& White (1996, 1997) profile.  The corresponding volume density profiles are shown in Figure 19.
The stellar mass distribution is derived from the stellar light profile using $M/L_V \simeq 4$ derived 
from stellar population models (Maraston \etal 2003).  
Results on the stellar velocity anisotropy are robust to changes in the
halo model:
$\sigma_r > \sigma_t$ near the center, where $\sigma \sim 300$ km s$^{-1}$ is dominated by the galaxy;
$\sigma_r < \sigma_t$ at intermediate radii, where $\sigma$ climbs to the cluster dispersion, and
$\sigma_r > \sigma_t$ at large radii. 
}
\end{figure}

\vskip -15pt

\eject

\cl{\null}


  \vskip 2.9truein

\begin{figure}[hb!]

\figurenum{19}

\includegraphics{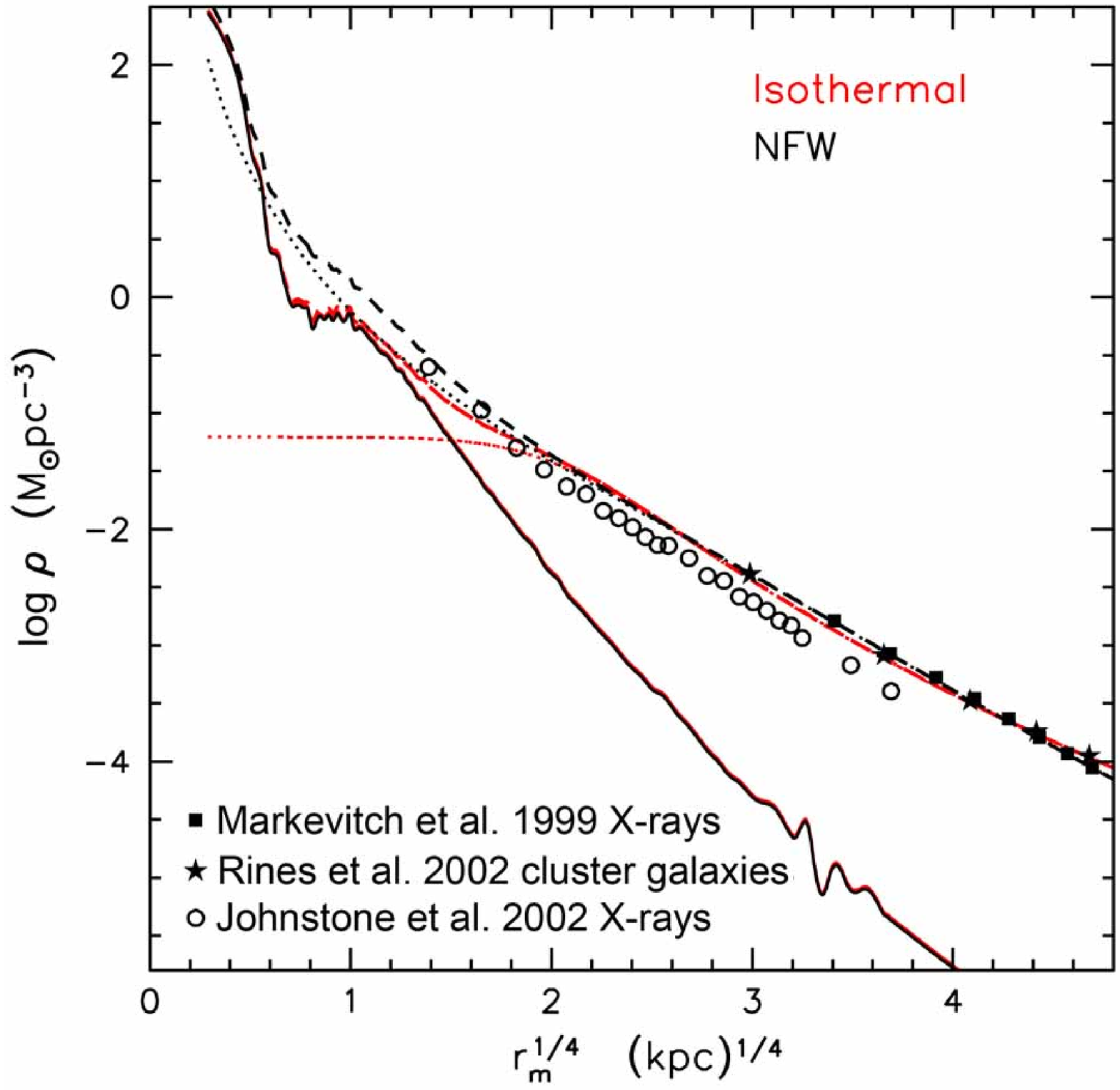}

\figcaption[]
{Volume mass densities in NGC 6166 given by the spherical Jeans models in Figure 18 for the 
isothermal and NFW dark matter halos.  Here, $r_m$ is the geometric mean of major- and minor-axis
radii.  The DM scale radii are $r_s = 150$ kpc for NFW and $r_c = 20$ kpc for the isothermal sphere.
The outer total density in our models is fitted to the filled stars, i.{\ts}e., the mass density
derived from the X-ray halo of the cluster Abell 2199 by Markevitch \etal (1999).  The density 
similarly derived from X-ray emission by Johnstone \etal (2002) and the mass density as derived from
the dynamics of cluster galaxies by Rines \etal (2002) are also shown for comparison.
}
\end{figure}

\cl{\null}

\vskip -35pt

\cl{\null}

\section{HEAVY ELEMENT ABUNDANCES}

      Our high $S/N$-ratio spectra also allow us to probe stellar population diagnostics out into the
part of the cD halo where the velocity dispersion is climbing to the cluster value.  In Figure 20,
we use the Lick Observatory spectral line indices 
(Faber \etal 1985;
Gorgas \etal 1993;
Worthey \etal 1994;
Trager \etal 1998;
Lee \& Worthey 2005;
Lee \etal 2009)
to estimate Fe abundances and [Mg/Fe] -- i.{\ts}e., $\alpha$-element -- overabundances in the main body
and cD halo of NGC 6166.

      Overabundances with respect to solar values of $\alpha$ elements such as Mg imply
short star formation~time~scales.  Rapid enrichment of $\alpha$ elements follows starbursts when high-mass 
stars die as supernovae of Type II.  Alpha elements then get diluted by Fe once there is time for lower-mass stars 
to die as white dwarfs and subsequently blow up as Type{\ts}I supernovae.  After that, [$\alpha$/Fe] can never 
be enhanced again.  Therefore super-solar [$\alpha$/Fe] abundances imply that essentially all star formation was 
completed in \lapprox \ts1{\ts}Gyr.
(Worthey \etal 1992;
Terndrup 1993;
Matteucci 1994;
Bender \& Paquet 1995;
Thomas \etal 1999, 2002, 2005).

      Kormendy \etal (2009) show that [$\alpha$/Fe] (over)abundance participates in the dichotomy
(see Kormendy \& Bender 1996; Kormendy 2009 for brief reviews) between giant, nonrotating, anisotropic
ellipticals that have boxy isophotes and cuspy cores and lower-luminosity ellipticals that rotate enough
to be more nearly isotropic and that have disky isophotes and (in general) central extra light components.
They argue that rotating-coreless-disky ellipticals formed via at least one wet merger in which a starburst
constructed the central extra component.  And they argue that nonrotating-core-boxy ellipticals -- which 
are embedded in large amounts of X-ray-emitting gas -- formed most recently via dry major mergers (plus, we
now believe,~minor-merger~addition~of~outer~halos), protected from late star formation by their X-ray gas halos.  
Kormendy \etal (2009) found that [$\alpha$/Fe] is enhanced in nonrotating-core-boxy ellipticals but not in
rotating-coreless-disky ellipticals (cf.~Thomas \etal 2005, 2010).  Essentially all star formation
was completed very early in these galaxies.  NGC\ts6166 is a giant core elliptical (Figures 9 -- 13).

      This machinery provides a partial test of our picture that cD halos consist of tidal debris torn from 
cluster galaxies.  If [$\alpha$/Fe] is super-solar in the main body of NGC 6166 but near-solar in its cD halo 
and in smaller cluster galaxies, then this strongly supports the idea that cD halos consist of tidal debris.  
In contrast, if [$\alpha$/Fe] is super-solar in both the main body and the cD halo of NGC 6166, then this is 
consistent with our picture but does not prove it.  Rather, that result is interesting because it suggests 
that star formation was switched off early in all galaxies that contribute to any part of NGC 6166.  If so,
then this result predicts that many (not necessarily all) smaller galaxies in the cluster are [$\alpha$/Fe]
enhanced, too.   We do not have such data.  But if spectroscopy of the smaller
galaxies shows that they have solar [$\alpha$/Fe] abundances whereas the cD halo has super-solar [$\alpha$/Fe],
then this argues {\it against\/} our picture and instead supports a picture in which all of the cD including
its halo forms early via some special process.  We carry out the first part of the test, measuring only
NGC 6166.

      Figure 20 shows our measurements in NGC 6166 of the Fe mean equivalent width versus that of Mg{\ts}b.  
The iron lines used are Fe $\lambda$ 5270 and 5335 \AA.  Colors encode radii
whose corresponding velocity dispersions are given in the key.  Thus, the red and orange points are dominated
by light from the central galaxy, whereas the green point and especially the blue point increasingly measure
stars in the cluster-$\sigma$ cD halo.

      Also shown are black points at specific metallicities and population ages ({\it lower key\/})
for three [$\alpha$/Fe] abundance~ratios.  The points are connected by solid lines for ages of 
$\sim$\ts10{\ts}Gyr and by dashed lines for ages of $\sim$\ts3{\ts}Gyr.  The models are from
Thomas, Maraston, \& Bender (2003);
Maraston \etal (2003); and
Thomas \& Maraston (2003).

\begin{figure}[hb!]

\figurenum{20}

\vskip 3.25truein

\includegraphics{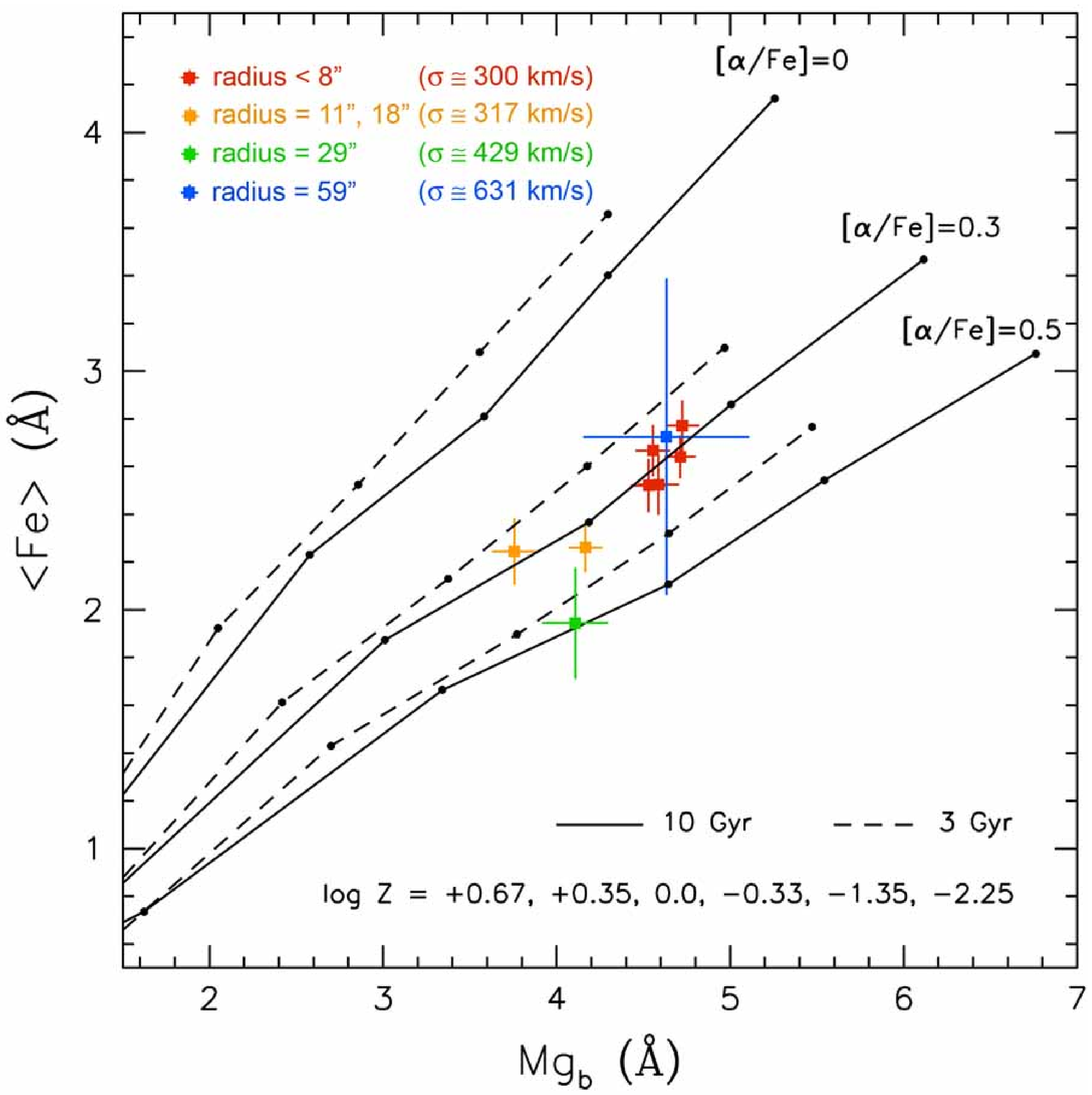}

\figcaption[]
{Correlation of $<$Fe$>$ equivalent width with that of the Mg{\ts}b lines along the central slit position
of NGC 6166 as a function of radius.  The measurements are on the Lick system.  Model lines for various
stellar population ages, metal abundances $Z$, and $\alpha$ element overabundances [$\alpha$/Fe] are
also shown (see the text for sources).
}
\end{figure}

\vfill\eject

      We conclude that the central, $\sigma \simeq 300$ km s$^{-1}$ parts of NGC\ts6166 are old
and substantially more metal-rich than solar.  They have [$\alpha$/Fe] $\simeq 0.3 \pm 0.1$.  These observations 
are consistent with the E{\ts}--{\ts}E dichotomy as discussed in KFCB.

      At radii $r \simeq 11^{\prime\prime}$ to 18$^{\prime\prime}$, where $\sigma$ begins to rise, 
the abundance is more nearly solar but [$\alpha$/Fe] remains high.

      In the inner cD halo, where rising $\sigma$ indicates that we see
substantial ({\it green point\/}) and mostly ({\it blue point\/}) cluster halo ($\sigma \sim 800$
km s$^{-1}$) stars, the metallicity remains at least as high as at intermediate radii and [$\alpha$/Fe]
remains at \gapprox \ts0.3.  This is consistent with but does not prove that the cD halo consists of 
tidally liberated galaxy debris. 

      Similar tests have been carried out in normal Es (e.{\ts}g., Coccato \etal 2010).  Greene \etal (2012, 2013) 
study 33 ellipticals with central $\sigma \geq 150$ km s$^{-1}$, not quite high enough to single out core galaxies.  
Quoting from the latter paper\ts: ``the typical star at $2 r_e$ is old ($\sim$10{\ts}Gyr), relatively \hbox{metal-poor}
([Fe/H] $\approx -0.5$), and $\alpha$-element enhanced ([Mg/Fe] $\approx 0.3$). \dots~Stars at large radii have 
different abundance ratio patterns from stars in the {\it center\/} of any present-day galaxy, but are similar 
to average Milky Way thick disk stars.  Our observations are consistent with a picture in which the stellar 
outskirts are built up through minor mergers with disky galaxies whose star formation is truncated early 
($z$\ts$\approx$\ts1.5\ts--\ts2).'' 

\section{Evolutionary History of NGC\ts6166 and Abell\ts2199}

      If the cD halo of NGC\ts6166 had its star formation quenched in \lapprox1{\ts{Gyr, then the environs 
of NGC\ts6166 have been special for a long time. This has implications for cD formation:

In recent years, there has been a substantial convergence from many lines of research on a consistent and plausible 
picture of what quenches star formation in general and especially in giant galaxies such as NGC 6166.  The essential
idea is often called ``$M_{\rm crit}$ quenching:'' a total galaxy or cluster mass 
$M$\gapprox$M_{\rm crit} \sim 10^{12}$ $M_\odot$  is required to hold gravitationally onto large amounts 
of hot, X-ray-emitting gas, and the hot gas quenches star formation.  Essentially equivalent pictures have been reached 
(1) via theoretical studies of cosmological gas accretion onto large potential wells
    (Dekel \& Birnboim 2006, 2008);
(2) via semi-analytic modeling
    (Cattaneo \etal 2006, 2008, 2009); 
(3) via studies of galaxies in the high-redshift universe (e.{\ts}g.,
    Faber \etal 2007;
    Peng \etal 2010;
    Knobel \etal 2014);
(4) via studies of physical differences between the two kinds of elliptical galaxies 
    (KFCB; Kormendy \& Bender 2012), and
(5) via studies of AGN feedback in relation to the demographics of supermassive black holes
    and the properties of their host galaxies (Kormendy \& Ho 2013). 
Note that the value of $M_{\rm crit}$ is somewhat higher at higher $z$ because of higher cold gas fractions there
(see the Dekel \& Birnboim papers).

      Peng \etal (2010) provide the clearest description: They distinguish mass-driven quenching from
environmentally-driven quenching and quenching related to bulge formation.  Like Knobel \etal (2014), 
we suggest that mass-driven~and environmentally-driven quenching are fundamentally the same process;
in mass-driven quenching, the quenched galaxy owns its own hot gas, whereas in environmentally-driven
quenching of satellite galaxies, the gas that does the work belongs to the parent giant galaxy or cluster.  
Both processes together are equivalent to the ``maintenance-mode AGN feedback'' discussed in Kormendy \& Ho (2013).
Again, the quenching is done by the hot gas, and the process that keeps it hot (AGN feedback is one possibility) 
is somewhat secondary.  Quenching by hot gas is the essential process that is relevant here.  (Peng's ``quenching 
associated with bulge formation'' is equivealent to Kormendy \& Ho's ``quasar-mode feedback''.)  

      The X-ray halo needed for $M_{\rm crit}$ quenching is present in Abell 2199 (e.{\ts}g.,
Markevitch \etal 1999;
Johnstone \etal 2002;
Kawaharada \etal 2010).  
However, the implications of our results are broader than this:

      In general, we expect that a cluster grows as galaxies and galaxy groupings fall into it
that are sufficiently sub-$M_{\rm crit}$ to have had prolonged star formation histories.  As they and their
stars get added to NGC 6166, it is natural to expect that the resulting halo would not be as 
$\alpha$-element enhanced as the main body of the galaxy.  Simulations suggest that cD halo stars are somewhat
older than typical stars in the galaxies that contribute to the halo 
(Murante \etal 2004;
Puchwein \etal 2010).
Also, simulations by Murante \etal (2007) suggest that the inner parts of cD halos -- this certainly includes
the parts of NGC 6166 that we have measured -- ``come from the [merger] family tree of the [parent galaxy]'';
that is, from galaxies that share the immediate history of the central galaxy.
And simulators agree that the halo tends to be contributed by the most massive cluster galaxies; their
star rofmation was presumably quenched early.
Still, if even the debris halo of NGC\ts6166 is $\alpha$-element enhanced, then this suggests that the environs 
of the galaxy -- including that of the progenitors that contributed to its cD halo -- constituted
a deep enough gravitational potential well to allow star formation to be quenched rapidly.  And this suggests a
solution to the following puzzle:

      Why does NGC 6166 have such a high-surface-brightness halo of intracluster
stars when apparently richer and denser clusters such as Coma have weaker cD characteristics?  Note that the
velocity dispersion has already risen significantly in NGC\ts6166 at $r \sim 30^{\prime\prime}$ (Figures 4 and 5), 
where the surface brightness is $\sim$\ts22.5\ts$V${\ts}mag{\ts}arcsec$^{-2}$ (Table\ts3).  Evidently~the 
processes freed the intracluster stars happened less strongly or for a shorter time in Coma than in
Abell\ts2199. Why? 

      Coma may have formed relatively recently\ts--{\ts}is, in fact, still forming now, with the imminent 
accretion of the NGC\ts4839 grouping.~In contrast, Abell\ts2199 looks less dense than Coma does now, but 
the central few hundred kpc evidently has been a massive enough environment to allow the early quenching 
of star formation.  It may also have been dense enough to allow cD halo formation
processes to operate efficiently for an unusually long time.  

      A more speculative remark follows from the large core radius of NGC\ts6166 (Figure 13).  There
is a tight correlation between the light and mass ``deficit'' that defines the core phenomenon and the
measured mass $M_\bullet$ of supermassive black holes (Kormendy \& Bender 2009).  The canonical interpretation is 
that cores are created when supermassive black hole binaries produced in major, dry mergers fling stars away from
the center as they decay toward an eventual merger (e.{\ts}g., 
Ebisuzaki \etal 1991; 
Faber \etal 1997;
Milosavljevi\'c \& Merritt 2001;
Milosavljevi\'c \etal 2002; 
Merritt 2006). 
If the $M_\bullet$--core correlation is valid for NGC\ts6166, then the core light deficit $M_{V,\rm def} \simeq -19.67$
corresponds to a BH mass $M_\bullet = 4.1^{+1.4}_{-1.1} \times 10^9$ $M_\odot$.  The core radius is unusually large,
but the core surface brightness is unusually small.  So the light deficit and $M_\bullet$ are almost the same as 
those of M{\ts}87.  Still, Abell 2199 is one of the most plausible environments in which episodic AGN feedback 
could help to keep its hot gas hot (Fabian 2012).  And the early quenching of star formation together with the 
long history of cluster dynamical evolution may be connected with the unusual properties (large radius but
low surface brightness) of the core of NGC\ts6166.

\section{Implications for \lowercase{c}D Formation Mechanisms}

      This observational paper does not fully review the large literature on possible formation
mechanisms for cD galaxies.  We restrict ourselves to the most basic conclusions from our new 
results and concentrate on formation~of~the~cD~halo.  Suggested mechanisms are divided 
into three categories:

\subsection{Star Formation in Cooling Flows in X-Ray Gas}

      Are cD halos made of stars that rain out of cooling flows in hot gas (see Fabian et al.\ts1991;
Fabian 1994 for reviews)?  This idea was entertained in the heyday of the cooling-flow problem, 
when we observed large amounts of X-ray-emitting, hot gas in clusters but could not measure temperature 
profiles.  Absent heating processes, hot-gas cooling times near the centers of many clusters 
and individual galaxies are short.  In clusters, $10^2$\ts--\ts$10^3$~$M_\odot$ yr$^{-1}$ of baryons should 
rain~out~of the hot gas, presumably by star formation.  To~escape~detection, the initial mass function 
would have to be truncated above $\sim$1\ts$M_\odot$ (Fabian \etal 1991).  We have never directly observed such 
star formation in any environment (Bastian \etal 2010). 

      This possibility is now regarded as a non-starter.  The~main reason is that we now can measure gas 
temperature profiles, and we find that temperatures decrease only modestly to a floor at 
$kT \sim 1${\ts}keV.  In particular,~we~do~not~see~the~strong emission lines from Fe XVII that would be 
our~signal~that gas has cooled below 0.7~keV~(see~Fabian~2012~for~review).  So the cooling flow problem has morphed
into a different question:  What keeps the gas hot?  At least three heating processes are hard to avoid.  
Most popular is heating by AGN feedback
(Fabian 2012;
Kormendy \& Ho 2013; and
Heckman \& Best 2014 provide reviews).  
Also, gas from the cosmological web that falls into objects with masses 
$M > M_{\rm crit} \sim 10^{12}$\ts$M_\odot$ accelerates so much 
that a shock forms where it impacts the static intergalactic or intracluster medium; this heats the hot gas 
from the outside inward 
(Birnboim \& Dekel 2003;
Kere\v s \etal 2005;
Dekel \& Birnboim 2006, 2008).  
This is an aspect of $M_{\rm crit}$ quenching of star formation.
Finally, dying stars eject large amounts of mass into the intracluster medium at the kinetic 
temperatures of stars in galaxies and galaxies in clusters (e.{\ts}g., Ostriker 2006).
All three mechanisms are likely to be important.
In this picture, episodic cooling fuels the AGN and switches it on long enough to allow
it to keep the center of the hot gas hot (Fabian 2012).  Small amounts of star formation may be connected
with these events, and small amounts of star formation are seen in brightest cluster galaxies (e.{\ts}g., Liu \etal
2012).  But no compelling argument suggests that large amounts of star formation occur in clusters at 
radii where we see cD halos.  Also, our observation that the cD halo of NGC 6166 is $\alpha$-element enriched
precludes the idea that prolonged, in-situ star formation made a significant fraction of the light that we see in the halo.

\subsection{Processes Intrinsic to the Origin of the Central Galaxy}

      Do cD halos originate as an integral part of the formation of the central galaxy?  For example,
could a specialized history of galaxy mergers make both the central and halo parts of a cD galaxy together?

      Our phrasing is somewhat different from the question that dominated work on brightest
cluster galaxies (``BCGs'') in the 1970s -- 1990s (see Tremaine 1990 for a review).~Then, the emphasis
was on observational~hints that BCGs in general (i.{\ts}e., including but not limited
to cDs) are inconsistent with statistical expectations based on the luminosity functions $\phi(L)$ of fainter
galaxies in the cluster.  If  $\phi \propto L^\alpha \exp(-L/L^*)$ with characteristic luminosity 
$L^*$ (Schechter 1976), then BCGs with $L \sim 10 L^*$ are statistically too bright to be drawn from the 
populations of other galaxies
in the clusters (see Figure 1 in Binggeli 1987 for an evocative illustration).  In many papers, cDs and non-cD 
BCGs were discussed together.  Given the observation that cD halos are approximately as bright as or brighter than the 
central parts of the galaxies (e.{\ts}g., Seigar \etal 2007), this essentially ensures that BCGs as a class 
will look especially luminous (Tremaine \& Richstone 1977).

      As some authors have done since the beginning of this subject, we differentiate between the main bodies of 
cDs and their halos.  In NGC 6166, we separate them operationally as having $\sigma$ $\simeq$300 km s$^{-1}$ 
and $\sigma \sim 832$ km s$^{-1}$, respectively.  How the main bodies of BCGs form and how cD halos form may be
separate questions.

      When their halos are inventoried separately, it is much less obvious that the main bodies of cDs are unusual
enough to imply formation physics that is different from that of other cluster galaxies.  The new observations
in this paper do not speak strongly to this issue, and we do not discuss it in detail.  Ways in
which the main body of NGC\ts6166 is {\it not\/} unusual are the subject of Sections 3.2\ts--\ts3.5.  Except for
its unusually large and low-surface-brightness core (discussed in the previous section), the main body of NGC\ts6166
is rather like M{\ts}87 (a marginal cD) but also like the other giant-core-boxy ellipticals in the Virgo cluster.
Quantitative differences (Section 3.6) are mainly due to the cD halo of NGC\ts6166.  However, we note here
one additional observation that {\it does\/} imply something special about cD-like galaxies:

      Prototypical of a compelling but mysterious phenomenon, M{\ts}87 has an unusually large number 
of globular clusters for its galaxy luminosity.  Harris \& van den Bergh (1981) introduced the
specific globular cluster frequency $S_N$ as the number of globulars per unit absolute magnitude $M_V$\ts=\ts$-15$
of galaxy luminosity.  Measurement of $S_N$ is tricky for many reasons (e.{\ts}g., galaxy distances are uncomfortably
large, so we see less deeply into cluster luminosity functions than we would~like), but the conclusion that 
$S_N \sim 10$ is factors of several larger for M{\ts}87 and for some other BCGs (e.{\ts}g.,
NGC 1399: Hanes \& Harris 1986; Harris \& Hanes 1987;
NGC 3311: Harris 1986;
see Harris, Harris, \& Alessi 2013 for the most recent summary) has withstood the test of time.  
The number of globular clusters in NGC\ts6166 is $N_{\rm GC} = 17,000 \pm 4000$ (Harris \etal 2013).  With
respect to the absolute magnitude of the main E-like part of NGC\ts6166, this implies that $S_N \simeq 12$.
If instead we normalize $N_{\rm GC}$ by the total luminosity including the cD halo, then $S_N \sim 4$.
This is still slightly higher than the canonical numeber of 1\ts--\ts2 for $L^*$ ellipticals.
As discussed, for example, in Burkert \& Tremaine (2010), this is one indication
that the early evolution of the objects that later assembled into these BCGs (some of which are
clearly cDs and others of which are just giant ellipticals) was already special.  
This theme of an early, special environment in which NGC\ts6166 and its cD halo formed was
discussed in \S\ts7.

\subsection{\lowercase{c}D Halo Formation by Stellar-Dynamical Processes Inherent to Clusters}

      Our observations are most consistent with the now favored picture that cD halos are constructed by 
stellar-dynamical processes that are inherent to cluster evolution.~The~main~body forms by the usual 
hierarchical clustering and galaxy merging, especially in smaller group precursors to present-day, rich clusters.  
In the process, violent relaxation splashes some stars to large radii.  But the cD halo is added as a result 
of cluster-related processes such as the stripping of stars off of member galaxies by dynamical harassment 
and the cannibalism and destruction of dwarf galaxies in minor mergers.  This picture was originated by 
Gallagher \& Ostriker (1972) and by Richstone (1975, 1976) and has now been greatly elaborated in many papers,
both observational (see the earlier papers on cluster background light and, e.{\ts}g.,
Bernstein \etal 1995;
Gonzalez, Zabludoff, \& Zaritsky 2005;
Arnaboldi \etal 2012;
Montes \& Trujillo 2014)
and theoretical (e.{\ts}g.,
Dubinski 1998;
Murante \etal 2004, 2007;
De Lucia \& Blaizot 2007;
Puchwein \etal 2010; and
Cui \etal 2014).

\subsection{Blurring the Distinction Between cD Galaxies and Elliptical Galaxies with Cores}

     Our observations (1) that the cD halo of NGC\ts6166 is more nearly at rest in Abell\ts2199 than is 
its central galaxy and (2) that this halo has the same velocity dispersion as the cluster galaxies 
support the idea that it consists of stars that were liberated from cluster members.
The high velocity dispersion implies that the cD halo is controlled by cluster gravity.  It is only by 
convention\ts--{\ts{and not because this is physically meaningful\ts--{\ts}that we call it the halo of NGC\ts6166.

      On the other hand, the outer parts of NGC\ts6166~and~the intracluster light merge seamlessly such
that the brightness profile outside the central core is well described by a single S\'ersic
function with index $n \simeq 8$.  In~this~sense, NGC\ts6166 qualitatively resembles other core-boxy-nonrotating
elliptical galaxies such as those studied in KFCB and emphasized in the SAURON/Atlas$^{\rm 3D}$ series of papers
(see Cappellari 2015 for a review).  The S\'ersic $n > 4$ halos of core-boxy-nonrotating ellipticals that are
not brightest cluster galaxies manifestly belong to the galaxy -- their velocity dispersions generally
decrease monotonically outward.

      This blurs the distinction between cDs and giant elliptical galaxies.  Perhaps they are more similar than we thought. 
The central puzzle about both kinds of galaxies is why~\hbox{$n > 4$}.  In contrast, many numerical simulations of major 
mergers of two similar galaxies robustly show that the scrambled-up remnants of the stars that were already present before 
the mergers have S\'ersic profiles with $n \sim 3 \pm 1$ (e.{\ts}g.,
van{\ts}Albada 1982;
Mihos\ts\&{\ts}Hernquist 1994;
Springel\ts\&{\ts}Hernquist 2005;
Naab \& Trujillo 2006;
Hopkins \etal 2009a, b).
These are precisely the S\'ersic indices observed for coreless-disky-rotating ellipticals, which are thought to be
formed in wet mergers during which starbursts grew the central extra light components (see Kormendy 1999 and KFCB
for observations and review and Hopkins \etal 2009a for the most detailed simulations).

      Maybe the main difference between cDs and core-boxy-nonrotating (but not cD) ellipticals is the degree 
to which clusters are dynamically old enough to have liberated enough stars from individual galaxies to make 
a detectable intracluster population.  It may also matter whether the large-$n$ halos formed in subgroups such
that the central galaxy controls their dynamics or conversely in high-$\sigma$, rich clusters at radii 
controlled by the cluster rather than the central galaxy.  An important goal of future work is to explore 
the reasons why cD galaxies and core-boxy-nonrotating ellipticals look so similar when their halo velocity 
dispersions point to significant differences in formation history.

\vfill\eject

\acknowledgments

      The spectra were taken with the Marcario Low-Resolution Spectrograph (LRS) and the Hobby-Eberly 
Telescope (HET).  LRS is named for Mike Marcario of High Lonesome Optics; he made the optics for the 
instrument but died before its completion.  LRS is a project of the HET partnership and the Instituto 
de Astronom\' \i a de la Universidad Nacional Aut\'onoma de M\'exico.  The HET is a project of the 
University of Texas at Austin, Pennsylvania State University, Stanford University, 
Ludwig-Maximilians-Universit\"at M\"unchen, and Georg-August-Universit\"at, G\"ottingen.
The HET is named in honor of its principal benefactors,William P. Hobby and Robert E. Eberly. 

      The images used for surface photometry came from the McDonald Observatory 0.8 m telescope, the 2 m telescope
of the Ludwig-Maximilians-Universit\"at at Wendelstein Observatory, and the Canada-France-Hawaii Telescope.

      We also used the digital image database of the Sloan Digital Sky Survey.  Funding for the SDSS and SDSS-II has been
provided by the Alfred P.~Sloan Foundation, the Participating Institutions, the National Science Foundation, the U.~S.~Department
of Energy, the National Aeronautics and Space Administration, the Japanese Monbukagakusho, the Max Planck Society, and the Higher 
Education Funding Council for England. The SDSS is managed by the Astrophysical Research Consortium for the Participating Institutions. 
The Participating Institutions are the American Museum of Natural History, Astrophysical Institute Potsdam, University of Basel,
University of Cambridge, Case Western Reserve University, University of Chicago, Drexel University, Fermilab, the Institute for 
Advanced Study, the Japan Participation Group, Johns Hopkins University, the Joint Institute for Nuclear Astrophysics, the 
Kavli Institute for Particle Astrophysics and Cosmology, the Korean Scientist Group, the Chinese Academy of Sciences (LAMOST), 
Los Alamos National Laboratory, the Max-Planck-Institute for Astronomy (MPIA), the Max-Planck-Institute for Astrophysics (MPA), 
New Mexico State University, Ohio State University, University of Pittsburgh, University of Portsmouth, Princeton University, 
the United States Naval Observatory, and the University of Washington. 

      This work makes use of the data products from the HST image archive and from the Two Micron All Sky Survey (2MASS).
2MASS is a joint project of the University of Massachusetts and the Infrared Processing and Analysis Center/California Institute 
of Technology, funded by NASA and the NSF.  STScI is operated by AURA under NASA contract NAS5-26555.

      We thank Tod Lauer for helpful conversations about cD galaxies.  He independently concluded
that NGC 6166 is well described by a single S\'ersic function.  We are also grateful to 
Magda Arnaboldi, 
Ortwin Gerhard, and 
Stella Seitz 
for helpful conversations.

      This work would not have been practical without extensive use of the NASA/IPAC Extragalactic Database (NED), which is 
operated by the Jet Propulsion Laboratory and the California Institute of Technology under contract with NASA.  We also used the 
HyperLeda electronic database (Paturel et al. 2003) at {\tt http://leda.univ-lyon1.fr} and the image display tool SAOImage DS9 
developed by Smithsonian Astrophysical Observatory.  Figure 1 was adapted from the WIKISKY image database at {\tt http://www.wikisky.org}.
Finally, we made extensive use of NASA's Astrophysics Data System bibliographic services.

      J.K.'s photometry work was supported by NSF grants AST-9219221 and AST-0607490.  M.E.C. was supported in part by a generous and 
much appreciated donation to McDonald Observatory by Mr.~Willis A.~Adcock.  Finally, J.K. and M.E.C were supported by the Curtis T.~Vaughan, 
Jr.~Centennial Chair in Astronomy. We are most sincerely grateful to Mr.~and Mrs.~Curtis T.~Vaughan, Jr.~for their support of Texas 
astronomy. 

Facilities: 
HET: Low-Resolution Spectrograph;
McDonald Observatory: 0.8 m telescope; 
Wendelstein 2 m telescope: Wide-field camera;
CFHT: Cassegrain camera; 
SDSS: digital image archive; 
HST: WFPC1;
HST: WFPC2;
HST: ACS; 
HST: NICMOS 

\end{document}